\documentclass[11pt, a4paper]{scrartcl}
\usepackage[]{natbib}
\setcitestyle{authoryear,open={(},close={)}}
\usepackage{indentfirst, setspace}
\usepackage{authblk}
\usepackage{bm, amsmath,amsthm,amssymb, graphics, graphicx, mathrsfs, multirow, lscape, pifont}

\usepackage[OT1]{fontenc}
\usepackage[colorlinks,citecolor=blue,urlcolor=blue,linkcolor=black]{hyperref}
\usepackage[font={footnotesize}]{caption}
\usepackage{lscape} 

\usepackage[skip=0.33\baselineskip]{caption}
\usepackage[flushleft]{threeparttable}
\usepackage{booktabs,rotating}

\makeatletter
\renewcommand\@biblabel[1]{#1.}
\makeatother

\usepackage[table]{xcolor}
\definecolor{lightgray}{gray}{0.9}

\begin{document}
\begin{center}
\textbf{\LARGE Intervention analysis for integer-valued autoregressive models
}\\[2ex]
\Large \textbf{Xanthi Pedeli}\footnote{Athens University of Economics and Business, Athens, Greece}  \textbf{and Roland Fried}\footnote{TU Dortmund University, Dortmund, Germany   \vspace{.2cm}

\noindent \textbf{Corresponding author:} \newline
Xanthi Pedeli, Athens University of Business and Economics, 76 Patision Street, 10434 Athens, Greece. E-mail: xpedeli@aueb.gr.}

\end{center} 
\noindent
\subsection*{Abstract}
We study the problem of intervention effects generating various types of outliers in an integer-valued autoregressive model with Poisson innovations. We concentrate on outliers which enter the dynamics and can be seen as effects of extraordinary events. We
consider three different scenarios, namely the detection of an intervention effect of a known type at a known time, the detection of an intervention effect of unknown type at a known time and
the detection of an intervention effect when both the type and the time are unknown. We develop $F$-tests and score tests for the first scenario. For the second and third scenarios we rely on the maximum of the different $F$-type or score statistics. The usefulness of the proposed approach is illustrated using monthly data on human brucellosis infections in Greece.\\
\noindent
\textbf{Keywords:} Count data; time series; innovation outlier; level shift; transient shift.

\section{Introduction}\label{sect:intro}
 
Detection and modelling of unusual events is crucial in time series analysis because their presence can strongly influence statistical inference, diagnostics and forecasting.
Following the seminal work of \cite{fox:1972}, several outlier detection and estimation methods have been proposed in the literature for both linear and non-linear time series, often under the assumption of Gaussian random variables, see for instance \cite{galeano:2013} and the references therein.
However, to the best of our knowledge, the topic has not been investigated thoroughly in the framework of integer-valued autoregressive (INAR) models. 

Integer valued autoregressive models have been introduced by \cite{mckenzie:1985} and \cite{alosh:1987} as a convenient way to capture the autoregressive structure of count time series while accounting for the discreteness of the data.  
Several extensions and generalizations of the first-order integer-valued autoregressive process have since been developed and are widely used nowadays \citep{davis:2016,weiss:2018}.

The integer-valued autoregressive process of order $p$, denoted briefly as INAR($p$), is defined as
\begin{equation}\label{eq:inarp}
Y_t=\sum_{i=1}^p\alpha_i\circ Y_{t-i}+e_t,\; t\in\mathbb{N},
\end{equation}
where $\{e_t\}$ is an innovation process consisting of a sequence of independent identically distributed nonnegative integer-valued random variables with finite mean and variance. 
Conditional on $Y_{t-i}$, $i\in\{1,\ldots,p\}$, the binomial thinning operator ``$\circ$" is defined as
\begin{equation*}
\alpha_i\circ Y_{t-i}=\left\{\begin{array}{ll}
\sum_{j=1}^{Y_{t-i}}X_{j,i}& X_{j,i}>0,\\
0,& \mbox{otherwise},
\end{array}\right.
\end{equation*}
where each counting series $\{X_{j,i}, \; j=1,\ldots,Y_{t-i}\}$ consists of independent identically distributed Bernoulli random variables, independent of $Y_{t-i}$, with success probability $\alpha_i$ \citep{steutel:1979}.
The counting series are assumed to be mutually independent for $i=1,\ldots,p$ \citep{du:1991} and the innovations $e_t$ are assumed to be independent of the thinning operations $\alpha_i\circ Y_{t-i}$ for all $t\in\mathbb{N}$. A unique stationary and ergodic solution of (\ref{eq:inarp}) exists if $\sum_{i=1}^p\alpha_i<1$, where $\alpha_i\in[0,1)$.

In the following, we focus on the parametric case that arises when the innovations are Poisson random variables with parameter $\lambda$. When $p=1$, 
the marginal stationary distribution of $Y_t$ is also Poisson with mean $E(Y_t)=\lambda/(1-\alpha)$. When $p>1$, the unconditional mean and variance of $Y_t$ are generally not equal so that the marginal stationary distribution of $Y_t$ is no longer Poisson although the innovations are Poisson distributed.

\cite{barczy:2010, barczy:2012} analyzed the effects of different types of outliers occurring at known time points on the conditional least squares estimators in case of Poisson INAR(1) models. Detection of additive outliers in Poisson INAR(1) time series has been treated by \cite{silva:2015} in a Bayesian framework. Additive outliers are often interpreted as effects of measurement errors as they change a single observation but do not enter the dynamics of the time series. We concentrate on other types of outliers which enter the dynamics and can be seen as effects of extraordinary events.  \cite{barczy:2010} consider an outlier model similar to ours, but their work is restricted to an analysis of conditional least squares estimation in the presence of innovation outliers, which are treated as deterministic effects at known time points. 
\cite{morina:2020} proposed an INAR($p$) model that allows for the quantification of an intervention while taking into account possible trends or seasonal behaviour. The suggested model assumes the special case that the intervention occurs at a known time point and affects all subsequent observations in the same way.

\begin{figure}
\centering
\includegraphics[scale=0.6]{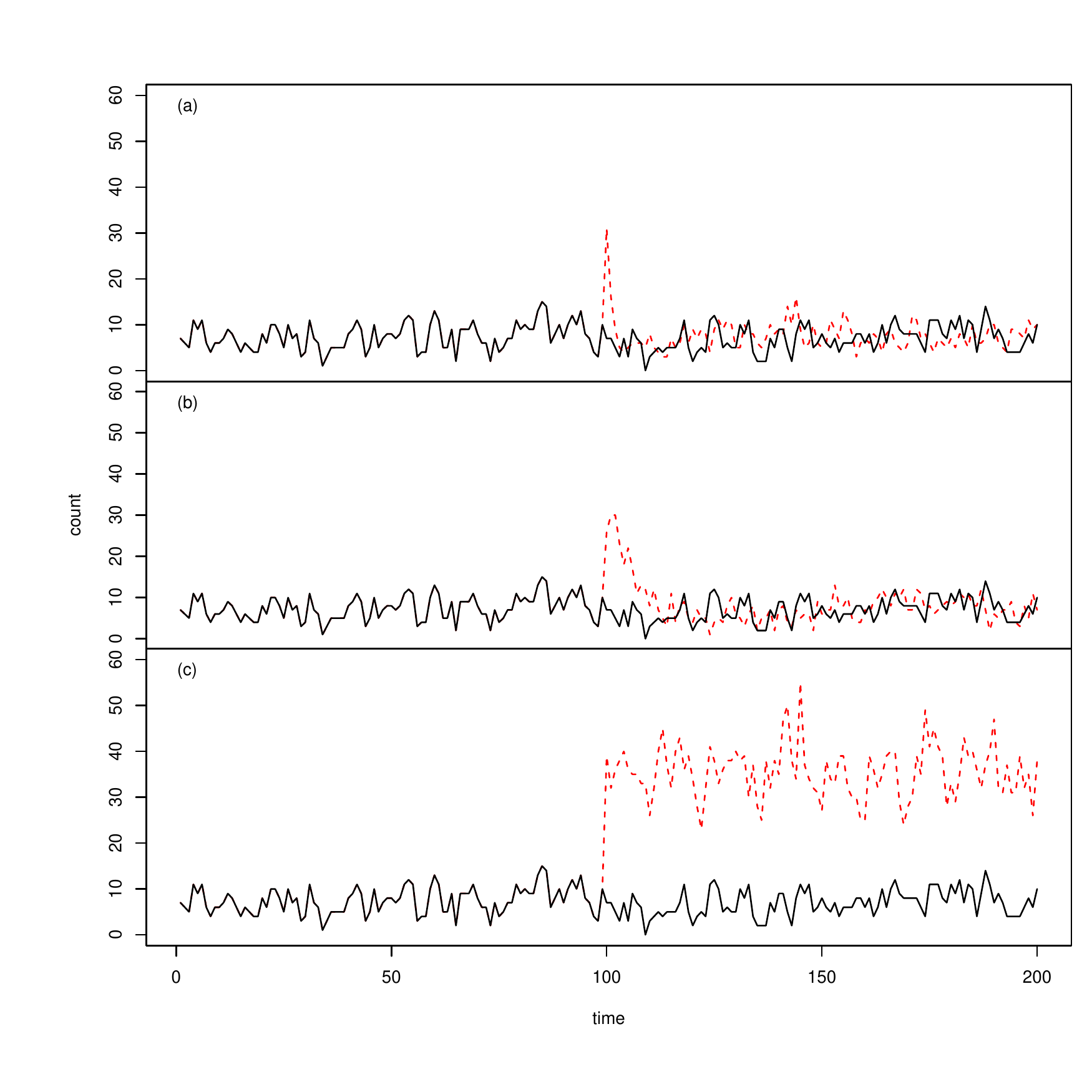}
\caption{Effects of different types of outliers of size $\kappa=20$ at time point $\tau=100$ on a realization of a Poisson INAR(1) process generated with $\alpha=0.3$, $\lambda=5$ and $n=200$. The solid black and dashed red lines correspond to the clean and contaminated processes (processes without and with contamination by outliers), respectively, where contamination is due to (a) an innovation outlier, (b) a transient shift with $\delta=0.8$ and (c) a level shift. }
\label{fig:effects}
\end{figure}

We aim at the detection of different types of effects including innovation outliers, transient shifts and level shifts at possibly unknown time points and use a somewhat different model formulation. More precisely, we extend model~(\ref{eq:inarp}) as follows:
\begin{equation}\label{eq:cont-inarp}
Y_t=\sum_{i=1}^p\alpha_i\circ Y_{t-i}+e_t+\sum_{j=1}^JU_{t,j},\; t\in\mathbb{N},\end{equation}
where $J$ is the number of intervention effects and $(U_{t,j}:t\in\mathbb{N})$, $j=1,\ldots,J$ are independent random variables denoting the effects of the different interventions on all time points. We assume that $(U_{t,j}:t\in\mathbb{N})$, $j=1,\ldots,J$ are independent of the thinning operations $\alpha_i\circ Y_{t-i}$ for all $t\in\mathbb{N}$. In addition, it is assumed that 
 $U_{t,j}\equiv 0$ for $t=0,\ldots,\tau_j-1$, and $U_{t,j}\sim Pois(\kappa_j\delta_j^{t-\tau_j})$ for $t=\tau_j,\tau_j+1,\ldots$, with $\tau_j$ and $\kappa_j$ denoting respectively the time point and the size of the $j$-th intervention and $\delta_j\in [0,1]$ controlling the effect of the intervention on the future of the time series after time $\tau_j$. For $\delta_j=1$ we get a permanent level shift starting at time $\tau_j$, for $\delta_j=0$ we get an innovation outlier, i.e., a single effect at time $\tau_j$ which spreads into the future according to the dynamics of the data generating process, and for $\delta_j\in (0,1)$ we get a transient shift in between the former two extremes which decays with rate $\delta_j$.
The effect of the above type of interventions on a realization of a stationary Poisson INAR(1) process is illustrated in Figure \ref{fig:effects}. 

The rest of the paper is organized as follows. Section~\ref{sect:cls-cml} discusses joint estimation of model parameters and intervention effects in the frameworks of conditional least squares and conditional maximum likelihood. Within the former framework, an $F$-test is developed for the detection of known types of interventions at known time points. For the same purpose, we suggest a score test in the framework of conditional maximum likelihood. The rejection rates and power of both tests are investigated through an extensive simulation study. In Sections~\ref{sect:unknowntype} and \ref{sect:unknowntime} we consider detection of intervention effects when either their type or both the type and time of intervention are unkown.
In the spirit of \cite{fokianos:2010}, we suggest in Section~\ref{sect:iterative} an iterative procedure for the detection, classification and elimination of multiple intervention effects. The procedure is illustrated through its application to simulated and real data series. Section~\ref{sect:discussion} concludes the article and outlines future research.

\section{Known types of intervention effects at known time points}
\label{sect:cls-cml}

If the number of interventions $J$, the time points $\tau_j$  of their occurrence and the types $\delta_j$ of the interventions $j=1,\ldots,J$ are known, then the conditional mean $E(Y_t|Y_{t-1},\ldots,Y_{t-p})$ in our intervention model is linear in the remaining parameters $\alpha_i$, $\lambda$ and $\kappa_j$, $i=1,\ldots,p$, $j=1,\ldots,J$, leading to simple formulae for the conditional least squares (CLS) or conditional maximum likelihood (CML) estimation. In this section we formulate the objective functions for CLS and CML estimation of model (\ref{eq:cont-inarp}) and suggest $F$-type and score statistics for the detection and identification of changes of known type at known time points.

\subsection{Conditional least squares estimation and the $F$-statistic}\label{sect:cls}

 The CLS estimates minimize the residual sum of squares,
\begin{equation*}
RSS(J)=\sum_{t=p+1}^n\left\{y_t-\lambda-\sum_{i=1}^p\alpha_i y_{t-i}-\sum_{j=1}^J\kappa_j\delta_j^{t-\tau_j}I(t\ge \tau_j)\right\}^2,\end{equation*}
and can be calculated using explicit formulae and software for ordinary least squares estimation in linear models.
The residual sum of squares can also be used when we want to decide whether a certain type of intervention effect is present at a given time point. A common measure for the goodness of fit of a linear model is the coefficient of determination, which is
$R^2=\{RSS(0)-RSS(1)\}/RSS(0)$ in case of a single intervention effect, i.e., $J=1$. $R^2$ always takes values in the interval $[0,1]$, which simplifies its interpretation. In Gaussian linear models one often prefers the $F$-type statistic
\begin{equation}
\label{eq:ftest}
F=\frac{RSS(0)-RSS(1)}{RSS(1)/(n-p-2)},
\end{equation}
since it is $F$-distributed with 1 and $n-p-2$ degrees of freedom if the model without the additional intervention effect holds \citep{hamilton:1994}. In our case, $n-p-2$ will usually be large so that such an $F$-distribution is close to the $\chi_1^2$-distribution and thereafter we shall consider this more convenient distribution. 

\subsection{Conditional maximum likelihood estimation and the score test statistic}\label{sect:cml}

The conditional log-likelihood function corresponding to model (\ref{eq:cont-inarp}) is given by
\begin{equation*}\ell(\boldsymbol{\theta})=\sum_{t=p+1}^n\log \mathnormal{p}(y_t|y_{t-1},\ldots, y_{t-p}),
\end{equation*}
where
\begin{eqnarray*}
&&\mathnormal{p}(y_t|y_{t-1},\ldots, y_{t-p})=\sum_{i_1=0}^{min(y_{t-1},y_t)}\left(\begin{array}{c}y_{t-1}\\i_1\end{array}\right)\alpha_1^{i_1}(1-\alpha_1)^{y_{t-1}-i_1}\\
&&\times\sum_{i_2=0}^{min(y_{t-2},y_t-i_1)}\left(\begin{array}{c}y_{t-2}\\i_2\end{array}\right)\alpha_2^{i_2}(1-\alpha_2)^{y_{t-2}-i_2}\cdots
\sum_{i_p=0}^{min\{y_{t-p},y_t-(i_1+\cdots+i_p)\}}\left(\begin{array}{c}y_{t-p}\\i_p\end{array}\right)\alpha_p^{i_p}(1-\alpha_p)^{y_{t-p}-i_p}\\
&&\quad \times \frac{\exp{[-\lambda-\sum_{j=1}^J\kappa_j\delta_j^{t-\tau_j}I(t\ge \tau_j)]}[\lambda+\sum_{j=1}^J\kappa_j\delta_j^{t-\tau_j}I(t\ge \tau_j)]^{y_t-(i_1+\cdots+i_p)}}{\{y_t-(i_1+\cdots+i_p)\}!},
\end{eqnarray*}
and $\boldsymbol{\theta}=(\alpha_1,\ldots, \alpha_p, \lambda, \kappa_1,\ldots, \kappa_J)^T$ is the vector of unknown model parameters.
It can be shown through differentiation \citep{freeland:2004, bu:2008} that the score function $V(\boldsymbol{\theta})=\partial\ell(\boldsymbol{\theta})/\partial\boldsymbol{\theta}$ is a $(J+p+1)$-dimensional vector with elements
\begin{eqnarray*}
\frac{\partial\ell(\boldsymbol{\theta})}{\partial\alpha_i} &=& \sum_{t=p+1}^{n}\frac{y_{t-i}[\mathnormal{p}(y_{t}-1|y_{t-1},\ldots,y_{t-i}-1,\ldots,y_{t-p})-\mathnormal{p}(y_t|y_{t-1},\ldots,y_{t-p})]}{(1-\alpha_i)\mathnormal{p}(y_t|y_{t-1},\ldots,y_{t-p})},\\
\frac{\partial\ell(\boldsymbol{\theta})}{\partial\lambda} &=& \sum_{t=p+1}^{n}\frac{\mathnormal{p}(y_{t}-1|y_{t-1},\ldots,y_{t-p})-\mathnormal{p}(y_t|y_{t-1},\ldots,y_{t-p})}{\mathnormal{p}(y_t|y_{t-1},\ldots,y_{t-p})},\\
\frac{\partial\ell(\boldsymbol{\theta})}{\partial\kappa_j} &=& \sum_{t=p+1}^{n}\frac{\delta^{t-\tau_j}I(t\geq\tau_j)[\mathnormal{p}(y_{t}-1|y_{t-1},\ldots,y_{t-p})-\mathnormal{p}(y_t|y_{t-1},\ldots,y_{t-p})]}{\mathnormal{p}(y_t|y_{t-1},\ldots,y_{t-p})},
\end{eqnarray*}
for $i=1,\ldots,p$ and $j=1\ldots, J$.
Provided that the solution of  $V(\boldsymbol{\theta})=0$ exists, it yields the conditional maximum likelihood estimate $\hat{\boldsymbol{\theta}}$ of $\boldsymbol{\theta}$. 
The conditional information for $\boldsymbol{\theta}$ is given by 
\begin{equation*}\mathcal I(\boldsymbol{\theta})=Cov\left(\left.\frac{\partial\ell(\boldsymbol{\theta})}{\partial\boldsymbol{\theta}}\right|y_{t-1},\ldots, y_{t-p}\right)
\end{equation*}
and under mild regularity conditions it can be written as
\begin{equation*}\mathcal{I}(\boldsymbol{\theta})=-E\left(\frac{\partial^2\ell(\boldsymbol{\theta})}{\partial\boldsymbol{\theta}\partial\boldsymbol{\theta}^T}\right),
\end{equation*}
where the Hessian matrix $\partial^2\ell(\boldsymbol{\theta})/\partial\boldsymbol{\theta}\partial\boldsymbol{\theta}^T$ has elements given in  Section~1 of the supplementary materials.

The availability of the score function $V(\boldsymbol{\theta})$ and conditional information matrix $\mathcal I(\boldsymbol{\theta})$ allows us to define the score test statistic
\begin{equation}
\label{eq:scoretest}
S=V^T(\tilde{\alpha}_1,\ldots,\tilde{\alpha}_p, \tilde{\lambda},0)\mathcal I^{-1}(\tilde{\alpha}_1,\ldots,\tilde{\alpha}_p,\tilde{\lambda},0)V(\tilde{\alpha}_1,\ldots,\tilde{\alpha}_p,\tilde{\lambda},0)
\end{equation}
for testing the presence of a single intervention effect ($J=1$) of known type and time of occurrence, i.e. testing the null hypothesis $H_0: \kappa=0$ against the alternative $H_{1}:\kappa\neq0$.
In formula (\ref{eq:scoretest}), $V(\tilde{\alpha}_1,\ldots,\tilde{\alpha}_p,\tilde{\lambda},0)$ and $\mathcal I(\tilde{\alpha}_1,\ldots,\tilde{\alpha}_p,\tilde{\lambda},0)$ are the score function and conditional information matrix evaluated at the maximum likelihood estimators $(\tilde{\alpha}_1,\ldots,\tilde{\alpha}_p,\tilde{\lambda},0)$ computed under the null hypothesis of a clean INAR($p$) process, i.e., an INAR($p$) process that does not include any intervention effects.  The fact that the score test statistic does not require fitting the model under the alternative hypothesis, gives a theoretical advantage over the $F$-type statistic that instead requires fitting the model under both the null and alternative hypotheses. Nevertheless, as discussed in Section~\ref{sect:unknowntype}, the fits in the $F$-type statistic are computationally much cheaper in practice.

Under the null hypothesis $H_0:\kappa=0$, (\ref{eq:cont-inarp}) reduces to a stationary INAR($p$) process with Poisson innovations. For such a process and under certain regularity conditions that are satisfied by the Poisson law \citep[see for instance][]{franke:1993, bu:2008}, the conditional maximum likelihood estimator is consistent and asymptotically normal,
\begin{equation*}
\sqrt{n}(\hat{\boldsymbol{\theta}}-\boldsymbol{\theta})\xrightarrow[]{d} N(0, \mathcal{I}^{-1}(\boldsymbol{\theta})).
\end{equation*}
Therefore, under $H_0$ and as $n\rightarrow\infty$, the score statistic (\ref{eq:scoretest}) converges to a $\chi^2_1$-distribution and derivation of critical values for an asymptotic test of the null hypothesis of no intervention against the alternative of an intervention of a certain type $\delta$ at known time $\tau$ is straightforward: we reject the null hypothesis at a given significance level $a$ if the value of $S$ is larger than the $(1-a)$-quantile of the $\chi^2_1$-distribution.

Although the suggested approach is general, its tractability is inevitably affected by the well-known computational difficulties with conditional maximum likelihood estimation in higher-order integer-valued autoregressive models. More specifically, even in the absence of intervention effects, maximization of (\ref{eq:inarp}) is cumbersome due to the nested summations appearing in the transition probabilities $\mathnormal{p}(y_t|y_{t-1},\ldots,y_{t-p})$ and the numerical difficulties that can arise when summing many small probabilities \citep[see e.g.][]{pedeli:2015, lu:2018}. In the following, we 
mainly focus on the Poisson INAR(1) model to avoid diverting the focus on computational aspects of CML estimation that are not related to the incorporation of intervention effects in the model specification.
For the first-order model, equations (\ref{eq:inarp}) and (\ref{eq:cont-inarp}) are simplified as
\begin{equation*}
Y_t=\alpha\circ Y_{t-1}+e_t
\quad \mbox{and} \quad
Y_t=\alpha\circ Y_{t-1}+e_t+\sum_{j=1}^JU_{t,j},\; t\in\mathbb{N},\end{equation*}
respectively. 
For higher-order models, we rely on the use of the $F$-type statistic whose performance is thoroughly investigated for both INAR(1) and INAR(2) processes. Results for the latter are summarized in Section~3 of the supplementary materials.


\subsection{Empirical results}
\label{subsect:empres}

Tables \ref{tab:sizes100} and \ref{tab:sizes200} of the supplementary materials report empirical rejection rates when testing for an intervention effect of known type $\delta\in \{0,0.8,1\}$ at a known time point $\tau\in \{0.25n,0.5n,0.75n\}$, using the 90\%, 95\% or 99\% quantile of the $\chi_1^2$-distribution as critical value for the $F$-type and score statistics based on CLS and CML estimation, respectively.
 The empirical rejection rates are obtained by analyzing 5000 time series of the same length $n\in \{100,200\}$ for each of
 different INAR(1) models with $\alpha\in\{0.3,0.6,0.9\}$ and $\lambda\in \{2,5\}$.

The $F$-type statistics for innovation outliers ($\delta=0$) achieve empirical rejection rates close to the target significance levels 1\%, 5\% and 10\% we aim at already in case of series of length $n=100$ and irrespective of the time $\tau$. The results are somewhat worse for larger values of $\delta$, particularly if $\alpha$ is large.  In the case of $n=100$, the $F$-test for a transient shift with $\delta=0.8$ achieves the target significance level if $\alpha=0.3$, but the rejection rate is about 2\%, 8\% and 13.5\% instead of 1\%, 5\% and 10\% if $\alpha=0.9$. For a permanent level shift ($\delta=1$), the results are worse, particularly if we test in the center of the series, $\tau=0.5n$.
 The value of $\lambda$ apparently has little effect on the rejection rates, and neither do we observe an obvious pattern whether the tests  have more problems when testing at early, central or late time points $\tau$, except for $\delta=1$.

 The results improve for larger values of the series length $n$. In the case of $n=200$, all $F$-type statistics achieve the target significance level well if $\alpha=0.3$, and the rejection rates are not much larger than the target significance level if $\alpha=0.6$. The $F$-type statistics for a transient shift are only slightly oversized even if $\alpha=0.9$, where only the test for a permanent shift shows serious size problems, particularly when testing in the center of the series $\tau=0.5n$.
We conclude that the $F$-type statistics allow simple yet promising tests for intervention effects of known type and time, and the quantiles of the $\chi_1^2$-distribution can be used as approximate critical values. Testing for permanent level shifts needs a rather long series length, particularly if the degree of autocorrelation $\alpha$ in the series is large.

The score tests achieve empirical rejection rates very close to the target significance levels 1\%, 5\% and 10\% even for $n=100$. The score test seems to be anti-conservative at the 1\% significance level but more conservative than the $F$-test at the 5\% and 10\% significance levels. Simulation results indicate that the score statistics perform better than the $F$-type statistics for transient shifts ($\delta=0.8$) and permanent level shifts ($\delta=1$), especially when the INAR(1) process is characterized by strong autocorrelation ($\alpha=0.9$). However, the $F$-type statistics achieve rejection rates closer to the targeted ones when the objective is to detect an innovation outlier ($\delta=0$). Similarly to the $F$-type statistic, the size of the score tests is little affected by the time $\tau$ of the occurrence of the intervention.

Next we examine the empirical power of these approximate significance tests for a single intervention effect at a known time point.
For this purpose we analyzed 2000 time series of length $n=200$ per simulation scenario.
The true size of the intervention effect is scaled to be $\kappa= 3\sqrt{\lambda}$, $\kappa= 2\sqrt{\lambda}$ or $\kappa= \sqrt{\lambda}$ for $\delta=0$, $\delta=0.8$ or $\delta=1$, since  the total effect on the series increases with $\delta$.
Table \ref{tab:power200-0} of the supplementary materials reports the empirical powers of the tests for the different types of intervention at a given time point $\tau\in\{0.25n,0.5n,0.75n\}$ when an innovation outlier occurs at the time point tested. We observe that both the $F$-type and score statistics for an innovation outlier possess larger power than the corresponding tests for other values of $\delta$. Nevertheless, the tests using a misspecified value of $\delta$ also have some power, particularly those using a value of $\delta$ not far from the true one.
 Moreover, the score test achieves larger power than the $F$-test for an innovation outlier especially when testing at early or central time points. 

In situations where a transient shift occurs, the $F$-type and score tests achieve similar empirical powers irrespective of the time $\tau$ of the intervention effect (Table \ref{tab:power200-0.8} of the supplementary materials). The tests using the correctly specified value of $\delta$ achieve the highest rejection rates, but other tests which use a similar value of $\delta$ are close. For instance, additional results not shown here suggest that the $F$-test using $\delta=0.8$ achieves almost the same rejection rate as that using $\delta=0.6$ if the latter is correct, and the same applies to the test for $\delta=0.9$ when there is a transient shift with $\delta=0.8$. The situation is slightly different if there is a transient shift with $\delta=0.9$, where besides the test with the correct $\delta=0.9$ usually only the test using $\delta=0.8$ reacts with a similarly large probability. The test for a permanent shift reacts with a high probability only if the transient shift occurs towards the end of the time series.  Similarly, a permanent shift is only detected with high probability by the tests for a transient shift if it occurs towards the end of the time series. Moreover, a permanent shift of a certain height is detected best by the test with the correctly specified $\delta=1$ if it occurs in the center of the series (Table \ref{tab:power200-1} of the supplementary materials). For misspecified values of $\delta\in[0, 1)$, the score test achieves higher rejection rates than the $F$-test and can then be recommended if we want to consider only a single value of $\delta$ different from $0$ and $1$.


\section{Unknown types of interventions at known time points}
\label{sect:unknowntype}

In the previous section we observed 
that the tests using the correct specification of $\delta$ usually give the largest power. This suggests that we can try to identify the type of an intervention at a known time point by comparing the $F$-type or score statistics for a selection of values of $\delta$, classifying a detected intervention according to the $F$-type or score statistic with the largest value.
We investigate the empirical detection rates of this classification rule by analyzing 2000 time series of length $n=200$ per simulation scenario obtained by setting $\alpha\in\{0.3,0.6,0.9\}$, $\lambda\in\{2,5\}$, and $\tau\in\{50,100,150\}$. We also consider 
$\delta\in\{0,0.6,0.8,0.9,1\}$ and we scale the true size of the intervention effect to be $\kappa=3\sqrt{\lambda}$, $\kappa=2.5\sqrt{\lambda}$, $\kappa=2\sqrt{\lambda}$, $\kappa=1.5\sqrt{\lambda}$ or $\kappa=\sqrt{\lambda}$ for $\delta=0$, $\delta=0.6$, $\delta=0.8$, $\delta=0.9$ and $\delta=1$, respectively. 

Applying the classification rule to data without interventions, any intervention is detected at a given time point in about $3\beta$\%-$4\beta$\% of the time series if all tests (either $F$-type statistics or score statistics) are applied with a nominal significance level of $\beta$, see Table \ref{tab:classif200} of the supplementary materials. The overall significance levels achieved by the score statistics when testing for each of the five values of $\delta\in\{0,0.6,0.8,0.9,1\}$ at a given nominal significance level are generally lower than the corresponding significance levels achieved by the $F$-type statistics. Such differences become most obvious when $\alpha=0.9$. In this case, the score test achieves an overall significance level of about 15\% at a nominal 5\% significance level while the corresponding significance level achieved by the $F$-test is about 20\%. The fact that the five tests performed for different values of $\delta$ add up to almost 20\% indicates that we are close to the case that a Bonferroni correction is useful.

Table \ref{tab:classif200-0} of the supplementary materials reports the classification results for the situation of an innovation outlier. Obviously, innovation outliers are identified correctly in most of the cases, although they are occasionally classified as a transient shift with $\delta=0.6$ with the missclassification rates being somewhat higher for the $F$-type statistics than for the score statistics.

The results look somewhat different for transient shifts, see Tables \ref{tab:classif200-6.1}-\ref{tab:classif200-9.1} of the supplementary materials. Moderately large transient shifts are classified by both the $F$-type statistic and score statistic into one of the two categories with adjacent values of $\delta$ with about the same probability as for the true value of $\delta$. That is, a moderately large transient shift with $\delta=0.6$ is often confused with an innovation outlier or a transient shift with $\delta=0.8$. Similarly, a moderately large transient shift with $\delta=0.8$ is often considered to be a transient shift with $\delta=0.6$ or $\delta=0.9$. It is worth noting that although confusion with values of $\delta$ which are either smaller or larger than the true one seems equally probable with the $F$-type statistic, missclassification is rather in favor of smaller values with the score statistic.
The situation is somewhat different for transient shifts with $\delta=0.9$. Simulation results not shown here suggest that transient shifts are confused with permanent shifts only if they occur late in the series, while Table \ref{tab:classif200-9.1} indicates that the confusion with a transient shift with $\delta=0.8$ is in the same line as before. 
The identification of permanent shifts seems not to pose missclassification issues according to the results displayed in Table \ref{tab:classif200-1} of the supplementary materials.

The problem of possible wrong classification is less pronounced when we consider transient shifts with a larger size, so that we can try to estimate a suitable value of $\delta$ by comparing the $F$-type or score statistics for a selection of values of $\delta$. Such a rule should work nicely at least for large effect sizes. Note that in the literature on the detection of intervention effects within ARMA or INGARCH models usually only the cases $\delta=0$ and $\delta=1$ corresponding to innovation outliers and permanent shifts are considered, along with a single value of $\delta$ like $\delta=0.8$ for transient shifts. This might be due to the misclassification problem outlined above and the additional complexity arising when considering multiple values of $\delta$. In our case, the $F$-type statistics based on CLS estimation are simple and do not cause computational efforts, so that we can consider several values of $\delta$ easily. The computational cost is larger with the score statistic, especially when the stationary mean of the process is large (see Table~\ref{tab:cost}). However, it can be reduced substantially by a  reasonable choice of the truncation parameter $m$ involved in the computation of the conditional information matrix as described in Section~1 of the supplementary materials.

\begin{table}
\caption{\label{tab:cost} Average CPU time required for the computation of the score and $F$-type statistics for the detection of a transient shift ($\delta=0.8$) at time $\tau=100$ in a clean INAR(1) process of length $n=200$. Results are based on 100 simulation experiments implemented on a Windows 10 Pro with a 3.6 GHz Intel Core i7-7700 processor and 16.0 GB RAM memory.}
\begin{center}
{\footnotesize
\begin{tabular}{cc|rr}
\hline
& & \multicolumn{2}{c}{Average computational cost (secs$\times 1000$)}\\
$\alpha$ & $\lambda$ & F-type statistic & Score statistic\\
\hline
0.3 & 2 & 0.11 & 45.25\\
0.3 & 5 & 0.25 & 91.19\\
0.6 & 2 & 0.14 & 67.83\\
0.6 & 5 & 0.15 & 175.53\\
0.9 & 2 & 0.15 & 324.69\\
0.9 & 5 & 0.14 & 1698.61\\
\hline
\end{tabular}}
\end{center}
\end{table}

\section{Unknown types of interventions at unknown time points}
\label{sect:unknowntime}
Now we take our considerations another step further and look at situations where we do not know neither the type nor the time of a possible intervention. For this scenario, we consider the maximum test statistics arising from calculating the $F$-type or score tests for a set of candidate time points $\tau$ and then selecting the maximum of the statistic.

\begin{figure}
\centering
\includegraphics[scale=0.6]{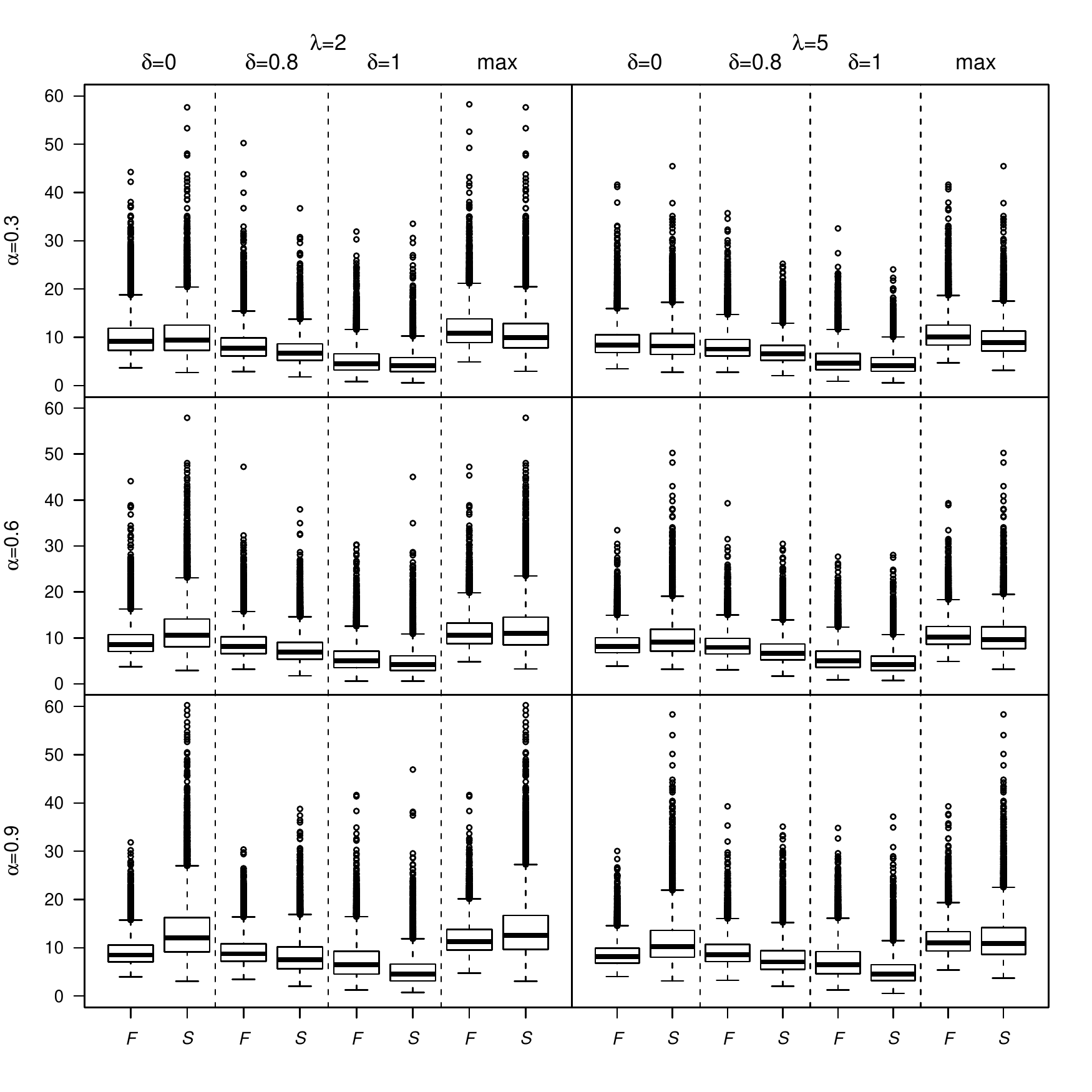}
\caption{\label{fig:maxstatistics0-100}Boxplots of the maximum $F$-type ($F$) and score ($S$) test statistics, maximized with respect to the candidate time point $\tau$ of a change when $n=100$.}
\end{figure}

\begin{figure}
\centering
\includegraphics[scale=0.6]{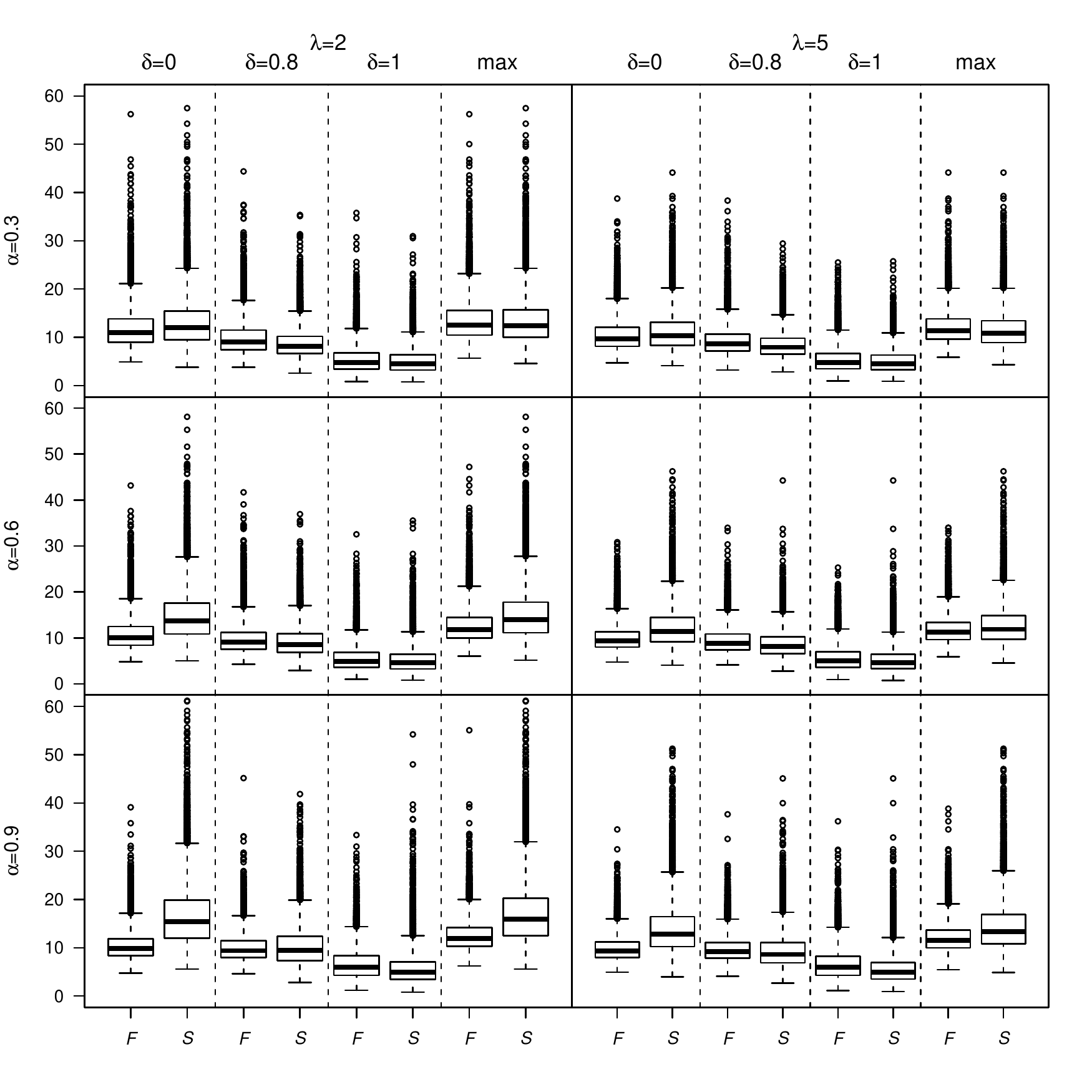}
\caption{\label{fig:maxstatistics0-200}Boxplots of the maximum $F$-type ($F$) and score ($S$) test statistics, maximized with respect to the candidate time point $\tau$ of a change when $n=200$.}
\end{figure}

 First we consider the results obtained from analyzing 10000 clean INAR(1) series for different parameter settings $\alpha\in\{0.3,0.6,0.9\}$, $\lambda\in\{2,5\}$, and series lengths \linebreak $n\in\{100,200\}$. Figures~\ref{fig:maxstatistics0-100} and~\ref{fig:maxstatistics0-200} display boxplots of these maximum statistics for each of several values of $\delta\in\{0,0.8,1\}$ individually as well as with an additional maximization with respect to $\delta$.
 Apparently the distributions of the maximum statistics under the null hypothesis are not very different for the different values of $\delta$. The main differences are that the maximum statistics take somewhat smaller values for larger values of $\delta$. 
This can be explained by the different degrees of dependence among the test statistics for the different time points $\tau$, with stronger dependencies for larger values of $\delta$. Nevertheless we expect the maximum test statistics to provide information on the type of an intervention, since these differences are not very large, especially for the $F$-type statistics.


Approximate critical values  for an overall test on any type of intervention effect can be derived from the empirical quantiles of the maximum $F$-type or score statistics with additional maximization with respect to $\delta$. In the case of $n=100$,  the 90\%, 95\% and 99\% quantiles of the overall maximum $F$-type statistics range from about 15.3 to 17.4, from 17.3 to 20.3, and from 22.4 to 26.6 for the different parameter combinations considered here, with the largest quantiles arising for $(\alpha,\lambda)=(0.3,2)$. We thus can use 17, 20 and 27 as critical values for approximate significance $F$-tests for an unknown intervention at an unknown time point at a 10\%, 5\% or 1\% significance level. 
The range of the 90\%, 95\% and 99\% quantiles of the overall maximum score statistics is wider with values between 14.3 and 21.9, between 16.7 and 25.6, and between 21.7 and 34.8, respectively. In this case, the largest quantiles arise from the somewhat extreme scenario $(\alpha,\lambda)=(0.9,2)$. To perform  approximate score tests for an unknown intervention at an unknown time point at a 10\%, 5\% or 1\% significance level, we can thus use 22, 26 and 35 as critical values.
In case of $n=200$, the empirical percentiles of the maximum $F$-type statistics range from 15.9 to 19.2, from 17.8 to 21.9, and from 22.2 to 27.8, so that we can use 19, 22 and 28 as approximate critical values. The corresponding empirical percentiles of the maximum score statistics range from 16.8 to 26, from 19.3 to 30.4 and from 25.3 to 40.2, so that 26, 30 and 40 can serve as approximate critical values. 
Additional simulation results in Section 3 of the supplementary materials indicate that the same critical values can be used for the $F$-type statistics in case of INAR(2) models.


Derivation of critical values based on the empirical percentiles of the maximum test statistics has obviously some drawbacks. Firstly, the resulting tests will usually be somewhat conservative within the range of situations that have been previously investigated. Secondly, their performance is not guaranteed, for different parameter configurations, sample sizes or higher-order models  outside this range. To confront such limitations we can employ a parametric bootstrap, as in \cite{fokianos:2010}. This procedure requires analyzing many artificial time series generated from the model fitted under the null hypothesis and comparing the values of the test statistics for the observed real data to those obtained for the artificial data. If the real data do not contain any interventions, the corresponding value of the maximum test statistic should be comparable to those of the bootstrapped series. This comes at the price of a much higher computational cost. The parametric bootstrap procedure is discussed further in Section~\ref{sect:iterative} where it is used in a couple of illustrative examples.

\begin{figure}\centering
\includegraphics[scale=0.6]{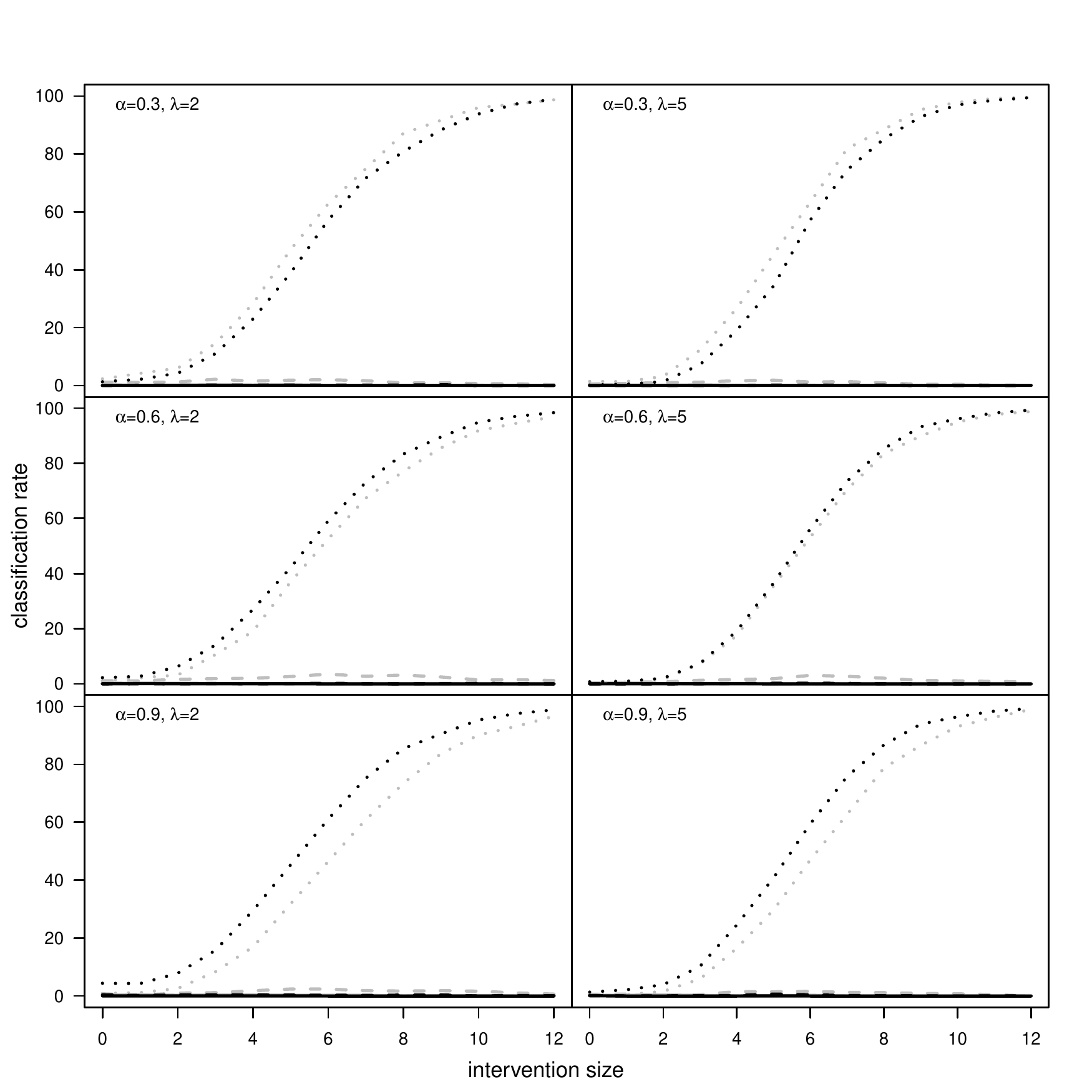}
\caption{\label{fig:classif100-0}Classification results when applying the maximum $F$-type (grey lines) and score test statistics (black lines) to time series of length $n=100$ containing an innovation outlier of increasing size $\kappa=0,\sqrt{\lambda},\ldots,12\sqrt{\lambda}$ at time point $\tau=50$.  Classification as $\delta=0$ (dotted), $\delta=0.8$ (dashed), $\delta=1$ (solid).}
\end{figure}

Next we inspect the performance of the classification rules when applied to time series containing an intervention effect. 
 Classification is based on the maximum test statistics whose significance is concluded according to the critical values derived previously.
Figure \ref{fig:classif100-0} depicts the classification results when being applied to time series of length $n=100$ containing an innovation outlier at time $\tau=50$. For this we generate 2000 time series for each of the parameter combinations $(\alpha,\lambda)$ considered before and each intervention size $\kappa=k\sqrt{\lambda}$, $k=0,\ldots,12$. Apparently, time series containing an innovation outlier are classified quite reliably by both the $F$-type and score test statistics, except if the outlier is very small. 
For INAR(1) processes characterized by a rather weak autocorrelation ($\alpha= 0.3$), the $F$-type and score test statistics attain very similar classification rates. As the autocorrelation of the series becomes larger, the score test statistic seems to be superior, but this is partly explained by the different empirical sizes achieved by the two tests. For example, for $\alpha=0.9$ and $\lambda=2$, the score test provides substantially higher classification rates than the $F$-test but this is in part due to the latter showing a quite conservative behavior as its empirical size is only about $0.6\%$ then.

\begin{figure}\centering
\includegraphics[scale=0.6]{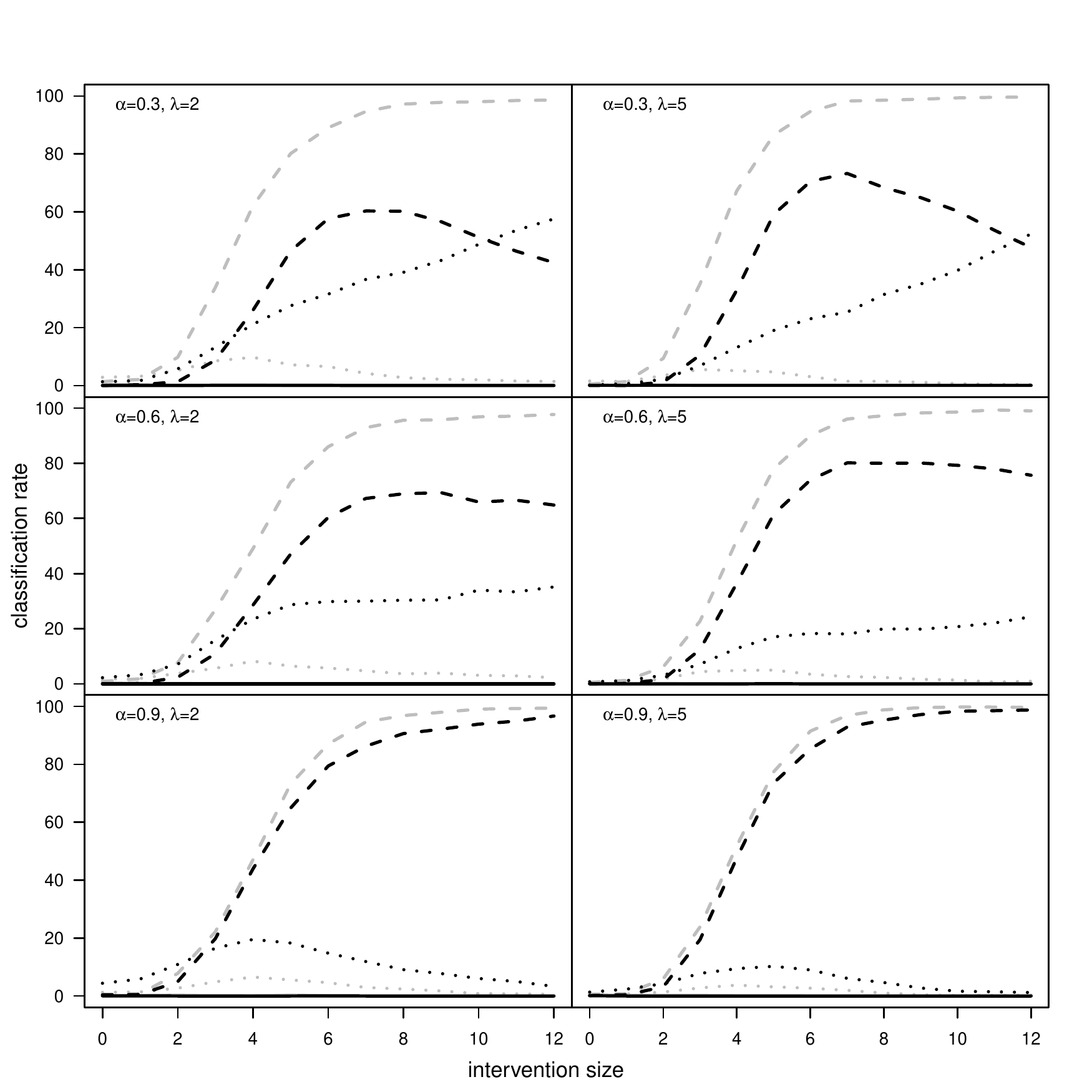}
\caption{\label{fig:classif100-08}Classification results when applying the maximum $F$-type (grey lines) and score test statistics (black lines) to time series  of length $n=100$ containing a transient shift with $\delta=0.8$ of increasing size $\kappa=0,\sqrt{\lambda},\ldots,12\sqrt{\lambda}$ at time point $\tau=50$.  Classification as $\delta=0$ (dotted), $\delta=0.8$ (dashed), $\delta=1$ (solid).}
\end{figure}

Figure \ref{fig:classif100-08} illustrates results for the classification of a transient shift with $\delta=0.8$. The $F$-type statistic provides results that look particularly well for 
all parameter configurations.
The performance of the score test statistic relates to the degree of autocorrelation in the series with better performance for strongly autocorrelated time series data.
In particular, the classification rates are very reliable for $\alpha=0.9$ but the number of misclassified cases increases as the autocorrelation weakens. For $\alpha=0.6$ and even more notably for $\alpha=0.3$, many cases are classified as innovation outliers instead of transient shifts. In such situations and especially for small values of $\alpha$, the classification rates tend to increase with the intervention size up to $\kappa=7\sqrt{\lambda}$. Thereafter, there is a decreasing tendency with  more than half of the cases being misclassified as innovation outliers when $\kappa=11\sqrt{\lambda}$ or $\kappa=12\sqrt{\lambda}$.

\begin{figure}\centering
\includegraphics[scale=0.6]{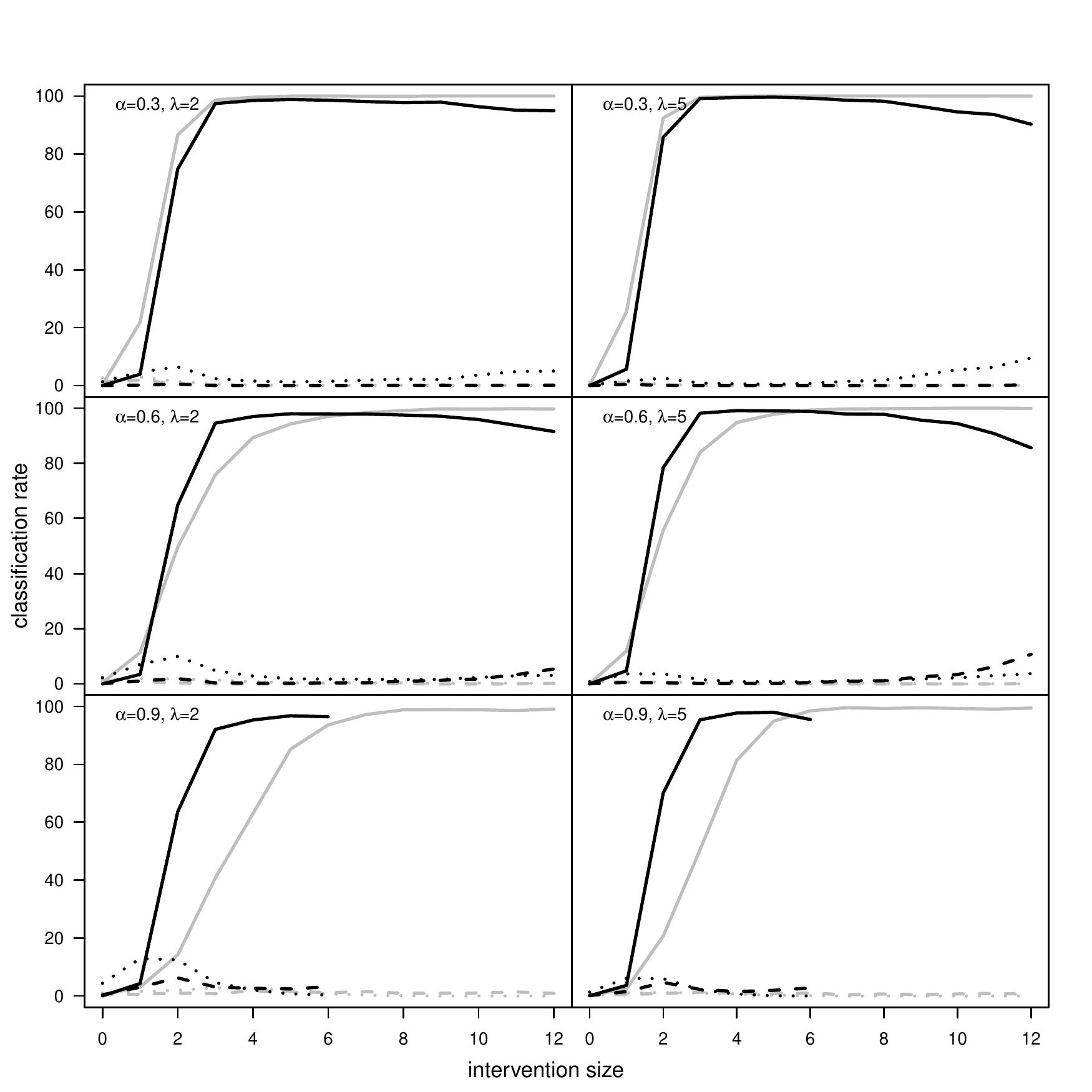}
\caption{\label{fig:classif100-1}Classification results when applying the maximum $F$-type (grey lines) and score test statistics (black lines) to time series of length $n=100$ containing a permanent shift with $\delta=1$ of increasing size $\kappa=0,\sqrt{\lambda},\ldots,12\sqrt{\lambda}$ at time point $\tau=50$.  Classification as $\delta=0$ (dotted), $\delta=0.8$ (dashed), $\delta=1$ (solid).}
\end{figure}

These problems do not exist for permanent shifts in the center of the series that are very rarely classified as one of the other types of intervention effects by both the $F$-type and score test statistics.
Figure \ref{fig:classif100-1} highlights the role of the autocorrelation parameter $\alpha$ and the intervention size $\kappa$ in the performance of the score test statistic and indicates the cases where it is preferable to the $F$-type statistic. Specifically, when $\alpha=0.3$, the $F$-type statistic achieves slightly higher classification rates than the score test statistic, independently of the intervention size. For $\alpha=0.6$, the score test statistic is preferable to the $F$-type statistic for medium intervention sizes, that is for $\kappa=2\sqrt{\lambda}$ to $\kappa=6\sqrt{\lambda}$. For lower or higher intervention sizes, the $F$-type statistic performs better. A similar pattern is observed for $\alpha=0.9$ but with a greater outperformance by the score test statistic for medium intervention sizes. The classification rates of the score test statistic are not available for $\kappa>6\sqrt{\lambda}$ and $\alpha=0.9$ (lower panel of Figure \ref{fig:classif100-1}) since the extremely large counts that appear in the time series render the Poisson distribution a poor parametric choice and 
cause the Fisher information matrix $\mathcal{I}(\boldsymbol{\theta})$ to be nearly singular. In other words, in such cases the Poisson model is seriously misspecified and the conditional maximum likelihood estimates are inconsistent with unreliable standard errors. Such problems do not occur with the $F$-type statistic which achieves classification rates close to $100\%$  when both the degree of autocorrelation and the intervention size are high.

Our empirical findings on the role of the degree of autocorrelation on the performance of the score and $F$-type statistics are consistent with previous results about the performance of the conditional least squares and maximum likelihood estimators of the Poisson INAR(1) model. In particular, \cite{alosh:1987} and \cite{brannas:1994} 
observed that the biases of the conditional least squares estimators increase as $\alpha$ takes higher values with a much higher increase in the bias of $\hat{\lambda}^{\mbox{\tiny{CLS}}}$ than that of $\hat{\alpha}^{\mbox{\tiny{CLS}}}$. In contrast, the conditional maximum likelihood estimates do not show such a behaviour. In particular, the bias of $\hat{\alpha}^{\mbox{\tiny{CML}}}$ increases as $\alpha$ takes values up to around 0.3 and then it starts to decrease until reaching a negligible bias when $\alpha=0.9$. The bias of $\hat{\lambda}^{\mbox{\tiny{CML}}}$ remains low and almost stable for different values of $\alpha$.
The observed tendency of our classification rules  to over- or underestimate $\delta$ can thus be better explained by accounting for the aforementioned results and the strong negative correlation between the estimators of $\alpha$ and $\lambda$, especially when $\alpha$ is large \citep[see][Figure 3]{alosh:1987}.


\section{Iterative detection of intervention effects}
\label{sect:iterative}
In real data problems, time series can contain more than one intervention throughout the observation period. For the detection, classification and elimination of multiple intervention effects we follow the stepwise procedure of \citet{fokianos:2010}, adapted to the INAR(1) framework. The steps of this iterative detection approach are described below, setting $j=1$ and $Y_t^{(j)}=Y_t$, $t=1,\ldots,n$, for initialization of the algorithm:
\begin{enumerate}
\item Fit an INAR(1) model to the data $\{Y_t^{(j)}, t=1,\ldots,n\}$.
\item Test for a single intervention of any type at any time point by employing (\ref{eq:cont-inarp}) and using the maximum of the $F$-type or score test statistics. At this step, we suggest using the parametric bootstrap procedure discussed briefly in Section~\ref{sect:unknowntime}. The individual steps for its implementation are described in detail in Table~\ref{tab:boot}.
\item If there is no significant result, the iterative detection procedure is terminated and the series $Y_1^{(j)},\ldots,Y_n^{(j)}$ is considered as clean. Otherwise:
\begin{enumerate}
\item Fit a contaminated INAR(1) model (\ref{eq:cont-inarp}) by choosing $\delta$ according to the type of intervention identified in the previous step. Let $\hat{\kappa}$ be the estimated size of the intervention effect and $\hat{\tau}$ its time point of occurrence.
\item For $t\geq\hat{\tau}$, sequentially estimate the effect of the intervention on the observation $Y_t^{(j)}$ by the rounded value
\[\hat{U}_t=\left\lfloor\frac{\hat{\kappa}\delta^{t-\hat{\tau}}}{\hat{\alpha}Y_{t-1}^{(j+1)}+\hat{\lambda}+\hat{\kappa}\delta^{t-\hat{\tau}}}Y_t^{(j)}\right\rfloor\]
and correct the corresponding observation for the estimated intervention effect by setting
\[Y_t^{(j+1)}=Y_t^{(j)}-\hat{U}_t.\]
Note that for $t<\hat{\tau}$, $Y_t^{(j+1)}=Y_t^{(j)}$ so that $\hat{U}_{\hat{\tau}}=\left\lfloor\hat{\kappa}Y_{\hat{\tau}}^{(j)}/(\hat{\alpha}Y_{\hat{\tau}-1}^{(j)}+\hat{\lambda}+\hat{\kappa})\right\rfloor$.
\end{enumerate} 
\item Set $j=j+1$ and return to step 1.
\end{enumerate}
The iterative procedure is continued until no further interventions are detected.
The correction in step 3.b) is adequate if the type of intervention and time point of its occurrence have been correctly identified. The estimated intervention effect $\hat{U}_t$ is actually the rounded estimate of the conditional expectation of the contaminating process $U_t$ in (\ref{eq:cont-inarp}) given $Y_t$ and the $\sigma$-field $\mathcal{F}_{t-1}=\{Y_{t-1}, U_{t-1}\}$:
\begin{eqnarray*}
E(U_t|Y_t=y,\mathcal{F}_{t-1})&=&\sum_{u=0}^{y}uP(U_t=u|Y_t=y,\mathcal{F}_{t-1})\\
&=&\sum_{u=0}^yu\frac{P(U_t=u, Y_t^{\mbox{\tiny{clean}}}=y-u|\mathcal{F}_{t-1})}{P(Y_t=y|\mathcal{F}_{t-1})}\\
&=&\sum_{u=0}^yu\frac{(\kappa\delta^{t-\tau})^u\exp(-\kappa\delta^{t-\tau})/u!(\alpha Y_{t-1}+\lambda)^{y-u}\exp(-\alpha Y_{t-1}-\lambda)/(y-u)!}{(\alpha Y_{t-1}+\lambda+\kappa\delta^{t-\tau})^y\exp(-\alpha Y_{t-1}-\lambda-\kappa\delta^{t-\tau})/y!}\\
&=&\sum_{u=0}^y u\left(\begin{array}{c}y\\u\end{array}\right)\left(\frac{\kappa\delta^{t-\tau}}{\alpha Y_{t-1}+\lambda+\kappa\delta^{t-\tau}}\right)^u\left(\frac{\alpha Y_{t-1}+\lambda}{\alpha Y_{t-1}+\lambda+\kappa\delta^{t-\tau}}\right)^{y-u}\\
&=&\left(\frac{\kappa\delta^{t-\tau}}{\alpha Y_{t-1}+\lambda+\kappa\delta^{t-\tau}}\right)y,
\end{eqnarray*}
where $Y_t^{\mbox{\tiny{clean}}}$ denotes the $t$-th observation from a clean INAR(1) process. 
 Note that $U_t|Y_t=y,\mathcal{F}_{t-1}$ is binomially distributed with  parameters $y$ and $\kappa\delta^{t-\tau}/(\alpha Y_{t-1}+\lambda+\kappa\delta^{t-\tau})$.
 
 \begin{table}
\caption{\label{tab:boot}The parametric bootstrap procedure for the identification of unknown types of interventions at unknown time points.}
{\small
\begin{center}
\begin{tabular}{|rl|}
\hline
1. & Fit an INAR model to the observed time series assuming that there are no interventions.\\
\hline
2. & Generate a large number of, say, $B=500$ bootstrap replicates from the fitted INAR \\
& model with the same parameters as those estimated for the observed real data.\\
\hline
3. & Calculate the maximum test statistics for the original and for the $B$ bootstrap series.\\
\hline
4. & Compute the number $N$ of bootstrap replicates for which the maximum test statistic\\
& is not smaller than its value computed by the original data.\\
\hline
5. &Compute the $p$-value $(N+1)/(B+1)$.\\
\hline
6. & Classify the type of the intervention according to the minimal $p$-value, with preference\\
& given to interventions with larger value of $\delta$ in case of equality.\\ 
\hline
\end{tabular}
\end{center}
}
\end{table}

\subsection{Simulation example}
\label{sect:simulex}

\begin{figure}\centering
\includegraphics[scale=0.6]{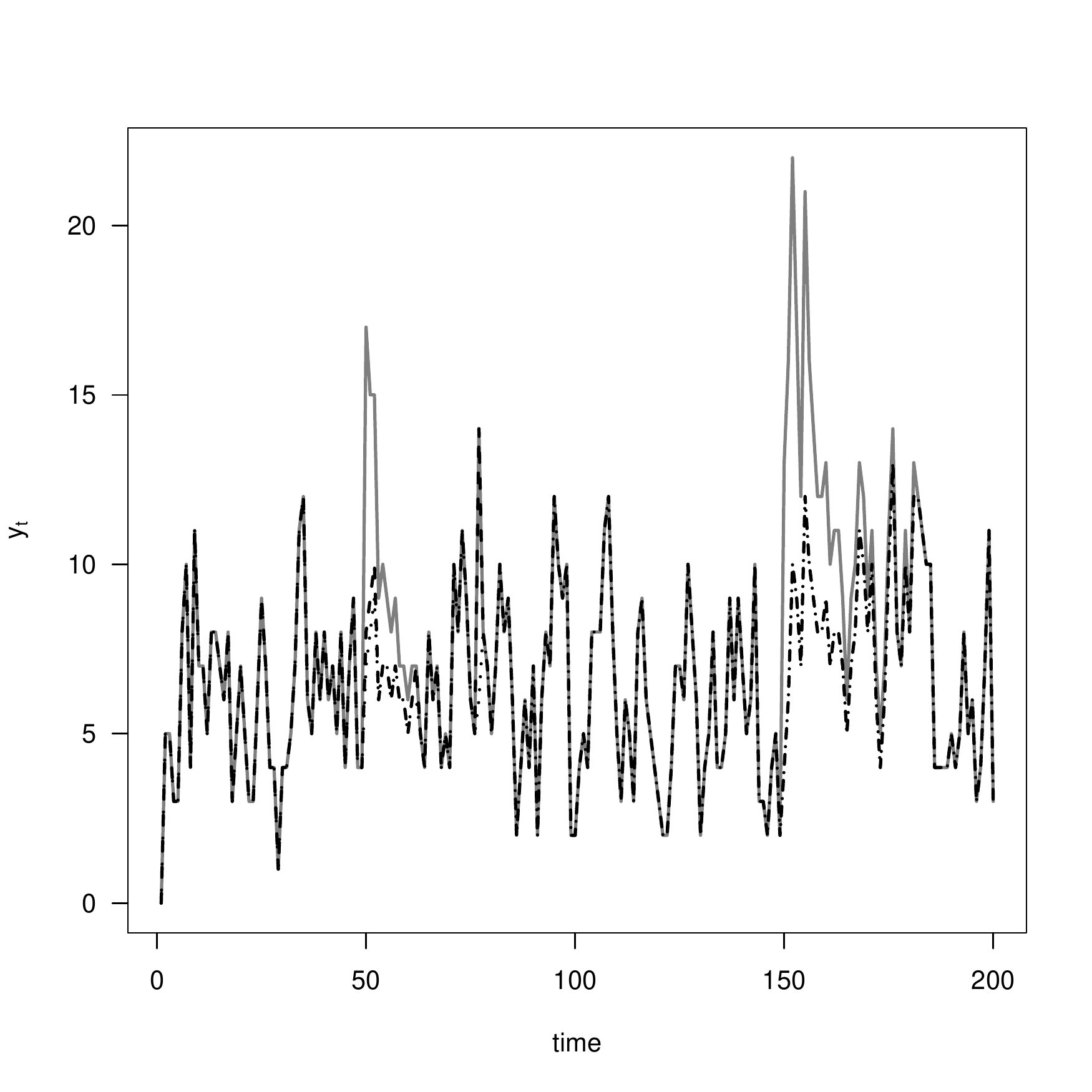}
\caption{\label{fig:sim}Simulated time series with two transient shifts at times $\tau_1=50$ and $\tau_2=150$ (solid line) and the series after correction for the intervention effects as estimated by the $F$-type statistic (dotted line) and the score test statistic (dashed line), which are quite similar here.}
\end{figure}

We consider a simulated time series of length $n=200$ generated from a contaminated Poisson INAR(1) model of the form
$Y_t=\alpha\circ Y_{t-1}+e_t+U_{t,1}+U_{t,2}$,
where $e_t\sim Pois(\lambda)$, $U_{t,j}\equiv0$ for $t=0,\ldots,\tau_j-1$ and $U_{t,j}\sim Pois(\kappa_j\delta_j^{t-\tau_j})$ for $t=\tau_j,\ldots,n$, $j=1,2$. 
We set $(\alpha,\lambda)=(0.5,3)$ and the interventions consisting of two transient shifts of the same size $\kappa_1=\kappa_2=\kappa=10$ at times $\tau_1=50$ and $\tau_2=150$ with $\delta_1=0.6$
and $\delta_2=0.9$, respectively (see Figure \ref{fig:sim}).  We apply the previously described stepwise detection algorithm to test for the existence of any type of outlier using $\delta=(0,0.6,0.8,0.9,1)$ at any time point. Step (2) of the iterative detection algorithm is implemented using the parametric bootstrap procedure of Table~\ref{tab:boot}. 
The conditional least squares and maximum likelihood estimates obtained at each step of the stepwise procedure are summarized in Table \ref{tab:sim-ex}. 

When we fit a Poisson INAR(1) model to the data assuming no interventions, we obtain that the conditional least squares and maximum likelihood estimates are $(\hat{\alpha}^{\mbox{\tiny{CLS}}},\hat{\lambda}^{\mbox{\tiny{CLS}}})=(0.60, 2.95)$ and $(\hat{\alpha}^{\mbox{\tiny{CML}}},\hat{\lambda}^{\mbox{\tiny{CML}}})=(0.46, 3.90)$, respectively. Then, we test for unknown types of interventions at unknown time points using both the $F$-type and score test statistics. At the first iteration, both test statistics correctly identify a transient shift with $\delta=0.9$ at time $\tau=150$. After correcting the data according to step 3.b), the second intervention corresponding to $\tau=50$ is also detected by the two test statistics and is correctly classified as a transient shift, although with $\delta=0.8$ instead of $\delta=0.6$. Correcting anew the data, the $F$-type statistic detects an additional innovation outlier at time $\tau=77$. The final conditional least squares and maximum likelihood estimates are $(\hat{\alpha}^{\mbox{\tiny{CLS}}},\hat{\lambda}^{\mbox{\tiny{CLS}}})=(0.45, 3.58)$ and $(\hat{\alpha}^{\mbox{\tiny{CML}}},\hat{\lambda}^{\mbox{\tiny{CML}}})=(0.42, 3.79)$, respectively.

Since the data are generated by a contaminated INAR(1) model, evaluation of the iterative detection procedure should be based on the ability of the test statistics to identify correct types of intervention at correct time points and on the efficiency of the corresponding parameter estimators.
For such an assessment, we repeated our experiment several times with data generated from the same model and with the same types and sizes of intervention effects. 
Our results indicate that the suggested stepwise detection procedure correctly identifies the intervention effects with some exceptions of additional outliers being occasionally identified. This is not surprising since even an uncontaminated INAR(1) process can occasionally present some relatively large values. Moreover, the effects $U_t$ of the same intervention on different time points are independent and thus it is hard to estimate all of them well.
In some instances, additional intervention effects were found right after the occurrence of the true ones. This can also happen if the size of the true intervention is underestimated or if the true intervention is not detected at all.

We also observed that conditional maximum likelihood and conditional least squares behave similarly in terms of efficiency of the stationary mean estimator.
Finally, we should note that  the larger is the persistence of a transient shift, the greater is the ability of both tests to correctly identify it. In our experiments, the test statistics usually identified the transient shift occurring at $\tau=50$, but they overestimated $\delta$. This overestimation can partly be explained by the convention in the sixth step of the bootstrap procedure (see Table~\ref{tab:boot}). 
%

\begin{table}
\caption{\label{tab:sim-ex} Conditional least squares and maximum likelihood estimates obtained at each step of the stepwise procedure for the detection and elimination of intervention effects in the simulated time series. The final estimates of the Poisson INAR(1) model parameters are shown in bold. The $F$-type and score test statistics are used with conditional least squares and maximum likelihood estimation, respectively. The true parameter values are $\alpha=0.5$ and $\lambda=3$ and there are outliers with $\kappa=10$ and $\delta=0.6$ at $\tau=50$ as well as $\kappa=10$ and $\delta=0.9$ at $\tau=150$.}
{\small
\begin{center}
\begin{tabular}{cc|cr|rr|rrr}
\hline
Iteration & Step & \multicolumn{2}{c|}{Test statistic} &\multicolumn{2}{c|}{Parameter estimates} & \multicolumn{3}{c}{Outlier}\\
&   & Type & \multicolumn{1}{c|}{Bootstrap p-value} & $\hat{\alpha}$ & $\hat{\lambda}$ & $\hat{\kappa}$ & $\hat{\tau}$ & $\hat{\delta}$\\
\hline
\rowcolor{lightgray}
1 &  1 & $F$-type & & 0.60 & 2.95 & & & \\
& & score & & 0.46 & 3.90 & & & \\
\rowcolor{lightgray}
 & 2-3 & $F$-type & $<0.001$ & 0.41 & 3.82 & 8.93 & 150 & 0.9\\
 & & score & $<0.001$ & 0.39 & 3.97 & 9.37 & 150 & 0.9\\
 \hline
 \rowcolor{lightgray}
 2 & 1 & $F$-type &  & 0.46 & 3.62 & & & \\
 & & score & & 0.42 & 3.89 & & & \\
 \rowcolor{lightgray}
 & 2-3& $F$-type & $<0.001$ & 0.38 & 3.97 & 7.19 & 50 & 0.8\\
 & & score & $<0.001$ & 0.40 & 3.86 & 6.66 & 50 & 0.8\\
 \hline
 \rowcolor{lightgray}
 3 & 1 & $F$-type &  & 0.43 & 3.72 & & & \\
 & & score & & \textbf{0.42} & \textbf{3.79} & & & \\
  \rowcolor{lightgray}
  & 2-3 & $F$-type & 0.02 & 0.44 & 3.62 & 8.18 & 77 & 0\\
 & & score & $0.148$ & - & - & - & - & -\\
\hline
 \rowcolor{lightgray}
 4 & 1 & $F$-type &  & \textbf{0.45} & \textbf{3.58} & & & \\
 & & score & & - & - & & & \\
  \rowcolor{lightgray}
  & 2-3 & $F$-type &  0.164 & - & - & - & - & -\\
 & & score & - & - & - & - & - & -\\
\hline

  \hline
\end{tabular}
\end{center}
}
\end{table}

\subsection{Brucellosis in Greece}

Brucellosis is a common disease worldwide, representing a serious public health problem in many countries,
especially those around the Mediterranean Sea. The infection can be directly transmitted from infected animals and contaminated tissues to humans via inhalation or through skin lesions which is an occupational risk for veterinarians, abattoir workers and farmers, particularly in endemic regions. However, the ingestion of contaminated raw milk and dairy products poses a major public health risk. In milk and products thereof, brucella is controlled most effectively by pasteurization or sterilization before marketing or by further processing into dairy products.
Despite intense efforts to eliminate brucellosis in Europe, the disease still occurs in Portugal, Spain, France, Italy, the Balkans, Bulgaria, and Greece \citep{karagiannis:2012, rossetti:2017}. 

\begin{figure}\centering
\includegraphics[scale=0.6]{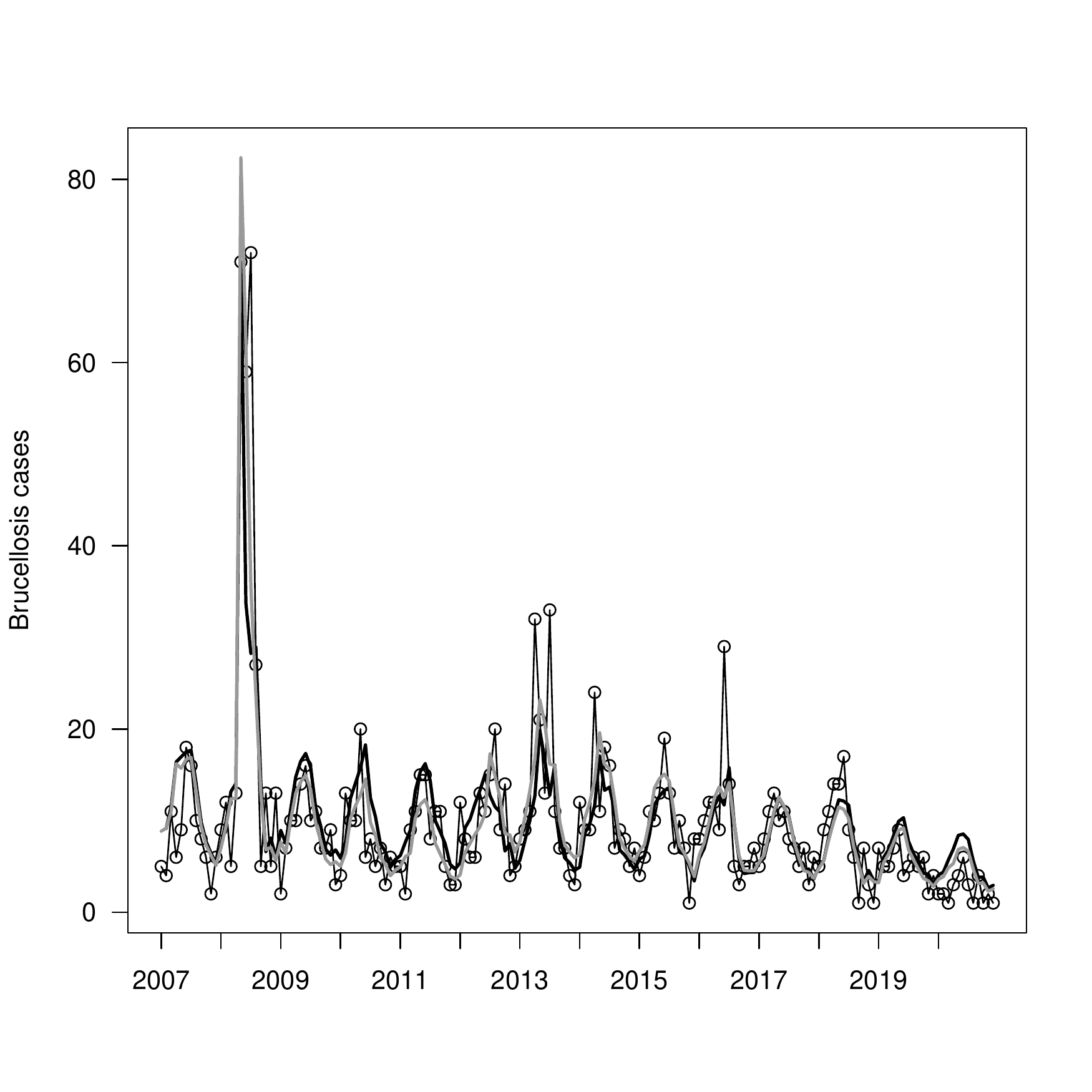}
\caption{\label{fig:grdat}Monthly number of brucellosis human cases in Greece for the years 2007–2020: time series (solid line), fitted Poisson INAR(1) regression model with interventions (bold solid in black) and fitted log-linear Poisson autoregressive model with interventions (bold solid in grey).}
\end{figure}

Figure \ref{fig:grdat} illustrates the monthly number of human brucellosis cases in Greece for the years 2007-2020 ($n=168$), as recorded by the European Center of Disease Control (ECDC) Surveillance Atlas.  The data display a seasonal pattern and an outbreak of disease cases from May to July 2008. Indeed, in spring 2008, the Hellenic Center for Disease Control and Prevention was notified about human brucellosis cases in Thassos, a Greek island that had been up to that point under a brucellosis eradication programme. During the subsequent days, more cases were notified from the island and an outbreak was verified \citep{karagiannis:2012}.

To investigate whether the suggested stepwise detection algorithm is able to effectively detect the intervention effects in the time series of brucellosis cases, we start by fitting a Poisson INAR(1) regression model of the form $Y_t=\alpha\circ Y_{t-1}+e_t$. The arrival process $(e_t)$ is Poisson distributed with parameter $\lambda_t$ accounting for annual seasonality and trend, that is
\[\log(\lambda_t)=\beta_0+\beta_1\sin\left(\frac{2\pi t}{12}\right)+\beta_2\cos\left(\frac{2\pi t}{12}\right)+\beta_3\frac{t}{168}, \quad t=1,\ldots,168.\]

After fitting the Poisson INAR(1) regression model to the data, we test for different types of interventions using the iterative procedure described earlier in Section~\ref{sect:iterative}. For this purpose, we employ the maximum score test statistic since, contrary to conditional least squares estimation, the non-linear form of $\lambda_t$ does not complicate conditional maximum likelihood estimation of the model parameters.

Table \ref{tab:grdat} summarizes the results of the iterative detection procedure. 
To decide about the approximate significance of the score test statistic we base ourselves on the parametric bootstrap procedure described in Section~\ref{sect:iterative}. In the first iteration, our classification rule decides in favor of an innovation outlier at time $t=17$ which corresponds to May 2008. The detected intervention effect is significant at $10\%$ significance level ($p$-value=0.078) and its estimated size is 52.828. After elimination of its effect from the time series, no further interventions are identified, since the score test statistic is not any more significant in the next step. 

Fitting the full model with the detected innovation outlier to the original data, we conclude with the enlarged Poisson INAR(1) regression model for the number of brucellosis human cases:

\vspace{-30pt}
\begin{eqnarray*}
Y_t&=&0.274\circ Y_{t-1}+e_t+U_{t}, \quad t=1,\ldots, 168,\\
e_t&\sim& Pois(\lambda_t), \quad \log(\lambda_t)=2.184+0.175\sin\left(2\pi t/12\right)-0.553\cos\left(2\pi t/12\right)-0.758 t/168\\
U_{t}&\sim& Pois(52.92 I(t=17))\\
\end{eqnarray*}

\vspace{-30pt}
The parameter estimates and the corresponding standard errors obtained with the enlarged (contamination) model are summarized in Table \ref{tab:grmod}. For comparison purposes we also report the results obtained by fitting 
a contaminated log-linear Poisson autoregressive model of order 1 \citep{fokianos:2012}. For the latter, we assume that $Y_t|\mathcal{F}_{t-1}\sim\mbox{Poisson}(\lambda_t)$, 
where
\[\log(\lambda_t)=\beta_0+\beta_1\sin\left(\frac{2\pi t}{12}\right)+\beta_2\cos\left(\frac{2\pi t}{12}\right)+\beta_3\frac{t}{168}+\gamma\log(Y_{t-1}+1)+\sum_{j=1}^J\kappa_j\delta_{j}^{t-\tau_j}I(t\geq\tau_j),\]
and we use the \texttt{R} package \texttt{tscount} for model fitting and detection of intervention effects \citep{liboschik:2017}. Starting from a first-order log-linear model, we detect a transient shift with $\delta=0.6$ at time 17 (May 2008) and a level shift at time 67 (July 2012), both being significant at $1\%$ significance level.  The log-intensity process of the fitted model with the detected interventions at their respective times is given by
\begin{eqnarray*}
\log(\lambda_t)&=&1.904+0.143\sin\left(\frac{2\pi t}{12}\right)-0.424\cos\left(\frac{2\pi t}{12}\right)-1.450\frac{t}{168}+0.234\log(Y_{t-1}+1)\\
&&+1.599\cdot 0.6^{t-17}I(t\geq17)+0.649I(t\geq67)
\end{eqnarray*}

The predictions from the two contamination models are also plotted in Figure~\ref{fig:grdat}, illustrating that they both fit the data well and successfully accommodate the disease outbreak.

\begin{figure}\centering
\includegraphics[scale=0.65]{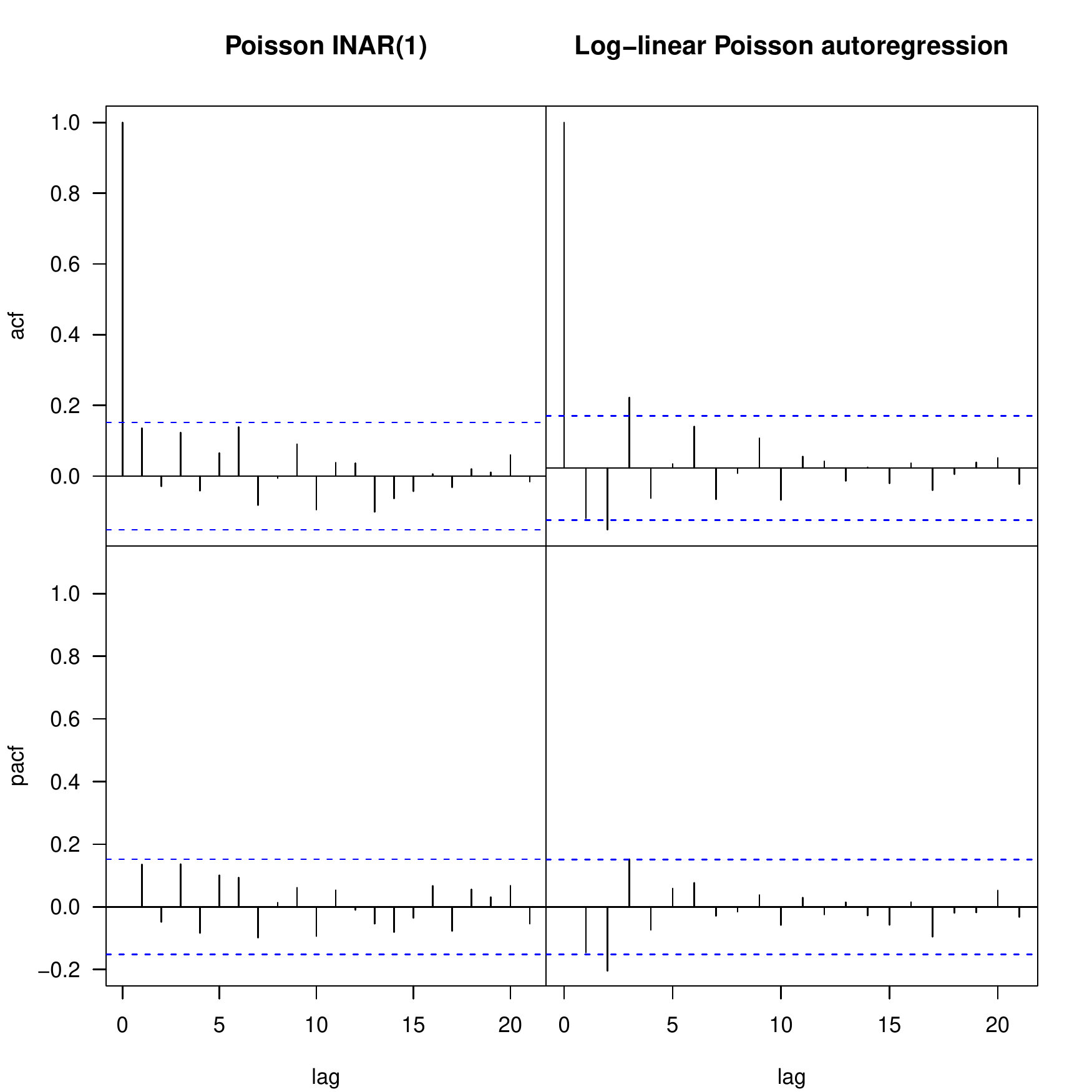}
\caption{\label{fig:grres} Correlograms and partial correlograms of the residuals obtained after fitting a contaminated Poisson INAR(1) (left panel) and a contaminated log-linear Poisson autoregressive model (right panel) to the brucellosis series.}
\end{figure}

The correlograms and partial correlograms of the residuals obtained after fitting the two contaminated time series regression models to the data are shown in Figure~\ref{fig:grres}. The empricial autocorrelation and partial autocorrelation functions of the residuals obtained by the Poisson INAR(1) process (left panel) do not exhibit any serial correlation which has not been taken into account by the model. In contrast, the log-linear Poisson autoregression fails to adequately account for serial correlation at lags 2 and 3 (right panel) indicating an improved fit of the contaminated INAR(1) model for this particular dataset.



\begin{table}
\caption{\label{tab:grdat} Iterative parameter estimates and intervention effects for the brucellosis data.}
{\small
\begin{center}
\begin{tabular}{cc|rrrr|rrr}
\hline
Iteration & Step  &\multicolumn{4}{c|}{Parameter estimates} & \multicolumn{3}{c}{Outlier}\\
&   & $\hat{\alpha}$ &  $\hat{\beta}_0$ & $\hat{\beta}_1$ & $\hat{\beta}_2$ & $\hat{\kappa}$ & $\hat{\tau}$ & $\hat{\delta}$\\
\hline
 1 & 1 & 0.290 & 1.824 & 0.251 & -0.625&\\
& 2-3 & 0.317  & 1.760 & 0.218 & -0.543 & 56.506 & 17 & 0 \\
\hline
\end{tabular}
\end{center}
}
\end{table}

\begin{table}
\caption{\label{tab:grmod}
 Parameter estimates (standard errors) obtained by fitting contaminated time series regression models to the monthly number of brucellosis human cases in Greece for the years 2007–2020.}
{\small
\begin{center}
\begin{tabular}{c|c|c}
\hline
& Poisson INAR(1) & Log-linear Poisson autoregression\\
\hline
$\hat{\alpha}$ &  0.274 (0.032) & -\\
$\hat{\gamma}$ & -  & 0.234 (0.049) \\
$\hat{\beta}_0$ & 2.184 (0.078)  & 1.904 (0.141)\\
$\hat{\beta}_1$ & 0.175 (0.050)  & 0.143 (0.038)\\
$\hat{\beta}_2$ & -0.553 (0.051) &  -0.424 (0.046)\\
$\hat{\beta}_3$ & -0.758 (0.117) &  -1.450 (0.194)\\
$\hat{\kappa}_1$ & 52.920 (8.419) &  1.599 (0.118)\\
$\hat{\kappa}_2$ & - &  0.649 (0.102)\\
\hline
\end{tabular}
\end{center}
}
\end{table}

\section{Discussion}
\label{sect:discussion}

We have developed a feasible procedure for the detection of intervention effects in integer-valued autoregressive models for count time series. The suggested procedure relies on the use of $F$-type or score test statistics. Extensive simulation experiments indicated that the elements that largely determine the performance of the two test statistics are the autocorrelation parameter $\alpha$ and the parameter $\delta$ identifying the intervention type. The $F$-type statistic is preferable when the INAR(1) process is characterized by a rather weak autocorrelation and the objective is to detect an innovation outlier ($\delta=0$). For transient or permanent level shifts ($\delta\in(0,1]$) and especially when the INAR(1) process is characterized by strong autocorrelation, the score test statistic performs better than the $F$-type statistic. The score test statistic is also preferable for misspecified values of $\delta$ although both tests pose some missclassification issues when transient shifts of a moderate size ($\delta=0.6$ or 0.8) are considered. The advantage of the $F$-type statistic is that it works reliably also in case of higher order models, see Section 3 of the supplementary materials.

Our modelling of intervention effects is additive for the INAR model and the same applies to the impact of intervention effects on the dynamics. Therefore, our model formulation allows for the detection and classification of different types of outliers, contrary to other approaches to intervention analysis in the context of INAR models which only consider one certain type of outlier. For instance, \cite{fried:2015} and \cite{silva:2015} have focused on additive outliers not entering the dynamics and have treated them through a Bayesian analysis. A possible interesting extension regards the possibility to distinguish and classify different intervention patterns when the uncontaminated process is unobserved. The flexibility of the Bayesian approach is promising in such a framework, but more work is necessary for this.

Another promising line for future research regards detection (and classification) of intervention effects by means of model selection criteria. The modified Bayesian information criterion developed by \cite{galeano:2012} or the modified Akaike information criterion and the average square standardized residuals used by \cite{fokianos:2012} are examples of such criteria.

\section*{Aknowledgements}
This project has received funding from the Athens University of
Economics and Business,  Action I Funding and the European Union's Horizon
2020 research and innovation programme under the Marie
Sk{\l}odowska-Curie grant agreement no. 699980.

\newpage
\section*{Supplementary Materials}

\subsection*{1. The conditional information matrix}

The Hessian matrix $\partial^2\ell(\boldsymbol{\theta})/\partial\boldsymbol{\theta}\partial\boldsymbol{\theta}^T$ has elements
\allowdisplaybreaks
\begin{eqnarray*}
&&\frac{\partial^2\ell(\boldsymbol{\theta})}{\partial\alpha_i^2}=\frac{1}{(1-\alpha_i)^2}\sum_{t=p+1}^{n}y_{t-i}\left\{\frac{2\mathnormal{p}(y_t-1|y_{t-1},\ldots,y_{t-i}-1,\ldots,y_{t-p})}{\mathnormal{p}(y_t|y_{t-1},\ldots,y_{t-p})}-1\right.\\
&&\quad+(y_{t-i}-1)\frac{\mathnormal{p}(y_t-2|y_{t-1},\ldots,y_{t-i}-2,\ldots,y_{t-p})}{\mathnormal{p}(y_t|y_{t-1},\ldots,y_{t-p})}\\
&&\quad\left.-y_{t-i}\left(\frac{\mathnormal{p}(y_{t}-1|y_{t-1},\ldots,y_{t-i}-1,\ldots,y_{t-p})}{\mathnormal{p}(y_t|y_{t-1},\ldots,y_{t-p})}\right)^2\right\}\\
&&\frac{\partial^2\ell(\boldsymbol{\theta})}{\partial\alpha_i\partial\alpha_j}=\frac{1}{(1-\alpha_i)(1-\alpha_j)}\sum_{t=p+1}^ny_{t-i}y_{t-j}\left\{\frac{\mathnormal{p}(y_t-2|y_{t-1},\ldots,y_{t-i}-1,\ldots,y_{t-j}-1,\ldots,y_{t-p})}{\mathnormal{p}(y_t|y_{t-1},\ldots,y_{t-p})}\right.\\
&&\quad\left.-\frac{\mathnormal{p}(y_t-1|y_{t-1},\ldots,y_{t-i}-1,\ldots,y_{t-p})\mathnormal{p}(y_t-1|y_{t-1},\ldots,y_{t-j}-1,\ldots,y_{t-p})}{\mathnormal{p}(y_t|y_{t-1},\ldots,y_{t-p})^2}\right\}\\
&&\frac{\partial^2\ell(\boldsymbol{\theta})}{\partial\lambda^2}=\sum_{t=p+1}^n\left\{\frac{\mathnormal{p}(y_t-2|y_{t-1},\ldots,y_{t-p})}{\mathnormal{p}(y_t|y_{t-1},\ldots,y_{t-p})}-\left(\frac{\mathnormal{p}(y_t-1|y_{t-1},\ldots,y_{t-p})}{\mathnormal{p}(y_t|y_{t-1},\ldots,y_{t-p})}\right)^2\right\}\\
&&\frac{\partial^2\ell(\boldsymbol{\theta})}{\partial\kappa_j\partial\kappa_s}=\sum_{t=p+1}^n\delta^{2t-\tau_j-\tau_s}I(t\geq\tau_j)I(t\geq\tau_s)\left\{\frac{\mathnormal{p}(y_t-2|y_{t-1},\ldots,y_{t-p})}{\mathnormal{p}(y_t|y_{t-1},\ldots,y_{t-p})}\right.\\
&&\quad\left.-\left(\frac{\mathnormal{p}(y_t-1|y_{t-1},\ldots,y_{t-p})}{\mathnormal{p}(y_t|y_{t-1},\ldots,y_{t-p})}\right)^2\right\}\\
&&\frac{\partial^2\ell(\boldsymbol{\theta})}{\partial\alpha_i\partial\lambda}=\frac{1}{1-\alpha_i}\sum_{t=p+1}^{n}y_{t-i}\left\{\frac{\mathnormal{p}(y_t-2|y_{t-1},\ldots,y_{t-i}-1,\ldots,y_{t-p})}{\mathnormal{p}(y_t|y_{t-1},\ldots,y_{t-p})}\right.\\
&&\quad\left.-\frac{\mathnormal{p}(y_t-1|y_{t-1},\ldots,y_{t-p})\mathnormal{p}(y_t-1|y_{t-1},\ldots,y_{t-i}-1,\ldots,y_{t-p})}{\mathnormal{p}(y_t|y_{t-1},\ldots,y_{t-p})^2}\right\}\\
&&\frac{\partial^2\ell(\boldsymbol{\theta})}{\partial\alpha_i\partial\kappa_j}=\frac{1}{1-\alpha_i}\sum_{t=p+1}^{n}y_{t-i}\delta^{t-\tau_j}I(t\geq\tau_j)\left\{\frac{\mathnormal{p}(y_t-2|y_{t-1},\ldots,y_{t-i}-1,\ldots,y_{t-p})}{\mathnormal{p}(y_t|y_{t-1},\ldots,y_{t-p})}\right.\\
&&\quad\left.-\frac{\mathnormal{p}(y_t-1|y_{t-1},\ldots,y_{t-p})\mathnormal{p}(y_t-1|y_{t-1},\ldots,y_{t-i}-1,\ldots,y_{t-p})}{\mathnormal{p}(y_t|y_{t-1},\ldots,y_{t-p})^2}\right\}\\
&&\frac{\partial^2\ell(\boldsymbol{\theta})}{\partial\lambda\partial\kappa_j}=\sum_{t=p+1}^{n}\delta^{t-\tau_j}I(t\geq\tau_j)\left\{\frac{\mathnormal{p}(y_t-2|y_{t-1},\ldots,y_{t-p})}{\mathnormal{p}(y_t|y_{t-1},\ldots,y_{t-p})}-\left(\frac{\mathnormal{p}(y_t-1|y_{t-1},\ldots,y_{t-p})}{\mathnormal{p}(y_t|y_{t-1},\ldots,y_{t-p})}\right)^2\right\}
\end{eqnarray*}
By convention $\mathnormal{p}(y_t-c_1|y_{t-1},\ldots,y_{t-i}-c_2,\ldots,y_{t-p})=0$ for $y_t<c_1$ or $y_{t-i}<c_2$.

The second-order derivatives are functions of $(y_t, y_{t-1},\ldots, y_{t-p})$ and so the elements of $\mathcal I(\boldsymbol{\theta})$ can be obtained as follows:
\allowdisplaybreaks
\begin{eqnarray*}
&&E\left\{\frac{\partial^2\ell(\boldsymbol{\theta})}{\partial\alpha_i^2}\right\}=\sum_{t=p+1}^nE\left\{h(y_t, y_{t-1},\ldots,y_{t-p})\right\}\\
&&\quad=\sum_{t=p+1}^n\sum_{y_t=0}^{\infty}\sum_{y_{t-1}=0}^{\infty}\cdots\sum_{y_{t-p}=0}^{\infty}\mathnormal{p}(y_t,y_{t-1},\ldots,y_{t-p})h(y_t, y_{t-1},\ldots,y_{t-p})\\
&&\quad=\frac{n-p}{(1-\alpha_i)^2}\sum_{y_t=0}^{\infty}\sum_{y_{t-1}=0}^{\infty}\cdots\sum_{y_{t-p}=0}^{\infty}\left(\prod_{k=1}^p\mathnormal{p}(y_{t-k})\right)\\
&&\quad \times y_{t-i}\bigg\{2\mathnormal{p}(y_t-1|y_{t-1},\ldots,y_{t-i}-1,\ldots,y_{t-p})\bigg.\\
&&\quad-\left.\mathnormal{p}(y_t|y_{t-1},\ldots,y_{t-p})+(y_{t-i}-1)\mathnormal{p}(y_t-2|y_{t-1},\ldots,y_{t-i}-2,\ldots,y_{t-p})\right.\\
&&\quad\left.-y_{t-i}\frac{\mathnormal{p}(y_t-1|y_{t-1},\ldots,y_{t-i}-1,\ldots,y_{t-p})^2}{\mathnormal{p}(y_t|y_{t-1},\ldots,y_{t-p})}\right\}\\
&&E\left\{\frac{\partial^2\ell(\boldsymbol{\theta})}{\partial\alpha_i\partial\alpha_j}\right\}=
\frac{n-p}{(1-\alpha_i)(1-\alpha_j)}\sum_{y_t=0}^{\infty}\sum_{y_{t-1}=0}^{\infty}\cdots\sum_{y_{t-p}=0}^{\infty}\left(\prod_{k=1}^p\mathnormal{p}(y_{t-k})\right)\\
&&\quad\times y_{t-i}y_{t-j}\bigg\{\mathnormal{p}(y_t-2|y_{t-1},\ldots,y_{t-i}-1,\ldots,y_{t-j}-1,\ldots,y_{t-p})\bigg.\\
&&\quad\left.-\frac{\mathnormal{p}(y_t-1|y_{t-1},\ldots,y_{t-i}-1,\ldots,y_{t-p})\mathnormal{p}(y_t-1|y_{t-1},\ldots,y_{t-j}-1,\ldots,y_{t-p})}{\mathnormal{p}(y_t|y_{t-1},\ldots,y_{t-p})}\right\}\\
&&E\left\{\frac{\partial^2\ell(\boldsymbol{\theta})}{\partial\lambda^2}\right\}=(n-p)\sum_{y_t=0}^{\infty}\sum_{y_{t-1}=0}^{\infty}\cdots\sum_{y_{t-p}=0}^\infty\left(\prod_{k=1}^p\mathnormal{p}(y_{t-k})\right)\\
&&\quad\times\left\{\mathnormal{p}(y_t-2|y_{t-1},\ldots,y_{t-p})-\frac{\mathnormal{p}(y_t-1|y_{t-1},\ldots,y_{t-p})^2}{\mathnormal{p}(y_t|y_{t-1},\ldots,y_{t-p})}\right\}\\
&&E\left\{\frac{\partial^2\ell(\boldsymbol{\theta})}{\partial\kappa_j\kappa_s}\right\}=\sum_{t=p+1}^n\delta^{2t-\tau_j-\tau_s}I(t\geq\tau_j)I(t\geq\tau_s)\sum_{y_t=0}^{\infty}\sum_{y_{t-1}=0}^{\infty}\cdots\sum_{y_{t-p}=0}^\infty\left(\prod_{k=1}^p\mathnormal{p}(y_{t-k})\right)\\
&&\quad\times\left\{\mathnormal{p}(y_t-2|y_{t-1},\ldots,y_{t-p})-\frac{\mathnormal{p}(y_t-1|y_{t-1},\ldots,y_{t-p})^2}{P(y_t|y_{t-1},\ldots,y_{t-p})}\right\}\\
&&E\left\{\frac{\partial^2\ell(\boldsymbol{\theta})}{\partial\alpha_i\partial\lambda}\right\}=\frac{n-p}{1-\alpha_i}\sum_{y_t=0}^{\infty}\sum_{y_{t-1}=0}^{\infty}\cdots\sum_{y_{t-p}=0}^\infty\left(\prod_{k=1}^p\mathnormal{p}(y_{t-k})\right)\\
&&\quad\times y_{t-i}\bigg\{\mathnormal{p}(y_t-2|y_{t-1},\ldots,y_{t-i}-1,\ldots,y_{t-p})\bigg.\\
&&\quad\left.-\frac{\mathnormal{p}(y_t-1|y_{t-1},\ldots,y_{t-p})\mathnormal{p}(y_t-1|y_{t-1},\ldots,y_{t-i}-1,\ldots,y_{t-p})}{\mathnormal{p}(y_t|y_{t-1},\ldots,y_{t-p})}\right\}\\
&&E\left\{\frac{\partial^2\ell(\boldsymbol{\theta})}{\partial\alpha_i\partial\kappa_j}\right\}=\frac{1}{1-\alpha_i}\sum_{t=p+1}^n\delta^{t-\tau_j}I(t\geq\tau_j)\sum_{y_t=0}^{\infty}\sum_{y_{t-1}=0}^{\infty}\cdots\sum_{y_{t-p}=0}^\infty\left(\prod_{k=1}^p\mathnormal{p}(y_{t-k})\right)\\
&&\quad\times y_{t-i}\bigg\{\mathnormal{p}(y_t-2|y_{t-1},\ldots,y_{t-i}-1,\ldots,y_{t-p})-\bigg.\\
&&\quad-\left.\frac{\mathnormal{p}(y_t-1|y_{t-1},\ldots,y_{t-p})\mathnormal{p}(y_t-1|y_{t-1},\ldots,y_{t-i}-1,\ldots,y_{t-p})}{\mathnormal{p}(y_t|y_{t-1},\ldots,y_{t-p})}\right\}\\
&&E\left\{\frac{\partial^2\ell(\boldsymbol{\theta})}{\partial\lambda\partial\kappa_j}\right\}=\sum_{t=p+1}^n\delta^{t-\tau_j}I(t\geq\tau_j)\sum_{y_t=0}^{\infty}\sum_{y_{t-1}=0}^{\infty}\cdots\sum_{y_{t-p}=0}^\infty\left(\prod_{k=1}^p\mathnormal{p}(y_{t-k})\right)\\
&&\quad\times\left\{\mathnormal{p}(y_t-2|y_{t-1},\ldots,y_{t-p})-\frac{\mathnormal{p}(y_t-1|y_{t-1},\ldots,y_{t-p})^2}{\mathnormal{p}(y_t|y_{t-1},\ldots,y_{t-p})}\right\}
\end{eqnarray*}
In practice, the elements of $\mathcal I(\boldsymbol{\theta})$ are calculated by truncating the infinite sums to some value $m$ selected such that $\mathnormal{p}(y_t>m)$ is approximately equal to zero. We select $m$ so that $\mathnormal{p}(y_t>m)\leq10^{-15}$.

\subsection*{2. Tables}

\setcounter{table}{0}
\renewcommand{\thetable}{SM2.\arabic{table}}

\begingroup
\vfill
\centering 
\begin{sideways}
  \setlength{\tabcolsep}{3pt}
  \renewcommand{\arraystretch}{0.8} 
  \begin{threeparttable}
\caption{\label{tab:sizes100} Empirical sizes (in percent) of the tests  based on the  $F$-type statistics and score statistics for a known type $\delta$ and time $\tau$ of intervention  in case of  INAR(1) series of length $n=100$ with different parameters $\alpha$ and $\lambda$. The nominal significance levels are 1\%, 5\% or 10\%.}
{\footnotesize
\begin{tabular}{rrr|rrr|rrr|rrr|rrr|rrr|rrr}
\hline
&& & \multicolumn{9}{c|}{$F$-type statistics} & \multicolumn{9}{c}{Score statistics}\\
\hline
 && & \multicolumn{3}{c|}{$\tau=0.25n$}& \multicolumn{3}{c|}{$\tau=0.5n$}& \multicolumn{3}{c|}{$\tau=0.75n$} & \multicolumn{3}{c|}{$\tau=0.25n$}& \multicolumn{3}{c|}{$\tau=0.5n$}& \multicolumn{3}{c}{$\tau=0.75n$} \\
$\alpha$ & $\lambda$ & $\delta$ &  1\% & 5\% & 10\%& 1\% & 5\% & 10\%& 1\% & 5\% & 10\% &  1\% & 5\% & 10\%& 1\% & 5\% & 10\%& 1\% & 5\% & 10\%\\ \hline
0.3&2&0 &   1.2	&4.1	&9.2	&1.8	&4.9	&9.9	&1.3	&4.9	&9.8 & 1.5 & 4.1 & 7.7 & 1.7 & 4.4 & 7.8 & 1.7 & 4.6 & 8.3 \\ 
&&0.8 &  1.4	& 5.5	&10.7	&1.6	&5.9	&11.2	&1.5	&5.9	&11.1& 1.0 & 5.0 & 9.6 & 1.1 & 4.9 & 10.3 & 1.0 & 5.4 & 10.5 \\ 
&&1 & 1.5	&6.4	&11.4	&1.6	&6.6	&11.7	&1.2	&5.6	&11.0& 0.9 & 5.0 & 9.6 & 0.6 & 4.8 & 9.4 & 1.0 & 5.4 & 10.3 \\  \hline
&5&0 &1.2	&5.0	&10.1	&1.2	&4.5	&9.4	&1.2	&5.0	&10.0 & 1.3 & 4.1 & 8.8 & 1.1 & 4.7 & 9.4 & 1.6 & 4.7 & 9.4 \\ 
&&0.8 & 1.5	&6.0	&10.9	&1.3	&5.7	&10.7	&1.7	&5.7	&11.0& 0.9 & 4.3 & 9.2 & 0.9 & 5.2 & 10.3 & 1.2 & 5.0 & 9.9\\
\hline
\end{tabular}}
  \end{threeparttable}
\end{sideways}
\vfill
\endgroup

\begin{landscape}
\begin{table}
\caption{\label{tab:sizes200} Empirical sizes (in percent) of the tests  based on the  $F$-type statistics and score statistics for a known type $\delta$ and time $\tau$ of intervention  in case of  INAR(1) series of length $n=200$ with different parameters $\alpha$ and $\lambda$. The nominal significance levels are 1\%, 5\% or 10\%.}
\begin{center}
{\footnotesize
\begin{tabular}{rrr|rrr|rrr|rrr|rrr|rrr|rrr}
\hline
&& & \multicolumn{9}{c|}{$F$-type statistics} & \multicolumn{9}{c}{Score statistics}\\
\hline
 && & \multicolumn{3}{c|}{$\tau=0.25n$}& \multicolumn{3}{c|}{$\tau=0.5n$}& \multicolumn{3}{c|}{$\tau=0.75n$} & \multicolumn{3}{c|}{$\tau=0.25n$}& \multicolumn{3}{c|}{$\tau=0.5n$}& \multicolumn{3}{c}{$\tau=0.75n$} \\
$\alpha$ & $\lambda$ & $\delta$ &  1\% & 5\% & 10\%& 1\% & 5\% & 10\%& 1\% & 5\% & 10\% &  1\% & 5\% & 10\%& 1\% & 5\% & 10\%& 1\% & 5\% & 10\%\\ \hline
0.3&2&0 & 1.2&	4.3	&9.0	&1.4	&4.8	&9.4	&1.6	&4.3&	8.9 & 1.4 & 3.5 & 7.3 & 1.6 & 4.3 & 7.9 & 1.7 & 4.2 & 7.7 \\ 
&&0.8 &1.3	&5.2	&10.2	&1.5	&5.1	&10.3	&1.3	&5.2	&9.9& 1.1 & 4.9 & 10.0 & 0.9 & 4.5 & 9.9 & 1.2 & 5.0 & 10.4 \\ 
&&1 &0.9	&4.9	&10.2	&1.2	&5.0	&10.2	&1.0	&4.6	&9.8& 0.9 & 4.5 & 9.4 & 0.9 & 4.7 & 9.7 & 0.9 & 5.2 & 10.2 \\  \hline
&5&0 &1.0	&5.4	&10.2	&0.9	&4.6	&10.0	&1.1	&4.5	&9.3& 1.3 & 4.6 & 8.7 & 1.0 & 4.1 & 8.5 & 1.4 & 4.7 & 9.6 \\ 
&&0.8 &1.2	&5.6	&10.4	&1.2	&5.1	&10.5	&1.4	&6.0	&10.8& 1.2 & 4.7 & 10.1 & 1.0 & 4.7 & 9.4 & 1.2 & 5.1 & 10.3 \\
\hline
\end{tabular}}
\end{center}
\end{table}

\begin{table}
\caption{\label{tab:power200-0} Empirical power (in percent) of the tests  based on the  $F$-type statistics and score statistics for a known time $\tau$ and known, but possibly misspecified type $\delta$ of intervention  in case of an innovation outlier $\delta=0$ of size $\kappa=3\sqrt{\lambda}$ at time $\tau$ in an INAR(1) series of length $n=200$ with different parameters $\alpha$ and $\lambda$. The nominal significance levels are 1\%, 5\% or 10\%.}
\begin{center}
{\footnotesize
\begin{tabular}{rrr|rrr|rrr|rrr|rrr|rrr|rrr}
\hline
&& & \multicolumn{9}{c|}{$F$-type statistics} & \multicolumn{9}{c}{Score statistics}\\
\hline
 && & \multicolumn{3}{c|}{$\tau=0.25n$}& \multicolumn{3}{c|}{$\tau=0.5n$}& \multicolumn{3}{c|}{$\tau=0.75n$} & \multicolumn{3}{c|}{$\tau=0.25n$}& \multicolumn{3}{c|}{$\tau=0.5n$}& \multicolumn{3}{c}{$\tau=0.75n$} \\
$\alpha$ & $\lambda$ & $\delta$ &  1\% & 5\% & 10\%& 1\% & 5\% & 10\%& 1\% & 5\% & 10\% &  1\% & 5\% & 10\%& 1\% & 5\% & 10\%& 1\% & 5\% & 10\%\\ \hline

0.3&2 & 0&48.3	&64.3	&71.7	&49.3	&65.3	&73.3	&50.4	&65.8	&72.4& 51.1 & 66.2 & 73.2 & 52.1 & 66.2 & 72.7 & 51.4 & 65.3 & 72.5 \\ 
&&0.8 &21.8	&37.1	&47.1	&23.4	&37.6	&47.1	&22.9	&37.1	&47.6& 24.9 & 42.2 & 51.2 & 25.8 & 41.2 & 49.3 & 23.7 & 39.7 & 49.0 \\ 
&&1 &1.0	&4.3	&10.2	&1.2	&5.8	&11.6	&1.6	&6.7	&11.9& 0.9 & 4.7 & 9.9 & 1.0 & 5.5 & 11.2 & 1.9 & 7.6 & 13.1 \\ \hline
&5& 0&49.3	&66.4	&74.5	&50.1	&65.8	&75.6	&49.9	&66.7	&75.3& 51.5 & 67.0 & 74.2 & 53.5 & 68.6 & 76.7 & 50.6 & 66.0 & 73.2 \\ 
&&0.8 &18.9	&36.1	&45.7	&20.9	&37.0	&45.3	&19.9	&36.3	&46.5& 21.6 & 38.6 & 48.6 & 22.9 & 40.0 & 49.3 & 21.1 & 39.7 & 48.9 \\
\hline
\end{tabular}}
\end{center}
\end{table}

\begin{table}
\caption{\label{tab:power200-0.8} Empirical power (in percent) of the tests  based on the  $F$-type statistics and score statistics for a known type $\delta$ and time $\tau$ of intervention  in case of a transient shift $\delta=0.8$ of size $\kappa=2\sqrt{\lambda}$ at time $\tau$ in an INAR(1) series of length $n=200$ with several parameters $\alpha$ and $\lambda$. The nominal significance levels are 1\%, 5\% or 10\%.}
\begin{center}
{\footnotesize
\begin{tabular}{rrr|rrr|rrr|rrr|rrr|rrr|rrr}
\hline
&& & \multicolumn{9}{c|}{$F$-type statistics} & \multicolumn{9}{c}{Score statistics}\\
\hline
 && & \multicolumn{3}{c|}{$\tau=0.25n$}& \multicolumn{3}{c|}{$\tau=0.5n$}& \multicolumn{3}{c|}{$\tau=0.75n$} & \multicolumn{3}{c|}{$\tau=0.25n$}& \multicolumn{3}{c|}{$\tau=0.5n$}& \multicolumn{3}{c}{$\tau=0.75n$} \\
$\alpha$ & $\lambda$ & $\delta$ &  1\% & 5\% & 10\%& 1\% & 5\% & 10\%& 1\% & 5\% & 10\% &  1\% & 5\% & 10\%& 1\% & 5\% & 10\%& 1\% & 5\% & 10\%\\ \hline

0.3&2&0& 24.4	&38.6	&47.3	&24.8	&39.6	&48.4	&25.1	&39.0	&47.6 &   28.4 & 41.4 & 49.2 & 29.3 & 41.9 & 48.9 & 31.9 & 44.8 & 52.0 \\ 
 &&0.8&56.1	&72.5	&79.3	&56.6	&74.0	&80.4	&53.4	&71.8	&79.8&  50.1 & 70.9 & 79.2 & 51.0 & 70.6 & 79.3 & 54.5 & 71.6 & 80.5 \\ 
&&1 &0.6	&4.7	&9.6	&2.2	&7.9	&14.1	&5.3	&16.0	&24.9& 0.9 & 4.7 & 9.8 & 2.4 & 8.1 & 14.1 & 5.5 & 16.6 & 26.3 \\  \hline
&5&0 &23.0	&38.6	&50.0	&23.4	&39.9	&48.3	&23.1	&40.0	&50.9&  26.9 & 42.6 & 51.9 & 26.1 & 40.6 & 48.6 & 28.6 & 45.0 & 52.9 \\ 
&&0.8 &57.1	&74.5	&82.2	&56.0	&74.9	&83.3	&57.5	&75.2	&82.8&  53.2 & 73.7 & 80.9 & 53.0 & 72.8 & 81.5 & 53.2 & 74.6 & 83.1 \\
\hline
\end{tabular}}
\end{center}
\end{table}

\begin{table}
\caption{\label{tab:power200-1} Empirical power (in percent) of the tests  based on the  $F$-type statistics and score statistics for a known type $\delta$ and time $\tau$ of intervention  in case of a level shift $\delta=1$ of size $\kappa=\sqrt{\lambda}$ at time $\tau$ in an INAR(1) series of length $n=200$ with several parameters $\alpha$ and $\lambda$. The nominal significance levels are 1\%, 5\% or 10\%.}
\begin{center}
{\footnotesize
\begin{tabular}{rrr|rrr|rrr|rrr|rrr|rrr|rrr}
\hline
&& & \multicolumn{9}{c|}{$F$-type statistics} & \multicolumn{9}{c}{Score statistics}\\
\hline
 && & \multicolumn{3}{c|}{$\tau=0.25n$}& \multicolumn{3}{c|}{$\tau=0.5n$}& \multicolumn{3}{c|}{$\tau=0.75n$} & \multicolumn{3}{c|}{$\tau=0.25n$}& \multicolumn{3}{c|}{$\tau=0.5n$}& \multicolumn{3}{c}{$\tau=0.75n$} \\
$\alpha$ & $\lambda$ & $\delta$ &  1\% & 5\% & 10\%& 1\% & 5\% & 10\%& 1\% & 5\% & 10\% &  1\% & 5\% & 10\%& 1\% & 5\% & 10\%& 1\% & 5\% & 10\%\\ \hline

0.3&2&0&1.7	&5.2	&9.8	&3.2	&8.9	&14.4	&5.5	&12.2	&18.0&  4.0 & 9.2 & 14.1 & 5.3 & 11.5 & 17.1 & 9.2 & 17.2 & 24.1 \\ 
&&0.8 &1.6	&6.2	&10.2	&4.7	&15.4	&24.8	&16.8	&34.2	&45.8 &2.8 & 9.2 & 15.1 & 8.8 & 20.2 & 30.4 & 22.1 & 42.8 & 55.0 \\
&&1 &96.1	&99.3	&99.7	&99.5	&100.0	&100.0&98.2	&99.5	&99.8&   97.3 & 99.3 & 99.8 & 99.8 & 100.0 & 100.0 & 98.2 & 99.5 & 100.0 \\  \hline
&5&0 &1.7	&5.0	&10.3	&2.6	&7.7	&12.3	&3.6	&11.1	&17.6&   2.7 & 9.2 & 14.8 & 5.8 & 12.4 & 18.7 & 7.0 & 16.1 & 23.1 \\ 
&&0.8 &1.5	&6.2	&11.2	&4.7	&13.7	&22.8	&15.7	&36.2	&47.6 & 2.7 & 9.3 & 15.3 & 9.1 & 23.8 & 31.9 & 20.6 & 43.1 & 54.4 \\
\hline
\end{tabular}}
\end{center}
\end{table}

\begin{table}
\caption{\label{tab:classif200} Classification rates (in percent) achieved by applying the F-type statistics and score statistics for a known type $\delta$ and time $\tau$ of intervention  to clean  INAR(1) series of length $n=200$ without outliers. The nominal significance levels are 1\%, 5\% or 10\%.}
\begin{center}
{\footnotesize
\begin{tabular}{rrr|rrr|rrr|rrr|rrr|rrr|rrr}
\hline
&& & \multicolumn{9}{c|}{$F$-type statistics} & \multicolumn{9}{c}{Score statistics}\\
\hline
 && & \multicolumn{3}{c|}{$\tau=0.25n$}& \multicolumn{3}{c|}{$\tau=0.5n$}& \multicolumn{3}{c|}{$\tau=0.75n$} & \multicolumn{3}{c|}{$\tau=0.25n$}& \multicolumn{3}{c|}{$\tau=0.5n$}& \multicolumn{3}{c}{$\tau=0.75n$} \\
$\alpha$ & $\lambda$ & $\delta$ &  1\% & 5\% & 10\%& 1\% & 5\% & 10\%& 1\% & 5\% & 10\% &  1\% & 5\% & 10\%& 1\% & 5\% & 10\%& 1\% & 5\% & 10\%\\ \hline

0.3&2&0 &     1.3 &   3.6 &   6.5 &   0.9 &   3.3 &   6.0   &   1.0   &   3.7 &   6.9 &  1.3 & 2.5 & 5.2 & 1.5 & 3.4 & 5.8 & 1.6 & 3.4 & 5.6 \\
&&0.6 &   0.6 &   2.5 &   4.5 &   0.6 &   1.9 &   3.7 &   0.6 &   2.4 &   4.6 &  0.6 & 1.8 & 3.8 & 0.8 & 2.2 & 4.0 & 0.5 & 1.5 & 3.4 \\ 
&&0.8 &   0.4 &   2.1 &   3.4 &   0.5 &   1.7 &   3.0   &   0.5 &   1.8 &   3.3 & 0.2 & 1.2 & 2.8 & 0.3 & 1.7 & 3.1 & 0.7 & 2.3 & 3.5 \\ 
&&0.9 &   0.9 &  2.9 &   5.8 &   0.9 &   3.5 &   6.7 &   0.6 &   2.6 &   4.6 &  0.6 & 3.3 & 6.5 & 0.5 & 2.5 & 5.1 & 0.7 & 2.5 & 5.5 \\ 
&&1 &   0.9 &    4.4 &   9.2 &   1.2 &   5.2 &   9.7 &   0.9 &   4.9 &   8.7 &  0.9 & 4.5 & 8.6 & 0.8 & 4.8 & 8.3 & 0.8 & 4.7 & 9.0 \\
\hline
\end{tabular}}
\end{center}
\end{table}

\begin{table}
\caption{\label{tab:classif200-0} Classification rates (in percent) achieved by applying the $F$-type statistics and score statistics for a known type $\delta$ and time $\tau$ of intervention to INAR(1) series of length $n=200$ with an innovation outlier ($\delta=0$) of size $\kappa=3\sqrt{\lambda}$ at time $\tau$. The nominal significance levels are 1\%, 5\% or 10\%.}
\begin{center}
{\footnotesize
\begin{tabular}{rrr|rrr|rrr|rrr|rrr|rrr|rrr}
\hline
&& & \multicolumn{9}{c|}{$F$-type statistics} & \multicolumn{9}{c}{Score statistics}\\
\hline
 && & \multicolumn{3}{c|}{$\tau=0.25n$}& \multicolumn{3}{c|}{$\tau=0.5n$}& \multicolumn{3}{c|}{$\tau=0.75n$} & \multicolumn{3}{c|}{$\tau=0.25n$}& \multicolumn{3}{c|}{$\tau=0.5n$}& \multicolumn{3}{c}{$\tau=0.75n$} \\
$\alpha$ & $\lambda$ & $\delta$ &  1\% & 5\% & 10\%& 1\% & 5\% & 10\%& 1\% & 5\% & 10\% &  1\% & 5\% & 10\%& 1\% & 5\% & 10\%& 1\% & 5\% & 10\%\\ \hline

0.3&2&0 &    40.6 &    51.6 &    56.4 &    40.7 &    52   &    56.1 &    40.7 &    51.7 &    56.4 & 47.4 & 58.9 & 63.6 & 47.9 & 58.0 & 62.5 & 47.1 & 58.9 & 63.7 \\ 
&&0.6 &      8.7  &    11.6 &    12.9 &    8.2  &    10.8 &    11.9 &    9.6  &    12.3 &    13.5 & 4.3 & 6.9 & 7.8 & 5.0 & 7.2 & 8.2 & 4.6 & 7.1 & 7.9 \\ 
&&0.8 &  1.5 &   2.5 &   3.0   &   1.8 &   2.6 &   3.0   &   1.7 &   2.6 &   2.9 &    0.8 & 1.9 & 2.4 & 1.0 & 1.6 & 1.9 & 0.4 & 1.2 & 1.8 \\ 
&&0.9 &    1.0   &   1.9 &   2.5 &   1.2 &   2.3 &   3.0   &   1.0   &   1.8 &   2.2 &   0.7 & 1.6 & 2.8 & 0.8 & 1.8 & 2.4 & 0.4 & 1.4 & 2.1 \\ 
&&1 &   0.7 &   2.4 &   3.8 &   0.7 &   2.4 &   3.8 &   0.6 &   2.0   &   3.2 &  0.6 & 1.9 & 3.1 & 0.3 & 1.8 & 3.1 & 0.4 & 1.5 & 2.6 \\
\hline
\end{tabular}}
\end{center}
\end{table}

\begin{table}
\caption{\label{tab:classif200-6.1} Classification rates (in percent) achieved by applying the F-type statistics and score statistics for a known type $\delta$ and time $\tau$ of intervention to INAR(1) series of length $n=200$ with a transient shift with $\delta=0.6$ and size $\kappa=2.5\sqrt{\lambda}$ at time $\tau$. The nominal significance levels are 1\%, 5\% or 10\%.}
\begin{center}
{\footnotesize
\begin{tabular}{rrr|rrr|rrr|rrr|rrr|rrr|rrr}
\hline
&& & \multicolumn{9}{c|}{$F$-type statistics} & \multicolumn{9}{c}{Score statistics}\\
\hline
 && & \multicolumn{3}{c|}{$\tau=0.25n$}& \multicolumn{3}{c|}{$\tau=0.5n$}& \multicolumn{3}{c|}{$\tau=0.75n$} & \multicolumn{3}{c|}{$\tau=0.25n$}& \multicolumn{3}{c|}{$\tau=0.5n$}& \multicolumn{3}{c}{$\tau=0.75n$} \\
$\alpha$ & $\lambda$ & $\delta$ &  1\% & 5\% & 10\%& 1\% & 5\% & 10\%& 1\% & 5\% & 10\% &  1\% & 5\% & 10\%& 1\% & 5\% & 10\%& 1\% & 5\% & 10\%\\ \hline

0.3&2&0  &    14.5 &    18.3 &    20.1 &    14.8 &    18.6 &    20.4 &    14.9 &    18.8 &    20.7 &  22.9 & 27.8 & 29.0 & 22.8 & 26.4 & 27.9 & 21.4 & 25.2 & 26.7 \\ 
&&0.6 &    28.5 &    35.3 &    37.7 &    27.6 &    33.6 &    36.1 &    27.0   &    33.5 &    35.8 &   22.8 & 28.9 & 30.6 & 23.9 & 30.9 & 33.2 & 23.1 & 29.7 & 31.9 \\ 
&&0.8 &     11.3 &    14.7 &    16.1 &    12.1 &    15.5 &    16.9 &    11.2 &    14.9 &    16.7 &  7.6 & 10.8 & 12.6 & 7.8 & 11.2 & 13.2 & 8.8 & 12.6 & 14.6 \\
&&0.9 &   4.4 &   7.2 &   8.3 &   4.3 &   6.5 &   7.8 &   4.6 &   6.4 &   7.6 & 2.8 & 5.8 & 7.1 & 2.6 & 4.9 & 5.6 & 2.6 & 4.5 & 5.8 \\
&&1 &   0.6 &   2.2 &   3.5 &   0.5 &   1.7 &   2.4 &   0.8 &   2.2 &   3.1 & 0.2 & 1.1 & 2.4 & 0.8 & 1.8 & 2.8 & 0.6 & 1.6 & 2.5 \\
\hline
\end{tabular}}
\end{center}
\end{table}

\begin{table}
\caption{\label{tab:classif200-8.1} Classification rates (in percent) achieved by applying the F-type statistics and score statistics for a known type $\delta$ and time $\tau$ of intervention to INAR(1) series of length $n=200$ with a transient shift with $\delta=0.8$ and size $\kappa=2\sqrt{\lambda}$ at time $\tau$. The nominal significance levels are 1\%, 5\% or 10\%.}
\begin{center}
{\footnotesize
\begin{tabular}{rrr|rrr|rrr|rrr|rrr|rrr|rrr}
\hline
&& & \multicolumn{9}{c|}{$F$-type statistics} & \multicolumn{9}{c}{Score statistics}\\
\hline
 && & \multicolumn{3}{c|}{$\tau=0.25n$}& \multicolumn{3}{c|}{$\tau=0.5n$}& \multicolumn{3}{c|}{$\tau=0.75n$} & \multicolumn{3}{c|}{$\tau=0.25n$}& \multicolumn{3}{c|}{$\tau=0.5n$}& \multicolumn{3}{c}{$\tau=0.75n$} \\
$\alpha$ & $\lambda$ & $\delta$ &  1\% & 5\% & 10\%& 1\% & 5\% & 10\%& 1\% & 5\% & 10\% &  1\% & 5\% & 10\%& 1\% & 5\% & 10\%& 1\% & 5\% & 10\%\\ \hline

0.3&2&0&    6.5 &   8.5 &   9.5 &   6.8 &   8.4 &   9.4 &   6.3 &   8.0   &   9.2 & 10.8 & 12.8 & 13.7 & 10.6 & 12.5 & 13.1 & 12.1 & 14.3 & 15.0 \\ 
&&0.6 &   15.9 &    19.0   &    20.4 &    14.4 &    17.3 &    18.4 &    14.5 &    18.1 &    19.4 &   16.1 & 19.2 & 20.3 & 16.8 & 20.2 & 21.6 & 18.1 & 20.9 & 22.2 \\ 
&&0.8 &    24.2 &    28.9 &    31.0   &    24.9 &    30.2 &    31.8 &    24.0   &    29.5 &    31.2 &  18.6 & 24.7 & 27.1 & 18.4 & 24.3 & 26.8 & 19.0 & 23.8 & 25.9 \\
&&0.9 &    15.5 &    22.1 &    24.9 &    16.9 &    23.0   &    25.1 &    16.4 &    22.0   &    23.8 &   12.6 & 19.8 & 22.4 & 13.0 & 19.9 & 21.9 & 13.1 & 18.7 & 21.6 \\ 
&&1 &    0.5 &   1.4 &   2.3 &   0.7 &   1.7 &   2.5 &   1.3 &   2.5 &   3.3 & 0.4 & 1.1 & 2.3 & 0.6 & 1.6 & 2.4 & 0.6 & 1.6 & 2.7 \\
\hline
\end{tabular}}
\end{center}
\end{table}

\begin{table}
\caption{\label{tab:classif200-9.1} Classification rates (in percent) achieved by applying the F-type statistics and score statistics for a known type $\delta$ and time $\tau$ of intervention to INAR(1) series of length $n=200$ with a transient shift with $\delta=0.9$ and size $\kappa=1.5\sqrt{\lambda}$ at time $\tau$. The nominal significance levels are 1\%, 5\% or 10\%.}
\begin{center}
{\footnotesize
\begin{tabular}{rrr|rrr|rrr|rrr|rrr|rrr|rrr}
\hline
&& & \multicolumn{9}{c|}{$F$-type statistics} & \multicolumn{9}{c}{Score statistics}\\
\hline
 && & \multicolumn{3}{c|}{$\tau=0.25n$}& \multicolumn{3}{c|}{$\tau=0.5n$}& \multicolumn{3}{c|}{$\tau=0.75n$} & \multicolumn{3}{c|}{$\tau=0.25n$}& \multicolumn{3}{c|}{$\tau=0.5n$}& \multicolumn{3}{c}{$\tau=0.75n$} \\
$\alpha$ & $\lambda$ & $\delta$ &  1\% & 5\% & 10\%& 1\% & 5\% & 10\%& 1\% & 5\% & 10\% &  1\% & 5\% & 10\%& 1\% & 5\% & 10\%& 1\% & 5\% & 10\%\\ \hline

0.3&2&0 &  3.6 &   5.1 &   6.1 &   3.5 &   4.9 &   5.8 &   3.7 &   5.3 &   6.1 &  6.0 & 7.6 & 8.4 & 6.3 & 7.6 & 8.4 & 6.8 & 8.1 & 9.0 \\ 
&&0.6 &  5.3 &   7.0   &   7.5 &   5.1 &   6.6 &   7.1 &   5.7 &   7.0   &   7.7 & 7.6 & 9.4 & 9.8 & 7.3 & 9.2 & 9.8 & 6.2 & 7.3 & 7.8 \\ 
&&0.8 &    13.7 &    16.5 &    17.6 &    14.0   &    16.9 &    17.7 &    13.1 &    16.0   &    17.2 & 11.8 & 14.8 & 16.7 & 13.9 & 16.9 & 17.8 & 11.1 & 13.8 & 15.0 \\
&&0.9 &   37.4 &    48.5 &    53.0   &    37.1 &    48.1 &    52.6 &    36.0   &    45.7 &    49.0   &  27.9 & 41.9 & 47.1 & 30.5 & 43.0 & 47.5 & 32.1 & 43.3 & 47.5 \\ 
&&1 &   0.4  &    1.7  &    2.5  &    1.1  &    2.7  &    3.8  &    3.8  &    6.6  &    7.8  &  0.4 & 1.2 & 2.2 & 1.0 & 2.3 & 3.2 & 3.1 & 5.5 & 6.8 \\
\hline
\end{tabular}}
\end{center}
\end{table}

\begin{table}
\caption{\label{tab:classif200-1} Classification rates (in percent) achieved by applying the F-type statistics and score statistics for a known type $\delta$ and time $\tau$ of intervention to INAR(1) series of length $n=200$ with a permanent shift with $\delta=1$ and size $\kappa=\sqrt{\lambda}$ at time $\tau$. The nominal significance levels are 1\%, 5\% or 10\%.}
\begin{center}
{\footnotesize
\begin{tabular}{rrr|rrr|rrr|rrr|rrr|rrr|rrr}
\hline
&& & \multicolumn{9}{c|}{$F$-type statistics} & \multicolumn{9}{c}{Score statistics}\\
\hline
 && & \multicolumn{3}{c|}{$\tau=0.25n$}& \multicolumn{3}{c|}{$\tau=0.5n$}& \multicolumn{3}{c|}{$\tau=0.75n$} & \multicolumn{3}{c|}{$\tau=0.25n$}& \multicolumn{3}{c|}{$\tau=0.5n$}& \multicolumn{3}{c}{$\tau=0.75n$} \\
$\alpha$ & $\lambda$ & $\delta$ &  1\% & 5\% & 10\%& 1\% & 5\% & 10\%& 1\% & 5\% & 10\% &  1\% & 5\% & 10\%& 1\% & 5\% & 10\%& 1\% & 5\% & 10\%\\ \hline

0.3&2&0&  0.1 &   0.1 &   0.1 &   0.0   &   0.0   &   0.0   &   0.3 &   0.3 &   0.3 &  0.4 & 0.5 & 0.5 & 0.2 & 0.2 & 0.2 & 0.8 & 0.8 & 0.8 \\ 
&&0.6 &   0.1 &   0.1 &   0.1 &   0.1 &   0.1 &   0.1 &   0.1 &   0.1 &   0.1 &  0.1 & 0.2 & 0.3 & 0.0 & 0.0 & 0.0 & 0.4 & 0.5 & 0.5 \\ 
&&0.8 &   0.1 &   0.1 &   0.1 &   0.0   &   0.0   &   0.0   &   0.1 &   0.2 &   0.2 &   0.0 & 0.0 & 0.0 & 0.0 & 0.0 & 0.0 & 0.1 & 0.1 & 0.1 \\ 
&&0.9 &    0.2  &    0.3  &    0.3  &    0.2  &    0.2  &    0.2  &    1.9  &    2.0    &    2.0    &  0.0 & 0.0 & 0.0 & 0.0 & 0.0 & 0.0 & 1.8 & 1.8 & 1.8 \\ 
&&1 &     95.4 &    98.8 &    99.2 &    99.2 &    99.6 &    99.6 &    95.7 &    97.0   &    97.3 &    96.9 & 98.6 & 99.0 & 99.5 & 99.7 & 99.7 & 95.1 & 96.3 & 96.8 \\
\hline
\end{tabular}}
\end{center}
\end{table}
\end{landscape}

\newpage
\subsection*{3. Simulation results for the INAR(2) process with Poisson innovations}

\setcounter{table}{0}
\renewcommand{\thetable}{SM3.\arabic{table}}

\setcounter{figure}{0}
\renewcommand{\thefigure}{SM3.\arabic{figure}}

In this section we study the performance of the $F$-type statistic (\ref{eq:ftest}) for the detection of outliers in an INAR(2) model with Poisson innovations. 
For $p=2$, the contaminated model (\ref{eq:cont-inarp}) takes the form
$$ Y_t=\alpha_1\circ Y_{t-1}+\alpha_2\circ Y_{t-2}+e_t +\sum_{j=1}^J U_{t,j}, \quad t\in\mathbb{N},$$
where $e_t\sim Pois(\lambda)$ and $(U_{t,j} : t\in\mathbb{N})$, $j=1,\ldots,J$ are independent random variables such that $U_{t,j}\equiv 0$ for $t=0,\ldots,\tau_j-1$
and $U_{t,j}\sim Pois(\kappa_j\delta_j^{t-\tau_j})$ for $t=\tau_j,\tau_j+1,\ldots$.

Tables \ref{tab3:sizes100} and \ref{tab3:sizes200} report empirical rejection rates when testing for an intervention effect of known type $\delta\in \{0,0.8,1\}$ at a known time point $\tau\in \{0.25n,0.5n,0.75n\}$, using the 90\%, 95\% or 99\% quantile of the $\chi_1^2$-distribution as critical value for the $F$-type statistic.
 The empirical rejection rates are obtained by analyzing 5000 time series of the same length $n\in \{100,200\}$ for each of
 different INAR(2) models with $\{\alpha_1,\alpha_2\}=\{(0.5,0.3), (0.3,0.4), (0.1,0.1)\}$ and $\lambda\in \{2,5\}$.

The $F$-type statistics for innovation outliers ($\delta=0$) achieve empirical rejection rates that are close to the target significance levels 1\%, 5\% and 10\% we aim at already in case of series of length $n=100$ and irrespective of the time $\tau$. The results are somewhat worse for larger values of $\delta$, particularly if the mean $\lambda/(1-\alpha_1-\alpha_2)$ of the series is large.  
 The results improve for larger values of the series length $n$. In the case of $n=200$, all $F$-type statistics achieve the target significance level well although they are slightly oversized when testing for a transient shift or a permanent level shift.

Next we examine the empirical power of these approximate significance tests for a single intervention effect at a known time point.
For this purpose we analyzed 2000 time series of length $n=200$ per simulation scenario.
The true size of the intervention effect is scaled to be $\kappa= 3\sqrt{\lambda}$, $\kappa= 2\sqrt{\lambda}$ or $\kappa= \sqrt{\lambda}$ for $\delta=0$, $\delta=0.8$ or $\delta=1$, since  the total effect on the series increases with $\delta$.
Table \ref{tab3:power200-0} reports the empirical powers of the tests for the different types of intervention at a given time point $\tau\in\{0.25n,0.5n,0.75n\}$ when an innovation outlier occurs at the time point tested. We observe that the $F$-type statistics for an innovation outlier possess larger power than the corresponding tests for other values of $\delta$. Nevertheless, the tests using a misspecified value of $\delta$ also have some power, particularly those using a value of $\delta$ not very far from the true one. Similar conclusions are drawn in case of transient shifts (see Table \ref{tab3:power200-08}).
Moreover, a  permanent shift of a certain height is detected best by the test with the correctly specified $\delta=1$ if it occurs in the center of the series (Table \ref{tab3:power200-1}).

Tables \ref{tab3:classif200}-\ref{tab3:classif200-1} report the classification results for situations that we can try to identify the type of an intervention at a known time point. For this purpose, we compare the $F$-type statistics for a selection of values of $\delta$, classifying a detected intervention according to the $F$-type statistic with the largest value.
We investigate the empirical detection rates of this classification rule by analyzing 2000 time series of length $n=200$ per simulation scenario. We use the same parameter configurations for $\alpha_1,\alpha_2$ and $\lambda$ as previously and we further set $\tau\in\{50,100,150\}$. We also consider 
$\delta\in\{0,0.6,0.8,0.9,1\}$ and we scale the true size of the intervention effect to be $\kappa=3\sqrt{\lambda}$, $\kappa=2.5\sqrt{\lambda}$, $\kappa=2\sqrt{\lambda}$, $\kappa=1.5\sqrt{\lambda}$ or $\kappa=\sqrt{\lambda}$ for $\delta=0$, $\delta=0.6$, $\delta=0.8$, $\delta=0.9$ and $\delta=1$, respectively. 
Overall, the type of an intervention is correctly identified in most of the cases with occassionally wrong classification especially for transient shifts of moderate size.

Next we look at situations where we do not know neither the type nor the time of a possible intervention. For this scenario, we consider the maximum test statistics for a set of candidate time points $\tau$ and then selecting the maximum of the statistic.

\begin{figure}
\centering
\includegraphics[scale=0.6]{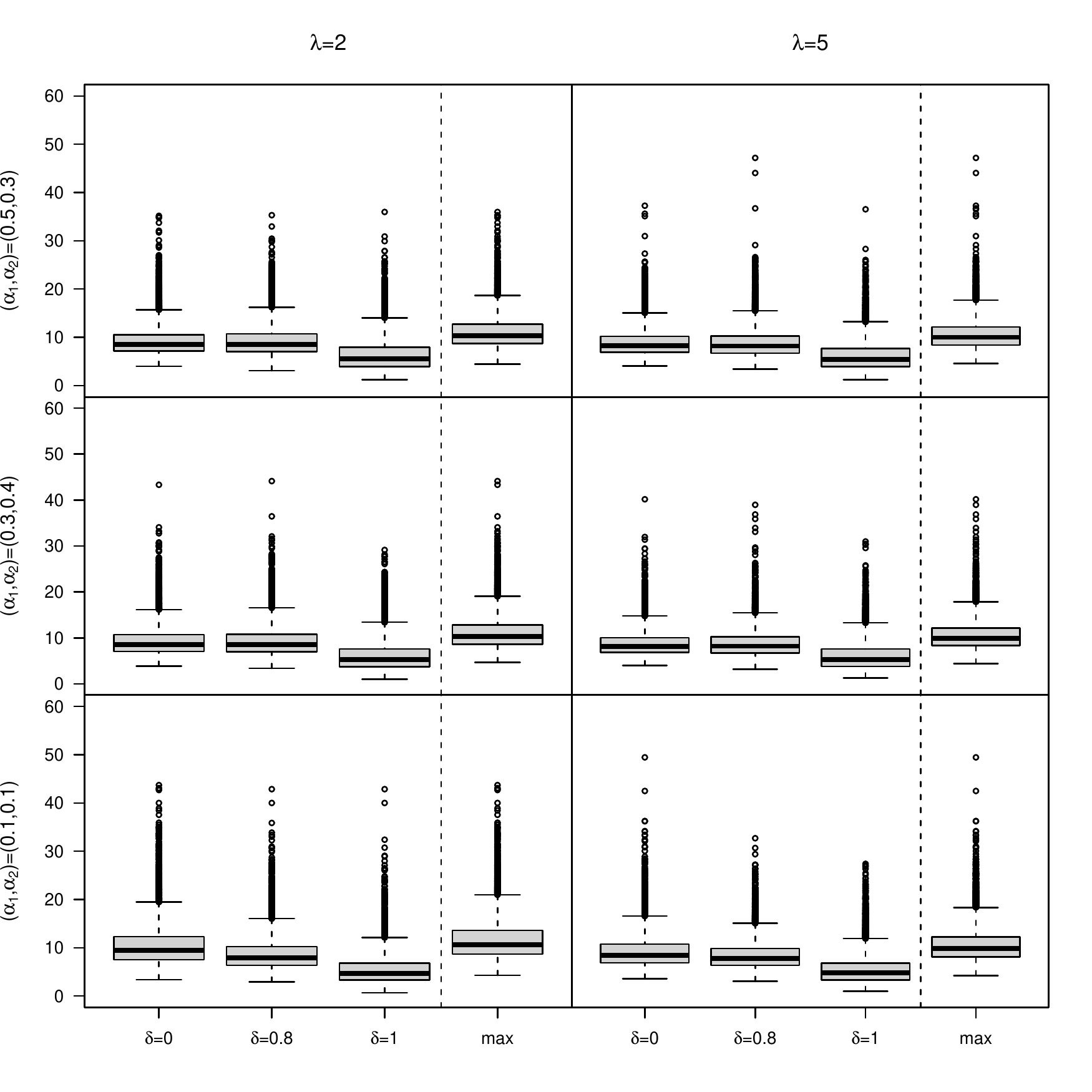}
\caption{\label{fig:maxstatistics0-100-INAR2}Boxplots of the maximum $F$-type statistic in case of INAR(2) models, maximized with respect to the candidate time point $\tau$ of a change when $n=100$.}
\end{figure}

\begin{figure}
\centering
\includegraphics[scale=0.6]{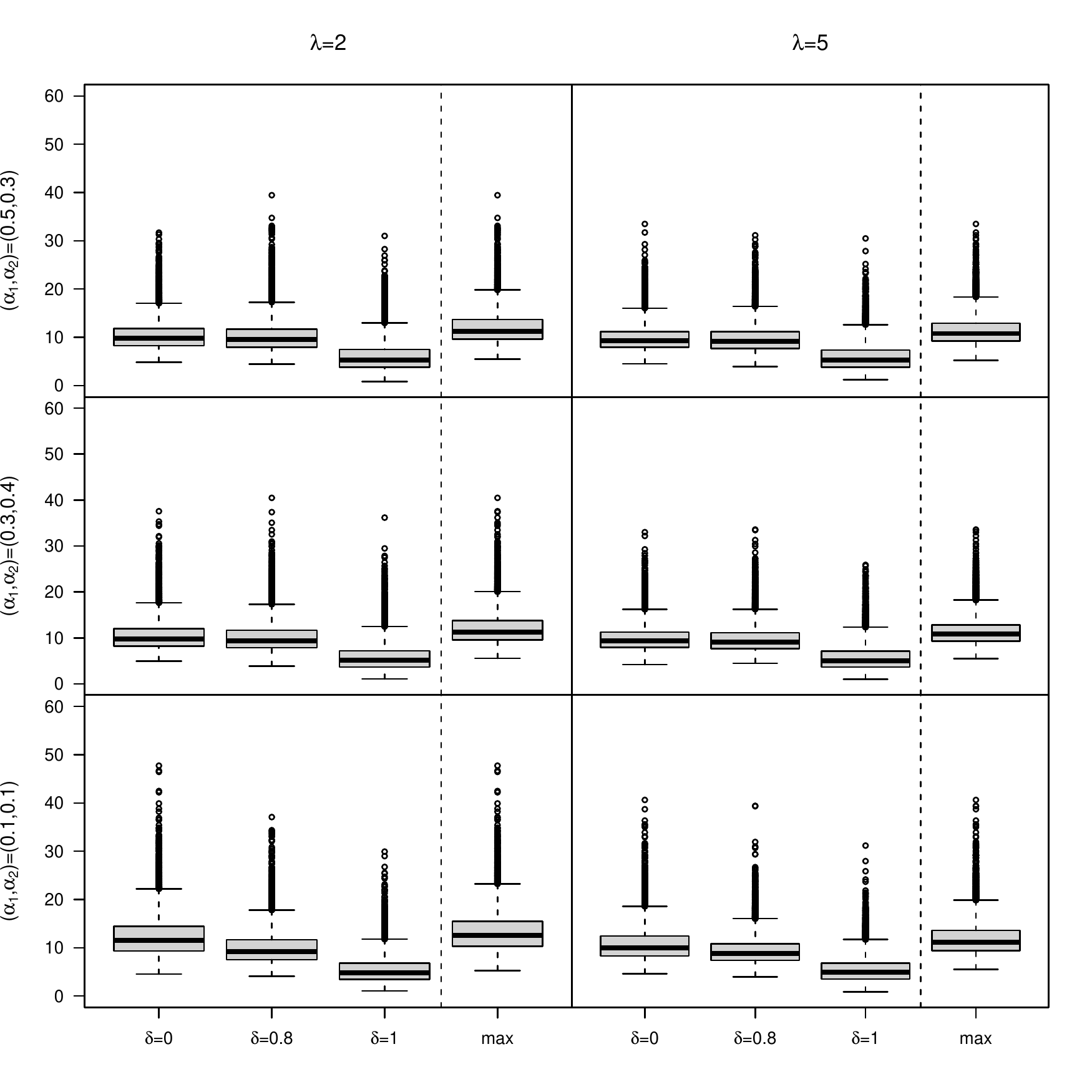}
\caption{\label{fig:maxstatistics0-200-INAR2}Boxplots of the maximum $F$-type statistic in case of INAR(2) models, maximized with respect to the candidate time point $\tau$ of a change when $n=200$.}
\end{figure}

 First we consider the results obtained from analyzing 10000 clean INAR(2) series for the different parameter settings and series lengths $n\in\{100,200\}$. Figures~\ref{fig:maxstatistics0-100-INAR2} and~\ref{fig:maxstatistics0-200-INAR2} display boxplots of these maximum statistics for each of several values of $\delta\in\{0,0.8,1\}$ individually as well as with an additional maximization with respect to $\delta$.


Approximate critical values  for an overall test on any type of intervention effect can be derived from the empirical quantiles of the maximum $F$-type statistic with additional maximization with respect to $\delta$. In the case of $n=100$,  the 90\%, 95\% and 99\% quantiles of the overall maximum $F$-type statistics range from about 14.8 to 17.3, from 16.8 to 20.0, and from 21.5 to 27 for the different parameter combinations considered here. We thus can use 17, 20 and 27 as critical values for approximate significance $F$-tests for an unknown intervention at an unknown time point at a 10\%, 5\% or 1\% significance level. 
In case of $n=200$, the empirical percentiles of the maximum $F$-type statistics range from 15.3 to 19, from 17.2 to 21.7, and from 21.6 to 27.9, so that we can use 19, 22 and 28 as approximate critical values. It is interesting to note that for both sample sizes $n=100$ and $n=200$, the approximate critical values derived for the INAR(2) model coincide with the corresponding critical values for the INAR(1) model.

\begin{figure}
\centering
\includegraphics[scale=0.6]{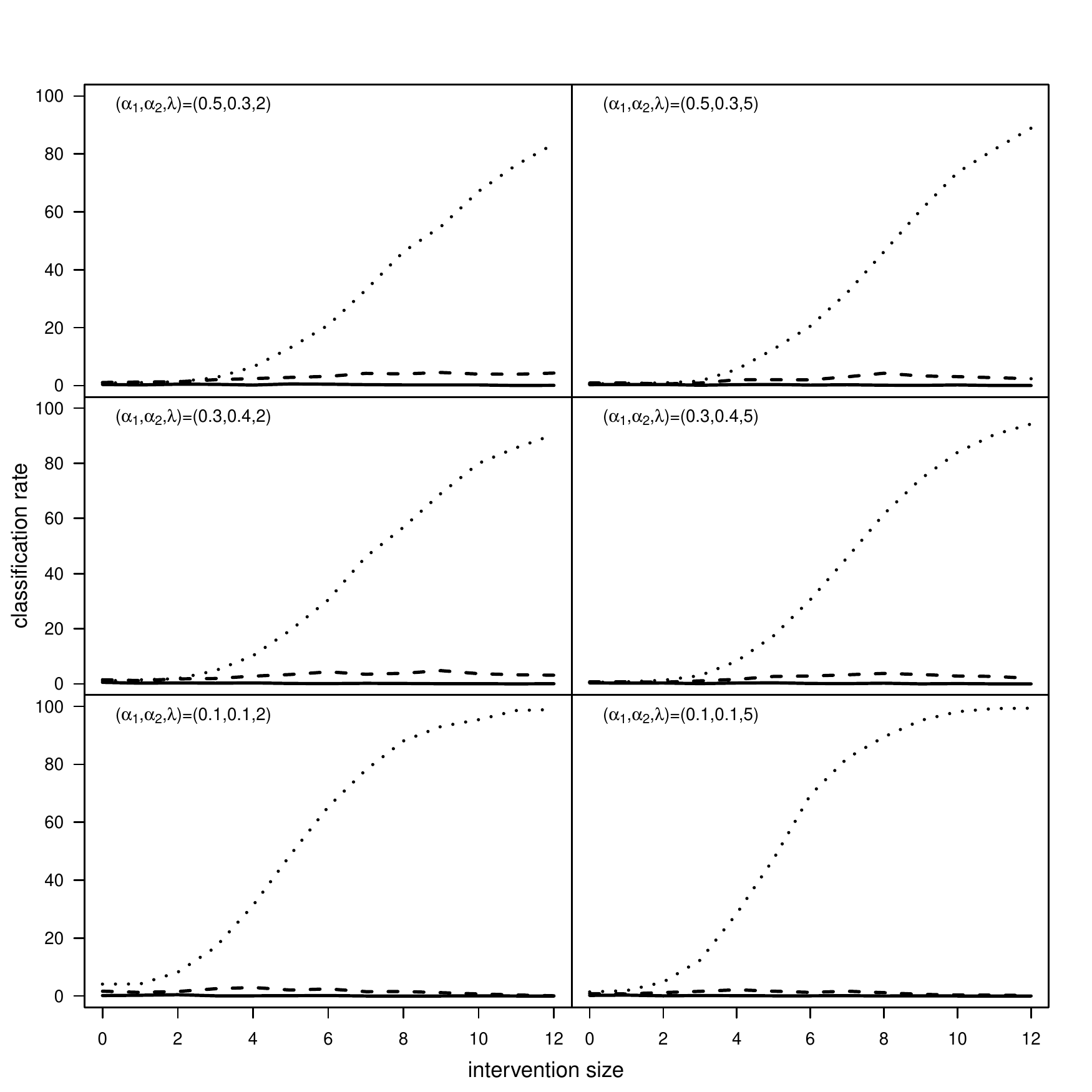}
\caption{\label{fig:classif100-0-inar2}Classification results when applying the maximum $F$-type statistics to INAR(2) time series of length $n=100$ containing an innovation outlier of increasing size $\kappa=0,\sqrt{\lambda},\ldots,12\sqrt{\lambda}$ at time point $\tau=50$.  Classification as $\delta=0$ (dotted), $\delta=0.8$ (dashed), $\delta=1$ (solid).}
\end{figure}

\begin{figure}
\centering
\includegraphics[scale=0.6]{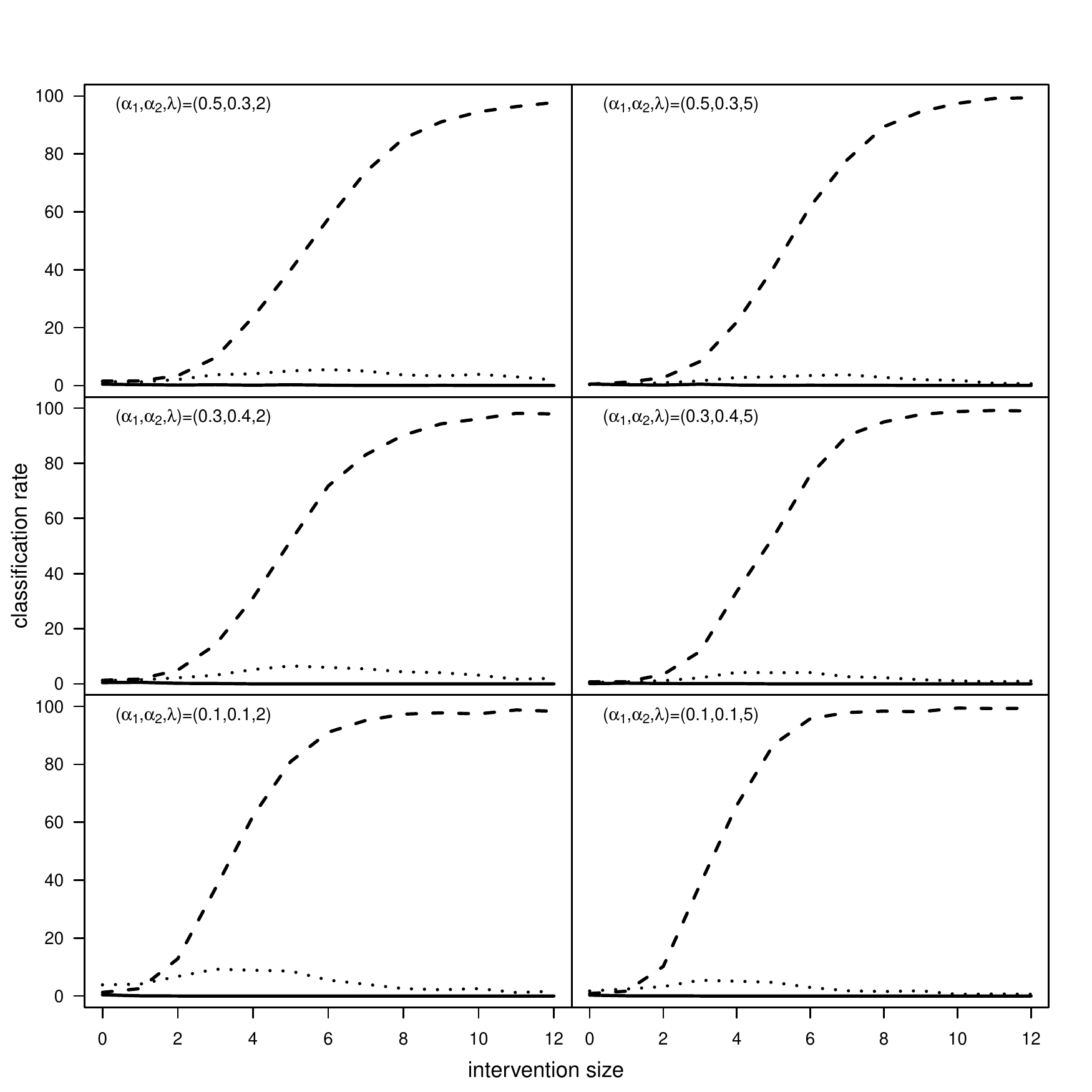}
\caption{\label{fig:classif100-08-inar2}Classification results when applying the maximum $F$-type statistics to INAR(2) time series of length $n=100$ containing a transient shift with $\delta=0.8$ of increasing size $\kappa=0,\sqrt{\lambda},\ldots,12\sqrt{\lambda}$ at time point $\tau=50$.  Classification as $\delta=0$ (dotted), $\delta=0.8$ (dashed), $\delta=1$ (solid).}
\end{figure}

\begin{figure}
\centering
\includegraphics[scale=0.6]{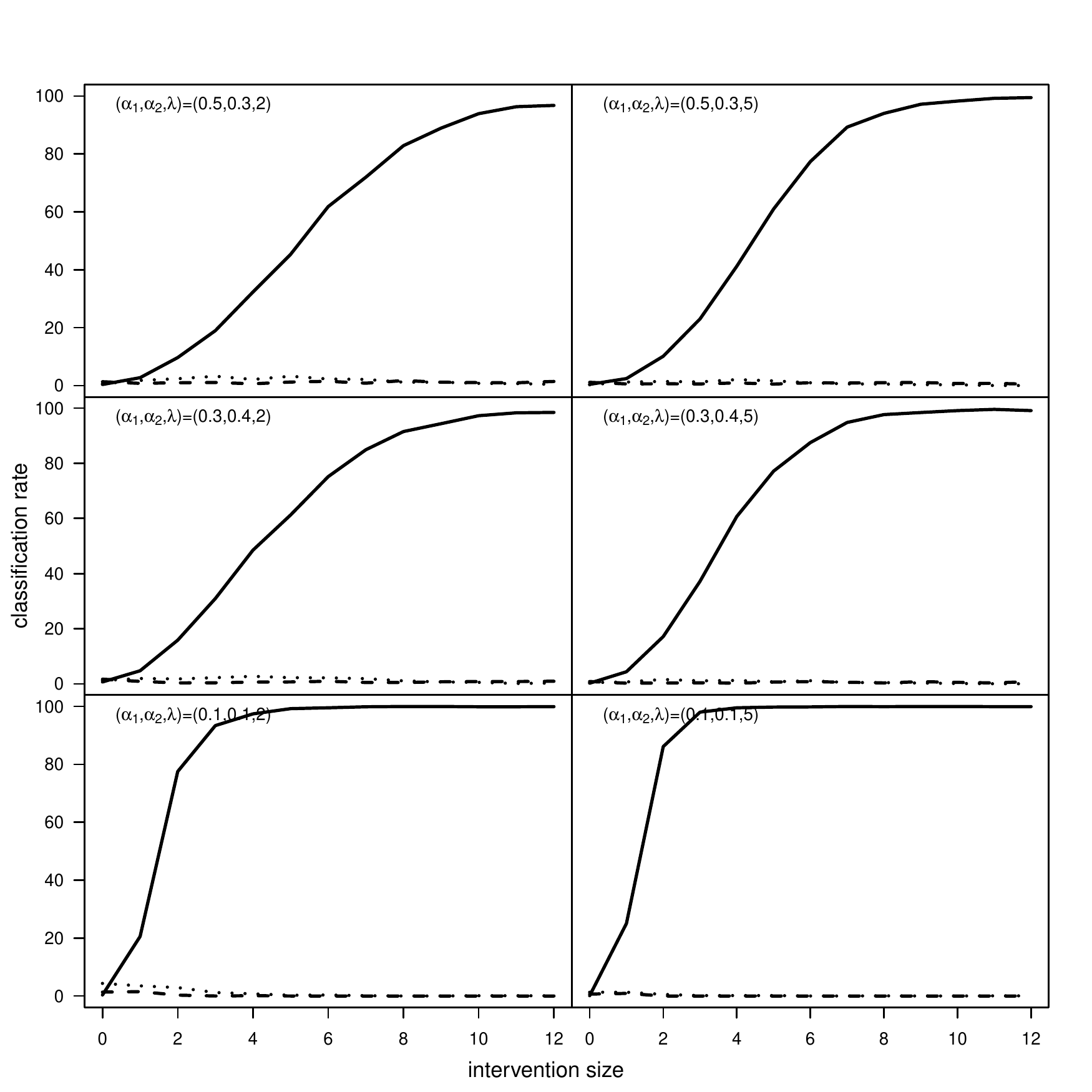}
\caption{\label{fig:classif100-1-inar2}Classification results when applying the maximum $F$-type statistics to INAR(2) time series of length $n=100$ containing a permanent shift with $\delta=1$ of increasing size $\kappa=0,\sqrt{\lambda},\ldots,12\sqrt{\lambda}$ at time point $\tau=50$.  Classification as $\delta=0$ (dotted), $\delta=0.8$ (dashed), $\delta=1$ (solid).}
\end{figure}

Next we inspect the performance of the classification rules when applied to time series containing an intervention effect. Figure \ref{fig:classif100-0-inar2} depicts the classification results when being applied to time series of length $n=100$ containing an innovation outlier at time $\tau=50$. For this we generate 2000 time series for each of the parameter combinations $(\alpha_1,\alpha_2,\lambda)$ considered before and each intervention size $\kappa=k\sqrt{\lambda}$, $k=0,\ldots,12$. Apparently, time series containing an innovation outlier are classified quite reliably, and the same holds for the classification of transient shifts with $\delta=0.8$ and permanent shifts (see Figures~\ref{fig:classif100-08-inar2} and \ref{fig:classif100-1-inar2} respectively). 

\begin{figure}\centering
\includegraphics[scale=0.6]{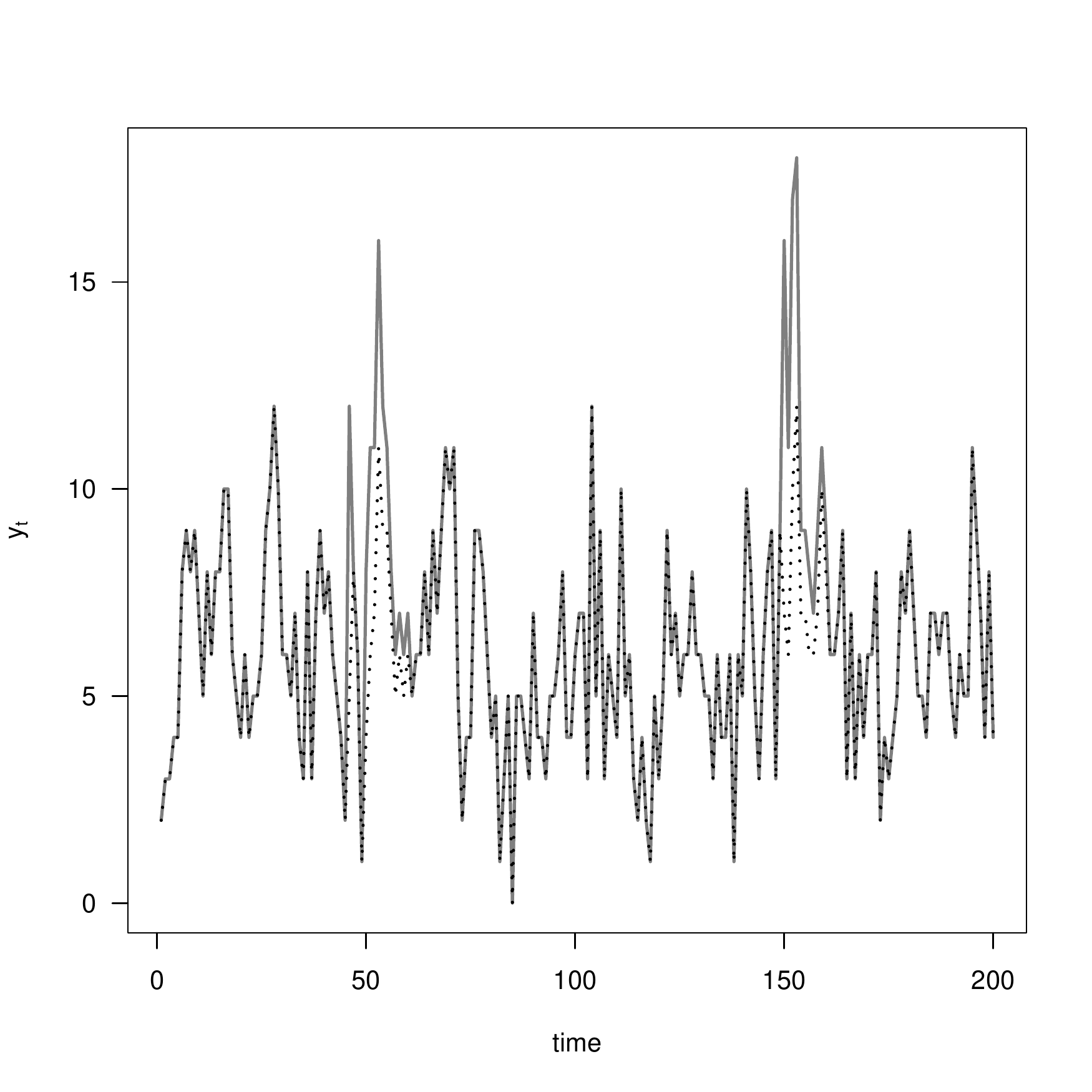}
\caption{\label{fig:sim-inar2}Simulated INAR(2) time series with two transient shifts at times $\tau_1=50$ and $\tau_2=150$ (solid line) and the series after correction for the intervention effects as estimated by the $F$-type statistic (dotted line).}
\end{figure}

In the following we adapt the stepwise detection algorithm of Section~\ref{sect:iterative} to test for the existence of any type of outlier using $\delta=(0,0.6,0.8,0.9,1)$ at any time point.
We consider a simulated time series of length $n=200$ generated from a contaminated Poisson INAR(2) model of the form 
\[Y_t=\alpha_1\circ Y_{t-1}+\alpha_2\circ Y_{t-2}+e_t+U_{t,1}+U_{t,2},\]
where $e_t\sim Pois(\lambda)$, $U_{t,j}\equiv 0$ for $t=0,\ldots,\tau_j-1$ and $U_{t,j}\sim Pois(\kappa_j\delta_j^{t-\tau_j})$ for $t=\tau_j,\ldots,n$, $j=1,2$. 
We set $(\alpha_1,\alpha_2,\lambda)=(0.3,0.2,3)$ and the interventions consisting of two transient shifts of the same size $\kappa_1=\kappa_2=\kappa=10$ at times 
$\tau_1=50$ and $\tau_2=150$ with $\delta_1=0.6$ and $\delta_2=0.9$, respectively (see Figure~\ref{fig:sim-inar2}). 

The iterative detection algorithm starts with fitting an INAR(2) model to the data assuming no interventions, at which step we obtain the initial conditional least squares estimates $(\hat{\alpha}_1,\hat{\alpha}_2,\hat{\lambda})=(0.38,0.18,2.81)$.
Then, we test for unknown types of interventions at unknown time points using the $F$-type statistic. At the first iteration, the test statistic correctly identifies a transient shift at time $\tau=150$, although with $\delta=0.8$ instead of $\delta=0.9$. Next, we correct the data according to step (3.b) of the algorithm. Note that in this specific step, the effect of the intervention is estimated as
\[\hat{U}_t=\left\lfloor\frac{\hat{\kappa}\delta^{t-\hat{\tau}}}{\hat{\alpha}_1Y_{t-1}^{(j+1)}+\hat{\alpha}_2Y_{t-2}^{(j+1)}+\hat{\lambda}+\hat{\kappa}\delta^{t-\hat{\tau}}}Y_t^{(j)}\right\rfloor.\]
After data correction, the second intervention corresponding to $\tau=50$ is also detected and is correctly classified as a transient shift but with $\delta=0.8$ instead of $\delta=0.6$. 
Correcting anew the data, an additional transient shift is detected at time $\tau=46$. The final conditional least squares estimates are $(\hat{\alpha}_1,\hat{\alpha}_2,\hat{\lambda})=(0.28, 0.17, 3.32)$ (see Table~\ref{tab:sim-ex-inar2}).

The above simulation experiment was repeated several times in order to evaluate the iterative detection procedure. Our findings are in line with those regarding the INAR(1) model, so that the discussion of Section~\ref{sect:simulex} also applies to the INAR(2) model.

\begin{table}
\begin{center}
\caption{\label{tab3:sizes100} Empirical sizes (in percent) of the tests  based on the  $F$-type statistic for a known type $\delta$ and time $\tau$ of intervention  in case of  INAR(2) series of length $n=100$ with different parameters $\alpha_1$, $\alpha_2$ and $\lambda$. The nominal significance levels are 1\%, 5\% or 10\%.}
{\footnotesize
\begin{tabular}{rrrr|rrr|rrr|rrr}
\hline
 && & & \multicolumn{3}{c|}{$\tau=0.25n$}& \multicolumn{3}{c|}{$\tau=0.5n$}& \multicolumn{3}{c}{$\tau=0.75n$}  \\
$\alpha_1$ & $\alpha_2$ & $\lambda$ & $\delta$ &  1\% & 5\% & 10\%& 1\% & 5\% & 10\%& 1\% & 5\% & 10\% \\ \hline
0.5 & 0.3 & 2 & 0 & 1.9 & 6.1 &10.9 & 1.4 & 5.8 &10.8 & 1.5 & 5.9 &10.7\\
 & & & 0.8 & 2.6 & 8.3 &14.4 & 2.1 & 7.4 &13.3 & 2.1 & 7.7 &13.8\\
 & & & 1 & 2.4 &  8.0 &15.3 & 3.3 &11.0& 18.2&  3.5& 11.2& 18.6\\ \hline
& & 5 & 0 & 1.2 & 5.8 &10.9&  1.7 & 5.9 &11.1 & 1.5 & 6.1& 11.9\\
& & & 0.8 & 1.9 & 7.3 & 12.7 & 2.1 & 8.0 &13.5&  1.5 & 6.5 &13.1\\
& & & 1 & 2.3 & 8.6 &14.5 & 2.7 & 9.7 &16.8 & 2.7 & 9.0 &15.4\\\hline
0.3 & 0.4 & 2 & 0 & 1.2 & 5.4& 11.1&  1.1&  5.5& 10.4 & 1.5 & 5.9 & 11.2\\
 & & & 0.8 & 2.0 & 7.3 & 13.2 & 1.6 & 7.4 & 13.4 & 2.4 & 7.9 & 13.5\\
 & & & 1 & 2.0 & 8.6 & 15.4 & 2.2 & 9.1 & 16.1 & 2.4 & 9.1 & 16.0\\\hline
 & & 5 & 0 & 1.6 & 5.9 & 10.9 & 1.4 & 6.3 & 11.7 & 1.2 & 5.6& 10.8\\
& & & 0.8 & 1.9 &  6.4 & 12.1 & 1.7 & 6.5 &12.2 & 2.2 & 6.8 & 11.9\\
& & & 1 &  2.3 & 8.7 & 15.2&  2.0&  7.9& 14.0&  2.8&  8.8& 15.6\\ \hline
0.1 & 0.1 & 2 & 0 &  1.5 & 4.5 &10.1 & 1.4 & 4.7 &10.3 & 1.8 &  4.7 & 9.5\\
& & & 0.8 & 1.1 & 5.8 & 10.7 & 1.4 & 6.3 & 11.2 & 1.4 & 6.1 & 11.4\\
& & & 1 & 2.3 & 7.1 & 12.6 & 1.4 & 6.6 & 12.2 & 1.5 & 6.1 & 12.0\\\hline
 & & 5 & 0 & 1.4 & 5.1 & 10.1 & 1.4 & 5.4 & 10.4 & 1.0 & 5.2 & 10.5\\
& & & 0.8 &  1.6 & 5.8 &11.5 & 1.5 & 5.8 & 11.5 & 1.5 & 6.4 & 11.3\\
& & & 1 & 1.3 & 6.5 & 11.7 & 1.4 & 6.2 & 11.7 & 1.7 & 6.4 & 12.6\\ \hline
\end{tabular}}
\end{center}
\end{table}

\begin{table}
\begin{center}
\caption{\label{tab3:sizes200} Empirical sizes (in percent) of the tests  based on the  $F$-type statistic for a known type $\delta$ and time $\tau$ of intervention  in case of  INAR(2) series of length $n=200$ with different parameters $\alpha_1$, $\alpha_2$ and $\lambda$. The nominal significance levels are 1\%, 5\% or 10\%.}
{\footnotesize
\begin{tabular}{rrrr|rrr|rrr|rrr}
\hline
 && & & \multicolumn{3}{c|}{$\tau=0.25n$}& \multicolumn{3}{c|}{$\tau=0.5n$}& \multicolumn{3}{c}{$\tau=0.75n$}  \\
$\alpha_1$ & $\alpha_2$ & $\lambda$ & $\delta$ &  1\% & 5\% & 10\%& 1\% & 5\% & 10\%& 1\% & 5\% & 10\% \\ \hline
0.5 & 0.3 & 2 & 0 & 1.0 & 5.3 & 10.3 & 1.4 & 5.9 & 10.9 & 1.6 & 5.6 & 10.6 \\ 
  & & & 0.8 & 1.6 & 5.9 & 11.1 & 1.3 & 6.2 & 11.7 & 1.4 & 5.9 & 11.1 \\ 
  & & & 1 & 1.8 & 7.3 & 12.7 & 1.9 & 7.5 & 13.5 & 1.8 & 7.1 & 13.0 \\ \hline
   & & 5 & 0 & 1.5 & 5.9 & 10.6 & 1.5 & 5.5 & 10.5 & 1.3 & 5.7 & 10.6 \\ 
   & & & 0.8 & 1.4 & 6.5 & 12.3 & 1.6 & 6.2 & 11.8 & 1.5 & 6.6 & 11.8 \\ 
   & & & 1 & 1.6 & 7.1 & 13.0 & 1.6 & 6.8 & 13.3 & 1.7 & 7.2 & 13.6 \\ \hline
  0.3 & 0.4 & 2 & 0 & 1.0 & 5.2 & 9.9 & 1.2 & 5.3 & 10.6 & 1.2 & 5.1 & 10.2 \\ 
   & & & 0.8 & 1.4 & 5.6 & 11.0 & 1.6 & 6.1 & 11.6 & 1.7 & 5.9 & 11.2 \\ 
   & & & 1 & 1.7 & 6.4 & 12.4 & 1.6 & 7.4 & 13.5 & 1.4 & 7.1 & 12.6 \\ \hline
   & & 5 & 0 & 1.5 & 5.9 & 10.5 & 1.1 & 5.3 & 10.4 & 1.2 & 5.3 & 10.4 \\ 
  & & & 0.8 & 1.7 & 6.5 & 11.6 & 1.5 & 6.1 & 11.1 & 1.5 & 6.1 & 11.5 \\ 
   & & & 1 & 1.5 & 6.8 & 12.3 & 1.6 & 7.5 & 13.8 & 1.7 & 7.1 & 13.2 \\ \hline
 0.1  &0.1 & 2 & 0 & 1.3 & 4.2 & 9.4 & 1.2 & 3.8 & 8.8 & 1.4 & 4.3 & 9.4 \\ 
   & & & 0.8 & 1.4 & 5.5 & 10.8 & 1.2 & 5.3 & 11.1 & 1.5 & 5.7 & 10.7 \\ 
  & & & 1 & 1.2 & 5.5 & 10.9 & 1.3 & 5.7 & 11.2 & 1.4 & 5.8 & 10.8 \\ \hline
   & & 5& 0 & 1.2 & 4.9 & 10.0 & 1.3 & 5.0 & 9.5 & 1.1 & 4.5 & 9.9 \\ 
   & & & 0.8 & 1.2 & 5.7 & 11.3 & 1.2 & 5.3 & 10.2 & 1.3 & 5.6 & 11.1 \\ 
   & & & 1 & 1.2 & 5.4 & 11.2 & 1.0 & 5.7 & 11.6 & 1.7 & 6.5 & 11.8 \\ 
   \hline
\end{tabular}}
\end{center}
\end{table}

\begin{table}
\caption{\label{tab3:power200-0} Empirical power (in percent) of the tests  based on the  $F$-type statistic for a known time $\tau$ and known, but possibly misspecified type $\delta$ of intervention  in case of an innovation outlier $\delta=0$ of size $\kappa=3\sqrt{\lambda}$ at time $\tau$ in an INAR(2) series of length $n=200$ with different parameters $\alpha_1$, $\alpha_2$ and $\lambda$. The nominal significance levels are 1\%, 5\% or 10\%.}
\begin{center}
{\footnotesize
\begin{tabular}{rrrr|rrr|rrr|rrr}
\hline
 && & & \multicolumn{3}{c|}{$\tau=0.25n$}& \multicolumn{3}{c|}{$\tau=0.5n$}& \multicolumn{3}{c}{$\tau=0.75n$}  \\
$\alpha_1$ & $\alpha_2$ & $\lambda$ & $\delta$ &  1\% & 5\% & 10\%& 1\% & 5\% & 10\%& 1\% & 5\% & 10\% \\ \hline
0.5 & 0.3 & 2 & 0 & 24.6 & 41.1 & 51.3 & 23.6 & 40.6 & 51.0 & 23.2 & 38.6 & 49.4 \\ 
  & & & 0.8  & 12.0 & 24.3 & 33.0 & 11.1 & 24.2 & 31.9 & 10.7 & 22.0 & 31.0 \\ 
  & & & 1 & 1.7 & 7.0 & 12.6 & 1.8 & 8.1 & 14.9 & 3.2 & 9.6 & 15.6 \\ \hline
  & & 5 & 0 & 21.3 & 37.8 & 48.5 & 21.7 & 38.6 & 48.6 & 21.6 & 38.2 & 48.6 \\ 
   & & & 0.8  & 9.0 & 20.0 & 27.3 & 9.8 & 20.5 & 28.8 & 8.9 & 20.4 & 28.8 \\ 
   & & & 1  & 1.8 & 6.9 & 12.9 & 2.0 & 8.1 & 14.1 & 3.0 & 8.2 & 15.2 \\ \hline
0.3 & 0.4 & 2 & 0  & 28.1 & 44.9 & 55.5 & 30.6 & 46.3 & 55.0 & 29.8 & 47.0 & 56.9 \\ 
   & & & 0.8 & 14.0 & 25.2 & 33.5 & 13.9 & 25.5 & 34.4 & 13.1 & 24.6 & 34.9 \\ 
  & & & 1 & 1.7 & 6.0 & 11.3 & 1.7 & 7.4 & 14.1 & 3.0 & 9.0 & 15.6 \\ \hline
  & & 5 & 0 & 28.4 & 46.0 & 55.8 & 27.5 & 46.7 & 56.6 & 29.0 & 45.4 & 56.6 \\ 
   & & & 0.8  & 11.6 & 24.1 & 33.2 & 11.1 & 23.2 & 32.1 & 11.8 & 24.6 & 34.0 \\ 
  & & & 1  & 1.6 & 7.4 & 13.0 & 2.1 & 7.8 & 13.8 & 2.3 & 7.8 & 14.2 \\ \hline
  0.1  & 0.1 & 2& 0  & 48.8 & 65.0 & 72.5 & 49.9 & 65.6 & 72.5 & 51.0 & 66.0 & 72.7 \\ 
    & & & 0.8  & 22.3 & 38.3 & 47.3 & 23.0 & 38.9 & 48.6 & 24.4 & 38.9 & 47.1 \\ 
    & & & 1 & 1.2 & 6.3 & 12.0 & 1.8 & 6.3 & 12.2 & 1.6 & 6.6 & 12.6 \\ \hline
    & &5 & 0  & 52.0 & 67.8 & 76.5 & 50.0 & 67.2 & 75.7 & 53.6 & 70.1 & 77.6 \\ 
    & & & 0.8  & 22.9 & 39.4 & 49.0 & 22.0 & 38.8 & 48.8 & 23.2 & 39.6 & 48.9 \\ 
    & & & 1  & 1.4 & 5.4 & 10.6 & 1.2 & 6.0 & 10.8 & 1.7 & 7.6 & 13.9 \\ 
\hline
\end{tabular}}
\end{center}
\end{table}

\begin{table}
\caption{\label{tab3:power200-08} Empirical power (in percent) of the tests  based on the  $F$-type statistic for a known time $\tau$ and known, but possibly misspecified type $\delta$ of intervention  in case of an innovation outlier $\delta=0.8$ of size $\kappa=2\sqrt{\lambda}$ at time $\tau$ in an INAR(2) series of length $n=200$ with different parameters $\alpha_1$, $\alpha_2$ and $\lambda$. The nominal significance levels are 1\%, 5\% or 10\%.}
\begin{center}
{\footnotesize
\begin{tabular}{rrrr|rrr|rrr|rrr}
\hline
 && & & \multicolumn{3}{c|}{$\tau=0.25n$}& \multicolumn{3}{c|}{$\tau=0.5n$}& \multicolumn{3}{c}{$\tau=0.75n$}  \\
$\alpha_1$ & $\alpha_2$ & $\lambda$ & $\delta$ &  1\% & 5\% & 10\%& 1\% & 5\% & 10\%& 1\% & 5\% & 10\% \\ \hline
0.5 & 0.3 & 2 & 0  & 9.9 & 21.9 & 30.8 & 11.6 & 22.9 & 32.0 & 10.5 & 21.7 & 30.0 \\ 
  & & & 0.8 & 25.2 & 45.1 & 55.6 & 28.3 & 45.9 & 55.3 & 25.9 & 43.8 & 54.3 \\ 
  & & & 1 & 1.8 & 6.9 & 13.4 & 2.6 & 9.4 & 16.0 & 4.8 & 13.5 & 20.1 \\ \hline
    & & 5 & 0 & 7.0 & 19.6 & 29.5 & 9.0 & 21.4 & 30.9 & 8.3 & 21.8 & 29.9 \\ 
    & & & 0.8 & 24.6 & 44.7 & 54.7 & 25.8 & 43.5 & 55.8 & 25.0 & 42.4 & 53.1 \\ 
   & & & 1 & 2.0 & 7.0 & 12.8 & 2.5 & 8.9 & 14.9 & 4.8 & 12.0 & 19.3 \\ \hline
  0.3 & 0.4 & 2 & 0 & 15.1 & 29.1 & 36.9 & 14.4 & 27.3 & 34.2 & 13.9 & 24.6 & 34.4 \\ 
   & & & 0.8 & 33.5 & 52.8 & 61.8 & 34.6 & 53.1 & 63.0 & 34.5 & 52.4 & 63.2 \\ 
  & & & 1 & 1.2 & 5.8 & 10.9 & 2.8 & 7.6 & 14.1 & 4.2 & 13.5 & 21.0 \\ \hline
   & & 5 & 0 & 13.2 & 25.9 & 35.6 & 11.5 & 25.4 & 35.5 & 12.8 & 25.6 & 35.6 \\ 
   & & & 0.8 & 31.3 & 53.1 & 64.6 & 33.0 & 55.4 & 64.2 & 32.8 & 52.5 & 64.5 \\ 
  & & & 1 & 1.2 & 5.3 & 11.2 & 2.1 & 7.9 & 14.6 & 4.7 & 13.0 & 20.8 \\ \hline
  0.1 & 0.1 & 2 & 0 & 25.6 & 40.4 & 48.4 & 26.2 & 40.6 & 48.3 & 26.9 & 43.0 & 50.5 \\ 
   & & & 0.8 & 56.3 & 73.0 & 81.0 & 55.9 & 74.5 & 82.4 & 57.1 & 73.2 & 80.1 \\ 
  & & & 1 & 1.1 & 5.1 & 10.0 & 2.5 & 8.1 & 14.4 & 5.1 & 15.2 & 25.0 \\ \hline
    & & 5 & 0 & 26.1 & 42.6 & 51.9 & 23.5 & 39.6 & 49.5 & 26.5 & 42.6 & 53.4 \\ 
    & & & 0.8 & 61.7 & 79.7 & 85.8 & 59.7 & 77.2 & 83.8 & 61.1 & 77.8 & 85.7 \\ 
    & & & 1 & 0.9 & 5.0 & 9.8 & 2.1 & 7.8 & 14.2 & 6.0 & 16.3 & 25.7 \\ 
  \hline
\end{tabular}}
\end{center}
\end{table}

\begin{table}
\caption{\label{tab3:power200-1} Empirical power (in percent) of the tests  based on the  $F$-type statistic for a known time $\tau$ and known, but possibly misspecified type $\delta$ of intervention  in case of an innovation outlier $\delta=1$ of size $\kappa=\sqrt{\lambda}$ at time $\tau$ in an INAR(2) series of length $n=200$ with different parameters $\alpha_1$, $\alpha_2$ and $\lambda$. The nominal significance levels are 1\%, 5\% or 10\%.}
\begin{center}
{\footnotesize
\begin{tabular}{rrrr|rrr|rrr|rrr}
\hline
 && & & \multicolumn{3}{c|}{$\tau=0.25n$}& \multicolumn{3}{c|}{$\tau=0.5n$}& \multicolumn{3}{c}{$\tau=0.75n$}  \\
$\alpha_1$ & $\alpha_2$ & $\lambda$ & $\delta$ &  1\% & 5\% & 10\%& 1\% & 5\% & 10\%& 1\% & 5\% & 10\% \\ \hline
0.5 & 0.3 & 2 & 0   & 1.0 & 3.5 & 8.1 & 1.5 & 5.1 & 9.8 & 2.1 & 7.6 & 12.8 \\ 
& & & 0.8  & 0.9 & 4.3 & 9.5 & 2.3 & 8.9 & 16.0 & 6.8 & 18.9 & 28.2 \\ 
& & & 1 & 41.5 & 73.5 & 84.9 & 66.4 & 88.7 & 94.2 & 63.7 & 84.2 & 89.8 \\ \hline
& & 5 & 0 & 0.5 & 4.2 & 8.6 & 1.5 & 6.1 & 11.8 & 2.1 & 7.8 & 14.4 \\ 
& & & 0.8  & 1.4 & 5.3 & 10.3 & 3.1 & 10.9 & 17.6 & 6.9 & 19.0 & 29.6 \\ 
& & & 1 & 46.5 & 75.7 & 86.4 & 69.0 & 90.6 & 95.6 & 64.6 & 86.8 & 93.2 \\ \hline
0.3 & 0.4 & 2 & 0 & 0.9 & 4.4 & 9.3 & 2.3 & 7.4 & 12.3 & 3.1 & 9.0 & 15.2 \\ 
& & & 0.8  & 0.8 & 5.3 & 10.4 & 3.1 & 11.3 & 19.0 & 9.0 & 22.4 & 34.0 \\ 
& & & 1 & 57.6 & 85.4 & 93.2 & 80.7 & 95.7 & 98.3 & 76.4 & 91.8 & 96.0 \\ \hline
& & 5 & 0 & 0.9 & 4.9 & 9.4 & 1.3 & 5.8 & 10.8 & 2.4 & 8.3 & 15.3 \\ 
& & & 0.8 & 1.4 & 5.3 & 11.2 & 2.6 & 10.2 & 18.6 & 7.8 & 21.3 & 31.7 \\ 
 & & & 1 & 62.5 & 87.1 & 93.0 & 83.0 & 95.7 & 98.2 & 77.4 & 93.8 & 97.0 \\ \hline
0.1 & 0.1 & 2 & 0 & 1.8 & 4.9 & 9.4 & 3.3 & 7.6 & 13.2 & 5.6 & 11.8 & 18.1 \\ 
& & & 0.8 & 1.3 & 5.5 & 10.2 & 4.6 & 14.1 & 22.4 & 14.4 & 32.8 & 45.8 \\ 
& & & 1 & 96.6 & 99.4 & 99.7 & 99.4 & 100.0 & 100.0 & 98.7 & 99.7 & 100.0 \\ \hline
& & 5 & 0 & 1.5 & 6.7 & 11.8 & 3.4 & 9.9 & 15.2 & 5.6 & 13.0 & 19.4 \\ 
& & & 0.8  & 1.2 & 5.3 & 10.8 & 4.2 & 14.6 & 24.6 & 16.0 & 34.5 & 48.0 \\ 
& & & 1 & 98.1 & 99.7 & 99.9 & 99.8 & 100.0 & 100.0 & 99.2 & 100.0 & 100.0 \\ 
  \hline
\end{tabular}}
\end{center}
\end{table}

\begin{table}
\caption{\label{tab3:classif200} Classification rates (in percent) achieved by applying the F-type statistic for a known type $\delta$ and time $\tau$ of intervention  to clean  INAR(2) series of length $n=200$ without outliers. The nominal significance levels are 1\%, 5\% or 10\%.}
\begin{center}
{\footnotesize
\begin{tabular}{rrrr|rrr|rrr|rrr}
\hline
 && & & \multicolumn{3}{c|}{$\tau=0.25n$}& \multicolumn{3}{c|}{$\tau=0.5n$}& \multicolumn{3}{c}{$\tau=0.75n$}  \\
$\alpha_1$ & $\alpha_2$ & $\lambda$ & $\delta$ &  1\% & 5\% & 10\%& 1\% & 5\% & 10\%& 1\% & 5\% & 10\% \\ \hline
0.5 & 0.3 & 2 & 0 & 1.1 & 4.3 & 7.0 & 0.8 & 3.3 & 6.6 & 1.1 & 4.2 & 7.1 \\ 
 &  & & 0.6 & 0.8 & 2.4 & 3.7 & 0.4 & 1.6 & 3.4 & 0.8 & 2.5 & 3.9 \\ 
 &  & & 0.8  & 0.5 & 2.2 & 3.5 & 0.7 & 2.0 & 3.2 & 0.8 & 1.9 & 3.6 \\ 
 &  & & 0.9 & 1.7 & 4.9 & 7.9 & 1.5 & 5.0 & 7.6 & 1.5 & 4.2 & 6.9 \\ 
 &  & & 1 & 1.8 & 7.5 & 11.7 & 1.8 & 7.3 & 13.7 & 1.6 & 6.3 & 11.2 \\ \hline
 &  & 5& 0  & 1.0 & 3.2 & 6.0 & 1.0 & 4.2 & 7.3 & 1.3 & 4.3 & 7.7 \\
\hline
\end{tabular}}
\end{center}
\end{table}

\begin{table}
\caption{\label{tab3:classif200-0} Classification rates (in percent) achieved by applying the F-type statistic for a known type $\delta$ and time $\tau$ of intervention  to INAR(2) series of length $n=200$ with an innovation outlier ($\delta=0$) of size $\kappa=3\sqrt{\lambda}$ at time $\tau$. The nominal significance levels are 1\%, 5\% or 10\%.}
\begin{center}
{\footnotesize
\begin{tabular}{rrrr|rrr|rrr|rrr}
\hline
 && & & \multicolumn{3}{c|}{$\tau=0.25n$}& \multicolumn{3}{c|}{$\tau=0.5n$}& \multicolumn{3}{c}{$\tau=0.75n$}  \\
$\alpha_1$ & $\alpha_2$ & $\lambda$ & $\delta$ &  1\% & 5\% & 10\%& 1\% & 5\% & 10\%& 1\% & 5\% & 10\% \\ \hline
0.5 & 0.3 & 2 & 0 & 18.4 & 29.1 & 35.0 & 18.0 & 29.4 & 35.2 & 18.6 & 28.6 & 35.5 \\ 
 &  &  & 0.6 & 5.2 & 8.6 & 10.2 & 5.1 & 8.6 & 10.2 & 6.1 & 8.8 & 10.4 \\ 
  &  &  & 0.8 & 2.4 & 3.5 & 4.5 & 1.5 & 2.6 & 3.7 & 1.8 & 3.0 & 3.6 \\ 
  &  &  & 0.9 & 2.7 & 5.7 & 7.6 & 1.9 & 4.0 & 5.7 & 2.9 & 5.3 & 6.5 \\ 
  &  &  & 1 & 1.4 & 4.5 & 8.0 & 1.5 & 5.1 & 8.2 & 1.7 & 4.4 & 7.0 \\ \hline
   &  &  5 & 0 & 17.1 & 28.3 & 34.8 & 19.3 & 31.1 & 37.0 & 18.6 & 30.2 & 36.9 \\
\hline
\end{tabular}}
\end{center}
\end{table}

\begin{table}
\caption{\label{tab3:classif200-06} Classification rates (in percent) achieved by applying the F-type statistic for a known type $\delta$ and time $\tau$ of intervention  to INAR(2) series of length $n=200$ with a transient shift with $\delta=0.6$ of size $\kappa=3\sqrt{\lambda}$ at time $\tau$. The nominal significance levels are 1\%, 5\% or 10\%.}
\begin{center}
{\footnotesize
\begin{tabular}{rrrr|rrr|rrr|rrr}
\hline
 && & & \multicolumn{3}{c|}{$\tau=0.25n$}& \multicolumn{3}{c|}{$\tau=0.5n$}& \multicolumn{3}{c}{$\tau=0.75n$}  \\
$\alpha_1$ & $\alpha_2$ & $\lambda$ & $\delta$ &  1\% & 5\% & 10\%& 1\% & 5\% & 10\%& 1\% & 5\% & 10\% \\ \hline
0.5 & 0.3 & 2 & 0 & 8.4 & 13.7 & 16.6 & 9.2 & 14.9 & 17.7 & 8.5 & 13.7 & 16.7 \\ 
 &  &  & 0.6 & 11.2 & 17.0 & 20.2 & 10.2 & 16.2 & 19.6 & 12.2 & 18.4 & 21.3 \\ 
  &  &  & 0.8 & 6.7 & 10.3 & 12.3 & 8.2 & 11.8 & 13.5 & 6.9 & 10.5 & 12.2 \\ 
  &  &  & 0.9 & 6.0 & 10.4 & 12.3 & 6.9 & 10.5 & 12.7 & 5.7 & 9.0 & 11.1 \\ 
  &  &  & 1 & 1.6 & 5.0 & 7.3 & 1.7 & 5.0 & 7.3 & 2.2 & 4.9 & 6.7 \\ \hline
  &  &  5 & 0 & 7.0 & 12.4 & 14.9 & 7.6 & 13.2 & 16.1 & 6.8 & 11.9 & 14.9 \\
\hline
\end{tabular}}
\end{center}
\end{table}

\begin{table}
\caption{\label{tab3:classif200-08} Classification rates (in percent) achieved by applying the F-type statistic for a known type $\delta$ and time $\tau$ of intervention  to INAR(2) series of length $n=200$ with a transient shift with $\delta=0.8$ of size $\kappa=2.5\sqrt{\lambda}$ at time $\tau$. The nominal significance levels are 1\%, 5\% or 10\%.}
\begin{center}
{\footnotesize
\begin{tabular}{rrrr|rrr|rrr|rrr}
\hline
 && & & \multicolumn{3}{c|}{$\tau=0.25n$}& \multicolumn{3}{c|}{$\tau=0.5n$}& \multicolumn{3}{c}{$\tau=0.75n$}  \\
$\alpha_1$ & $\alpha_2$ & $\lambda$ & $\delta$ &  1\% & 5\% & 10\%& 1\% & 5\% & 10\%& 1\% & 5\% & 10\% \\ \hline
0.5 & 0.3 & 2 & 0 & 4.5 & 8.0 & 10.3 & 4.9 & 7.1 & 8.8 & 4.3 & 7.8 & 9.8 \\ 
 & & & 0.6 & 7.7 & 11.8 & 13.8 & 7.0 & 10.9 & 12.7 & 5.9 & 9.9 & 11.9 \\ 
 & & & 0.8 & 9.2 & 14.6 & 17.3 & 9.2 & 14.2 & 16.6 & 9.8 & 15.4 & 17.9 \\ 
 & & & 0.9 & 13.4 & 20.5 & 24.2 & 14.2 & 21.6 & 25.6 & 13.6 & 19.9 & 22.8 \\ 
 & & & 1 & 1.1 & 3.7 & 5.6 & 1.8 & 5.0 & 7.6 & 2.7 & 6.4 & 8.6 \\ \hline
  & & 5 & 0 & 3.4 & 6.3 & 8.2 & 4.1 & 8.2 & 10.1 & 4.0 & 6.9 & 8.4 \\
\hline
\end{tabular}}
\end{center}
\end{table}

\begin{table}
\caption{\label{tab3:classif200-09} Classification rates (in percent) achieved by applying the F-type statistic for a known type $\delta$ and time $\tau$ of intervention  to INAR(2) series of length $n=200$ with a transient shift with $\delta=0.9$ of size $\kappa=2\sqrt{\lambda}$ at time $\tau$. The nominal significance levels are 1\%, 5\% or 10\%.}
\begin{center}
{\footnotesize
\begin{tabular}{rrrr|rrr|rrr|rrr}
\hline
 && & & \multicolumn{3}{c|}{$\tau=0.25n$}& \multicolumn{3}{c|}{$\tau=0.5n$}& \multicolumn{3}{c}{$\tau=0.75n$}  \\
$\alpha_1$ & $\alpha_2$ & $\lambda$ & $\delta$ &  1\% & 5\% & 10\%& 1\% & 5\% & 10\%& 1\% & 5\% & 10\% \\ \hline
0.5 & 0.3 & 2 & 0 & 2.1 & 4.6 & 6.2 & 2.4 & 5.2 & 6.8 & 2.8 & 5.7 & 7.2 \\ 
& & & 0.6 & 3.4 & 6.3 & 7.2 & 3.5 & 5.7 & 7.1 & 3.4 & 5.7 & 7.3 \\ 
& & & 0.8 & 6.5 & 9.3 & 10.8 & 4.9 & 8.5 & 10.0 & 5.5 & 8.9 & 10.7 \\ 
 & & & 0.9 & 19.6 & 31.6 & 38.2 & 18.9 & 31.0 & 37.2 & 17.0 & 26.6 & 30.5 \\ 
 & & & 1 & 1.1 & 4.4 & 7.2 & 2.0 & 5.8 & 8.8 & 5.8 & 10.6 & 13.4 \\ \hline
 & & 5 & 0 & 1.8 & 4.6 & 6.2 & 2.2 & 5.6 & 7.4 & 2.2 & 4.9 & 6.3 \\
\hline
\end{tabular}}
\end{center}
\end{table}

\begin{table}
\caption{\label{tab3:classif200-1} Classification rates (in percent) achieved by applying the F-type statistic for a known type $\delta$ and time $\tau$ of intervention  to INAR(2) series of length $n=200$ with a permanent shift ($\delta=1$) of size $\kappa=\sqrt{\lambda}$ at time $\tau$. The nominal significance levels are 1\%, 5\% or 10\%.}
\begin{center}
{\footnotesize
\begin{tabular}{rrrr|rrr|rrr|rrr}
\hline
 && & & \multicolumn{3}{c|}{$\tau=0.25n$}& \multicolumn{3}{c|}{$\tau=0.5n$}& \multicolumn{3}{c}{$\tau=0.75n$}  \\
$\alpha_1$ & $\alpha_2$ & $\lambda$ & $\delta$ &  1\% & 5\% & 10\%& 1\% & 5\% & 10\%& 1\% & 5\% & 10\% \\ \hline
0.5 & 0.3 & 2 & 0 & 0.3 & 1.4 & 2.0 & 0.7 & 1.3 & 1.5 & 0.5 & 1.4 & 1.7 \\ 
 & & & 0.6 & 0.4 & 0.9 & 1.1 & 0.4 & 0.8 & 0.8 & 0.5 & 1.0 & 1.1 \\ 
 & & & 0.8 & 0.2 & 0.6 & 0.9 & 0.3 & 0.4 & 0.5 & 0.4 & 0.6 & 0.9 \\ 
 & & & 0.9 & 0.7 & 1.7 & 2.2 & 0.5 & 1.6 & 1.8 & 4.9 & 6.4 & 6.8 \\ 
  & & & 1 & 40.1 & 69.4 & 80.6 & 65.5 & 86.2 & 90.8 & 59.5 & 77.5 & 82.4 \\ \hline
  & & 5 & 0 & 0.4 & 1.2 & 1.8 & 0.7 & 0.9 & 1.1 & 0.8 & 1.6 & 1.8 \\
\hline
\end{tabular}}
\end{center}
\end{table}

\begin{table}
\caption{\label{tab:sim-ex-inar2} Conditional least squares estimates obtained at each step of the stepwise procedure for the detection and elimination of intervention effects in the simulated INAR(2) time series. The final estimates of the Poisson INAR(2) model parameters are shown in bold. The true parameter values are $\alpha_1=0.3$, $\alpha_2=0.2$ and $\lambda=3$ and there are outliers with $\kappa=10$ and $\delta=0.6$ at $\tau=50$ as well as $\kappa=10$ and $\delta=0.9$ at $\tau=150$.}
{\small
\begin{center}
\begin{tabular}{cc|c|rrr|rrr}
\hline
Iteration & Step & Bootstrap  &\multicolumn{3}{c|}{Parameter estimates} & \multicolumn{3}{c}{Outlier}\\
& & p-value& $\hat{\alpha}_1$ & $\hat{\alpha}_2$ & $\hat{\lambda}$ & $\hat{\kappa}$ & $\hat{\tau}$ & $\hat{\delta}$\\
\hline
1 & 1 & & 0.38 & 0.18 & 2.81 & & & \\
& 2-3 & 0.002 & 0.28 & 0.15 & 3.47 & 7.39 & 150 & 0.8\\
\hline
2 & 1 & & 0.31 & 0.18 & 3.20 & & & \\
& 2-3 & 0.012 & 0.28 & 0.16 & 3.39 & 5.52 & 50 & 0.8\\
\hline
3 & 1 & & 0.27 & 0.16 & 3.49 & & &\\
& 2-3 & 0.036 & 0.29 & 0.16 & 3.27 & 7.49 & 46 & 0\\
\hline
4 & 1 & & \textbf{0.28} & \textbf{0.17} & \textbf{3.32} & & &\\
& 2-3 & 0.082 & - & - & - & - & - & -\\
\hline
\end{tabular}
\end{center}
}
\end{table}


\begin{thebibliography}{}
\bibitem[\protect\citeauthoryear{Al-Osh and Alzaid}{1987}]{alosh:1987}
Al-Osh, M.A. and Alzaid, A.A. (1987). First-order integer-valued autoregressive (INAR(1)) process. \textit{Journal of Time Series Analysis}, \textbf{8:} 261--275.

\bibitem[\protect\citeauthoryear{Barczy et al.}{2010}]{barczy:2010} 
 Barczy, M., Ispany, M., Pap, G., Scotto, M, and Silva M.E.  (2010).  Innovational outliers in INAR(1) Models. \textit{Communications in Statistics - Theory and Methods}, \textbf{39:} 3343--3362.

\bibitem[\protect\citeauthoryear{Barczy et al.}{2012}]{barczy:2012} 
 Barczy, M., Ispany, M., Pap, G., Scotto, M, and Silva M.E.  (2012).  Additive outliers in INAR(1) Models. \textit{Statistical Papers}, \textbf{53:} 935--949.
 
 \bibitem[\protect\citeauthoryear{Brännäs}{1994}]{brannas:1994}
Brännäs, K. (1994). Estimation and testing in integer valued AR(1) models. Umeå  Economic Studies, vol. 355, University of Umeå.
 
 \bibitem[\protect\citeauthoryear{Bu et al.}{2008}]{bu:2008}
 Bu, R., McCabe, B. and Hadri, K. (2008). Maximum likelihood estimation of higher-order integer-valued autoregressive processes. \textit{Journal of Time Series Analysis}, \textbf{29:} 973--994. 
 
\bibitem[\protect\citeauthoryear{Davis et al.}{2016}]{davis:2016}
Davis, R.A., Holan, S.H., Lund, R., and Ravishanker, N. (2016). \textit{Handbook of Discrete-valued Time Series}. CRC Press.

 \bibitem[\protect\citeauthoryear{Du and Li}{1991}]{du:1991}
 Du, J. and Li, Y. (1991). The integer-valued autoregressive (INAR($p$)) model. \textit{Journal of Time Series
Analysis}, \textbf{12:} 129--142.
 
\bibitem[\protect\citeauthoryear{Fokianos and Fried}{2010}]{fokianos:2010}
Fokianos, K. and Fried, R. (2010). Interventions in INGARCH processes. \textit{Journal of Time Series Analysis}, \textbf{31:} 210--225.

\bibitem[\protect\citeauthoryear{Fokianos and Fried}{2012}]{fokianos:2012}
Fokianos, K. and Fried, R. (2012). Interventions in log-linear Poisson autoregression. \textit{Statistical Modeling}, \textbf{12:} 299--322.

\bibitem[\protect\citeauthoryear{Fox}{1972}]{fox:1972}
Fox, A. J. (1972). Outliers in Time Series. \textit{Journal of the Royal Statistical Society, Series B}, \textbf{34:} 350--363.

 \bibitem[\protect\citeauthoryear{Franke and Seligmann}{1993}]{franke:1993}
 Franke, J. and Seligmann, T. (1993). \textit{Conditional maximum likelihood estimates for INAR(1) processes and their application to modelling epileptic seizure counts}. In: Developments in Time Series Analysis (ed. T. Subba Rao). Chapman and Hall, London. pp. 310--330.
 
 
\bibitem[\protect\citeauthoryear{Freeland and McCabe}{2004}]{freeland:2004}
Freeland, R.K. and McCabe, B.P.M. (2004). Analysis of low count time series data by Poisson autoregression.
\textit{Journal of Time Series Analysis}, \textbf{25:} 701--722.

\bibitem[\protect\citeauthoryear{Fried}{2015}]{fried:2015}
Fried, R., Aguesop, I., Bornkamp, B., Fokianos, K., Fruth, J., Ickstadt, K. (2015). Retrospective Bayesian outlier detection in INGARCH series. \textit{ Statistics and Computing}, \textbf{25:} 365--374.

\bibitem[\protect\citeauthoryear{Galeano and Peña}{2012}]{galeano:2012}
Galeano, P., and Peña, D. (2012). \textit{Additive outlier detection in seasonal ARIMA models by a modified
Bayesian information criterion}. In W. R. Bell, S. H. Holan, and T. S. McElroy (Eds.), Economic
time series: modeling and seasonality (pp. 317--336). Boca Raton: Chapman \& Hall.

\bibitem[\protect\citeauthoryear{Galeano and Peña}{2013}]{galeano:2013}
Galeano, P., and Peña, D. (2013). \textit{Finding outliers in linear and nonlinear time series}. In: Becker, C., Fried, R., Kuhnt, S. (eds.) Robustness and Complex Data Structures, pp. 243--260. Springer, Heidelberg.

\bibitem[\protect\citeauthoryear{Hamilton}{1994}]{hamilton:1994}
Hamilton, J. D. (1994). \textit{Time series analysis},  pp. 206--207. Princeton, N.J: Princeton University Press.



\bibitem[\protect\citeauthoryear{Karagiannis et al.}{2012}]{karagiannis:2012}
Karagiannis, I., Mellou, K., Gkolfinopoulou, K., Dougas, G., Theocharopoulos, G., Vourvidis, D., Ellinas, D., Sotolidou, M., Papadimitriou, T., Vorou, R. (2012). Outbreak investigation of Brucellosis in Thassos, Greece, 2008.
\textit{Euro Surveillance}, \textbf{17(11):} 20116. 


\bibitem[\protect\citeauthoryear{Liboschik et al.}{2017}]{liboschik:2017}
Liboschik, T., Fokianos, K.,  Fried, R. (2017). tscount: An R Package for Analysis of Count Time Series Following Generalized Linear Models. \textit{Journal of Statistical Software}, \textbf{82(5):} 1--51.

\bibitem[\protect\citeauthoryear{Lu}{2021}]{lu:2018}
Lu, Y. (2021). The predictive distributions of thinning‐based count processes. \textit{Scandinavian Journal of Statistics}, \textbf{48:} 42--67.  

\bibitem[\protect\citeauthoryear{McKenzie}{1985}]{mckenzie:1985}
McKenzie, E. (1985). Some simple models for discrete variate time series. \textit{Journal of the American Water Resources Association}, \textbf{21:} 645--650.

\bibitem[\protect\citeauthoryear{Moriña et al.}{2020}]{morina:2020}
Moriña, D., Leyva-Moral, J.M. and Feijoo-Cid, M. (2020). Intervention analysis for low-count time series with applications in public health. \textit{Statistical Modelling}, \textbf{20:} 58--70.

\bibitem[\protect\citeauthoryear{Pedeli et al.}{2015}]{pedeli:2015}
Pedeli, X., Davison, A. C., and Fokianos, K. (2015). Likelihood Estimation for the INAR($p$) Model by Saddlepoint Approximation. \textit{Journal of the American Statistical Association}, \textbf{110:} 1229--1238.

\bibitem[\protect\citeauthoryear{Rossetti et al.}{2017}]{rossetti:2017}
Rossetti, C.A., Arenas-Gamboa, A.M., Maurizio, E. (2017). Caprine Brucellosis: A historically neglected disease with significant
impact on public health. \textit{PLoS Neglected Tropical Diseases} \textbf{11(8):} e0005692. 

\bibitem[\protect\citeauthoryear{Silva and Pereira}{2015}]{silva:2015}
Silva M.E., and Pereira, I. (2015). \textit{Detection of Additive Outliers in Poisson INAR(1) Time Series}. In: Bourguignon, J.P. et al. (eds.) Mathematics of Energy and Climate Change. CIM Series in Mathematical Sciences, pp. 377–388. Springer, Berlin.

\bibitem[\protect\citeauthoryear{Steutel and van Harn}{1979}]{steutel:1979}
Steutel, F. W., and van Harn, K. (1979). Discrete Analogues of Self–Decomposability and Stability. \textit{The Annals of Probability}, \textbf{7:} 893--899.

\bibitem[\protect\citeauthoryear{Weiss}{2018}]{weiss:2018}
Weiss, C.H. (2018). \textit{An Introduction to Discrete-Valued Time Series}. Chichester: Wiley.

\end{thebibliography}
\end{document}